%% file: eptcs3.tex
\newif\ifarxiv
\arxivtrue

\documentclass[submission,copyright,creativecommons]{eptcs}
\providecommand{\event}{QPL 2017}

\input{header-eptcs}

\usepackage{underscore}

    \tikzset{every picture/.style={scale=1.3}}
    \def\figuretopsuck{\vspace{-10pt}}
    \def\figurecaptionsuck{\vspace{-15pt}}
        \def\figurecaptionpostsuck{\vspace{-15pt}}

    \newcommand\justarxiv[1]{#1}
    \newcommand\justconf[1]{}
    \newcommand{\beginfig}{\begin{figure}[b!]}
        \allowdisplaybreaks[1]
\begin{document}

\title{\justarxiv{\bf} Shaded Tangles for the Design and\\Verification of Quantum Programs (Extended Abstract)}

\author{
\ifarxiv
\begin{tabular}{cc}
David Reutter & Jamie Vicary
\\
Department of Computer Science & Department of Computer Science
\\
University of Oxford & University of Oxford
\\
\texttt{david.reutter@cs.ox.ac.uk}
&
\texttt{jamie.vicary@cs.ox.ac.uk}
\end{tabular}
\else
\IEEEauthorblockN{David Reutter}
\IEEEauthorblockA{Department of Computer Science,
University of Oxford
\\
\texttt{david.reutter@cs.ox.ac.uk}}
\and
\IEEEauthorblockN{Jamie Vicary}
\IEEEauthorblockA{Department of Computer Science, University of Oxford
\\
\texttt{jamie.vicary@cs.ox.ac.uk}}
\fi}

\date{\today}

\def\titlerunning{Shaded Tangles for the Design and Verification of Quantum Programs}
\def\authorrunning{D. Reutter \& J. Vicary}

\maketitle 

\begin{abstract}
We give a scheme for interpreting shaded tangles as quantum programs, with the property that isotopic tangles yield equivalent programs. We analyze many known quantum programs in this way---including entanglement manipulation and error correction---and in each case present a fully-topological formal verification, yielding in several cases substantial new insight into how the program works. We also use our methods to identify several new or generalized procedures.
\end{abstract}

%\tableofcontents

\section{Introduction}

\subsection{Overview}

\begin{figure}[b]
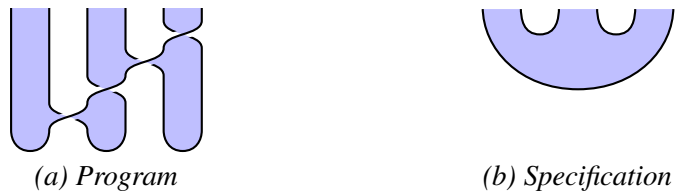

\figuretopsuck
\figuretopsuck
\figuretopsuck
\def\scl{0.65}
\def\yscl{0.9}
\def\maskscl{1.2}
\begin{calign}
\nonumber
\begin{tz}[scale=0.6,yscale=0.8,scale=\scl,yscale=\yscl]
\path[surface,even odd rule] (0,5.5) to (0,1) to [out=down, in=left] (0.5,0.25) to [out=right, in=down] (1,1) to [out=up, in=down] (2,2) to [out=up, in=down] (3,3) to [out=up, in=down] (4,4) to [out=up, in=down] (5,5) to (5,5.5)
 (1,5.5) to (1,2) to [out=down, in=up] (2,1) to [out=down, in=left] (2.5,0.25) to [out=right, in=down] (3,1)to (3,2) to [out=up, in=down] (2,3) to (2,5.5)
 (3,5.5) to (3,4) to [out=down, in=up] (4,3) to (4,1) to [out=down, in=left] (4.5,0.25) to [out=right, in=down] (5,1) to (5,4) to [out=up, in=down] (4,5) to (4,5.5);
\draw[edge] (0,5.5) to (0,1) to [out=down, in=left] (0.5,0.25) to [out=right, in=down] (1,1) to [out=up, in=down]node[mask point, scale=\maskscl, pos=0.5] (1){} (2,2) to [out=up, in=down] node[mask point, scale=\maskscl, pos=0.5] (2){} (3,3) to [out=up, in=down] node[mask point, scale=\maskscl, pos=0.5] (3){} (4,4) to [out=up, in=down] node[mask point, scale=\maskscl, pos=0.5] (4){} (5,5) to (5,5.5); 
\cliparoundtwo{1}{2}{\draw[edge] (1,5.5) to (1,2) to [out=down, in=up] (2,1) to [out=down, in=left] (2.5,0.25) to [out=right, in=down] (3,1)to (3,2) to [out=up, in=down] (2,3) to (2,5.5);}
\cliparoundtwo{3}{4}{
\draw[edge] (3,5.5) to (3,4) to [out=down, in=up] (4,3) to (4,1) to [out=down, in=left] (4.5,0.25) to [out=right, in=down] (5,1) to (5,4) to [out=up, in=down] (4,5) to (4,5.5) ;}
\tidythetop
\end{tz}
&
\begin{tz}[scale=0.6,yscale=0.8,scale=\scl,yscale=\yscl]
\clip (-0.1,0.25) rectangle (5.1,5.5);
\path[surface,even odd rule,edge] (0,5.5) to [out=down, in=left] (2.5,2.5) to [out=right, in=down] (5,5.5)
(1,5.5) to [out=down, in=left](1.5,4.5) to [out=right, in=down](2,5.5 )
(3,5.5) to [out=down, in=left](3.5,4.5) to [out=right, in=down](4,5.5 );
\tidythetop
\end{tz}
\\[-5pt]
\nonumber
\vc{\textit{(a) Program}}
&
\vc{\textit{(b) Specification}}
\end{calign}

\figurecaptionsuck
\caption{Shaded tangles giving the program and specification for constructing a tripartite GHZ state.\label{fig:exampletangles}}
\figurecaptionpostsuck
\figurecaptionpostsuck
\end{figure}

\noindent
In this paper we introduce a new knot-based language for designing and verifying quantum programs. Terms in this language are \textit{shaded tangles}, which look like traditional knot diagrams, possibly involving multiple strings and strings with open ends, and decorated with a shading pattern. Examples of shaded tangles are given in \autoref{fig:exampletangles}. 

%Terms in this language are \textit{tangle diagrams}, two-dimensional images that represent the embedding of a finite number of closed or open wires into $\mathbb R^3$, equipped with a \textit{shading} satisfying a simple property. Examples of shaded tangles are given in \autoref{fig:exampletangles}. 
\begin{figure}[b]
\figuretopsuck
\def\lscl{0.475}
\[
\begin{array}{@{}c@{}c@{}c@{}c@{}}
\nonumber
\begin{tz}[scale=\lscl]
\draw [surface] (0,0) rectangle (1.5,2);
\draw [edge] (0,0) to +(0,2);
\draw [edge] (1.5,0) to +(0,2);
\end{tz}
&
\begin{tz}[scale=\lscl]
\draw [surface, edge] (0,0) to [out=down, in=down, looseness=2] (1.5,0);
\path (0,0) rectangle (1.5,-2);
\end{tz}
&
\begin{tz}[scale=\lscl,yscale=1]
\path[surface] (0.25,0) to [out=up, in=-135] (1,1) to [out=-45, in=up] (1.75,0);
\path[surface] (0.25,2) to [out=down, in=135] (1,1) to [out=45, in=down] (1.75,2);
\draw[edge] (0.25,0) to [out= up, in=-135] node[mask point, pos=1](1){} (1,1) to[out=45, in=down] (1.75,2);
\cliparoundone{1}{\draw[edge](1.75,0) to [out=up, in=-45] (1,1) to [out=135, in=down] (0.25,2);}
\end{tz}
&
\begin{tz}[scale=\lscl]
\path[surface,even odd rule] 
(0.25,0) to [out= up, in=-135]  (1,1) to [out= 45, in=down] (1.75,2) to (-0.75,2) to (-0.75,0) to (0.25,0)
(1.75,0) to [out= up, in=-45] (1,1) to [out= 135, in=down] (0.25,2) to (2.75,2) to (2.75,0) to (1.75,0);
\draw[edge] (0.25,0) to [out= up, in=-135] node[mask point, pos=1](1){} (1,1) to[out=45, in=down] (1.75,2);
\cliparoundone{1}{\draw[edge](1.75,0) to [out=up, in=-45] (1,1) to [out=135, in=down] (0.25,2);}
\draw[edge] (-0.75,0) to +(0,2);
\draw[edge] (2.75,0) to + (0,2);
\tidythetop
\end{tz}
\ignore{\\
\nonumber
\C^2
&
\ket 0 + \ket 1
&
\scriptsize
{\scriptstyle e ^{-i \pi/8}}
\left(
\begin{array}{@{}c@{\,\,\,}c@{}}
1 & i \\ i & 1
\end{array}
\right)
&
\scriptsize
{\scriptstyle e ^{-i \pi/8}}
\left(
\begin{array}{@{}c@{\,\,\,}c@{\,\,\,}c@{\,\,\,}c@{}}
1 & 0 & 0 & 0
\\
0 & i & 0 & 0
\\
0 & 0 & i & 0
\\
0 & 0 & 0 & 1
\end{array}
\right)}
\\\nonumber\vc{\textit{(a) Qudit}}&\vc{\textit{(b) Qudit preparation}}&\vc{\textit{(c) 1-qudit gate}}&\vc{\textit{(d) 2-qudit gate}}
\end{array}
\]

\figurecaptionsuck
\caption{Part of the graphical language along with its interpretation in terms of quantum structures.\label{fig:language}}
\figurecaptionpostsuck
\end{figure}

We give an operational semantics in which a shaded tangle is interpreted as a linear map between Hilbert spaces. Since this is the basic mathematical foundation for quantum information, this enables us to interpret our shaded tangles as \textit{quantum programs}. Under this interpretation, we read our shaded tangles as quantum circuits, with time flowing from bottom to top, and with individual geometrical features of the diagrams---such as shaded regions, cups and caps, and crossings---interpreted as distinct quantum circuit components, such as qudits\footnote{A \emph{qudit} is a $d$-dimensional quantum system; a \textit{qubit} is a qudit for $d=2$.}, qudit preparations, and 1- and 2\-qudit gates (see \autoref{fig:language} for this part of the graphical language.)

Given two shaded tangles with the same shading pattern on their boundaries, we say they are \textit{isotopic} just when, ignoring shading and considering them as ordinary knotted strings, one can be deformed topologically into the other. We show that our semantics is \textit{sound} with respect to this isotopy relation: that is, if two shaded tangles are isotopic, then they have equal interpretations as quantum programs.

This yields a powerful method for the design and verification of quantum procedures. We draw one shaded tangle for the \textit{program}, describing the exact steps the quantum computer would perform, and another shaded tangle for the \textit{specification}, describing the intended computational effect. The program is then verified simply by showing that the two shaded tangles are isotopic. Since humans have an innate skill for visualizing knot isotopy, this verification procedure can often be performed immediately by eye, even in sophisticated cases.  We illustrate this idea in \autoref{fig:exampletangles}, which illustrates the program and specification for constructing a \textit{GHZ state}, an important primitive resource in quantum information. It can be seen by inspection that, ignoring shading, the tangles are isotopic, and hence the program is correct. By reference to \autoref{fig:language}, we see that the program \autoref{fig:exampletangles}(a) involves three qudit preparations, two 1-qudit gates, and two 2-qudit gates.

\subsection{Main results}
\label{sec:mainresults}

\noindent
In our main results, we apply this new high-level technique to represent and verify 9 quantum programs, some generalized from their form in the literature, and some completely new. \ignore{We give a summary here of the programs we analyze.}
\begin{itemize}
\item \Autoref{sec:creatingghz}. A generalization of a program due to Uchida et al~\cite{Uchida:2015} for constructing GHZ states.
\ignore{
\item \Autoref{sec:creatingcluster}. Programs due to Raussendorf and others~\cite{Raussendorf:2001, Cui:2015} for constructing \textit{cluster chains}~\cite{Hein:2006}, resources of central importance in quantum information.
}%
\item \Autoref{sec:localequivalence}. Programs due to Briegel et al~\cite{Briegel:2001,Cui:2015} for converting certain GHZ and cluster states.\looseness=-1
\item \Autoref{sec:cutting, sec:splicing}. Programs related to those of van den Nest and others~\cite{vandenNest:2004} for cutting and splicing cluster chains, which play an important role in measurement-based quantum computation~\cite{Hein:2006}. We  present novel versions of  these programs for qudits.\ignore{ and new classes of cluster chains.}
\ignore{
\item \autoref{sec:statetransfer}. A new program for robust state transfer in a cluster state--based quantum computer.}
%\item \autoref{sec:GHZteleportation}. A program due to X~\cite{} for teleportation along a GHZ state.
\ignore{
\item \autoref{sec:MBGHZteleportation}. A generalization of programs due to Karlsson, Hillery, Grudka and others~\cite{Karlsson:1998,Hillery:1999,Grudka:2002} for measurement-based teleportation along a GHZ state.
\item \autoref{sec:clusterteleportation}. A generalization of a program due to Raussendorf and Briegel~\cite{Raussendorf:2001} for measurement-based teleportation along a cluster chain.
\item \autoref{sec:robustteleportation}. A program for teleportation along an $n$\-party GHZ state, which we believe is folklore, with a new  robustness property against a broad family of errors.
\item \autoref{sec:compressedteleportation}. A program due to Yu, Jaffe and others~\cite{Yu:2010, Jaffe:2016b} which reduces resource requirements for executing a distributed controlled quantum gate.
}%
\item \Autoref{sec:phasecode, sec:shorcode}. The phase code and the Shor code~\cite{Shor:1995,Ke:2010,Nielsen:2009}, important error correcting codes in quantum information, which are built from Hadamard matrices.\footnote{A \textit{Hadamard} is a unitary matrix with all coefficients having the same absolute value. Hadamard matrices are important primitive structures in quantum information, playing a central role in quantum key distribution and many other phenomena~\cite{Durt:2010}.}
\item \Autoref{sec:uebcodes}. New generalizations of the phase code and Shor code, based on unitary error bases\footnote{A \textit{unitary error basis} is a basis of unitary operators on a finite-dimensional Hilbert space, orthogonal with respect to the trace inner product. They provide the basic data for all quantum teleportation and dense coding procedures~\cite{Werner:2001}, and some error correction procedures~\cite{Knill:1996-2, Shor:1996}.} rather than Hadamard matrices.
\end{itemize}

\subsection{Significance}

\noindent
We outline some areas of potential significance of our work.

\paragraph{Novelty.} Some of the programs we verify are generalizations of those described in the literature, or are completely new. Perhaps most significantly, we highlight the new constructions of error correcting codes based on unitary error bases (\autoref{sec:uebcodes}), and our identification of the $\frac \pi 2$-rotations around the $X$\-axis on the Bloch sphere as having privileged topological properties  among all qubit Hadamards (\autoref{sec:programsandspecifications}).

\paragraph{Insight.} Throughout, the shaded tangle syntax gives considerable new insight into why each procedure works. For example, in our verification of error correcting codes the errors are literally `trapped by bubbles' and removed from the diagram, and in our verification of cluster chain surgery procedures the qubits are literally untangled from the chain. In both cases, this gives a powerful intuition for these schemes which we believe to be new.% This stands in contrast to traditional verification methods in quantum computer science~\cite{Nielsen:2009}, where a program is often given as a series of linear maps encoded algebraically (for example, as matrices of complex numbers), and verification involves composing the maps and examining the result; from this perspective, high-level structure can be difficult to perceive, and it may be unclear whether a program can be generalized.

\paragraph{Efficiency.} Where our methods apply, we can often give the program, specification and verification in a concise way; compare for example our discussion of \autoref{fig:exampletangles} above with the traditional verification of a related program due to Uchida et al~\cite{Uchida:2015}, which requires  a page of algebra, and is also less general. As a consequence, even in this short extended abstract, we are able to give detailed analyses of 9 distinct procedures. We suggest that our methods would therefore be suitable for reasoning about large-scale quantum programs, such as architectures for quantum computers.

\subsection{Criticism}

\firstparagraph{Completeness.}
We define our semantics to be \textit{sound} if topological isotopy implies computational equivalence, and \textit{complete} if computational equivalence implies topological isotopy. The main semantics we give is sound, allowing the verification method for quantum programs that we use throughout the paper. However, it is not complete, meaning that there exist quantum programs that cannot be verified by our methods.\footnote{Our language is also not \textit{universal}, meaning that not all quantum programs can be constructed. It would be easy to make it universal by adding additional 1-qubit generators; however, without completeness, this has limited value.} Achieving completeness will be a focus of future work. We note that the \textit{ZX calculus} (see~\autoref{sec:relatedwork}), a dominant existing high-level approach to quantum information, shares this property of being sound but not complete in general~\cite{Witt:2014}, although it is complete for the stabilizer fragment~\cite{Backens:2014}.

\paragraph{Algorithms.}
All of our examples are in the broad area of quantum communication; we do not study quantum algorithms, such as Grover's or Shor's algorithms~\cite{Nielsen:2009}. These algorithms have been analyzed in the related CQM approach~\cite{Vicary:2013}; in future work we aim to analyze them using our new syntax.

\subsection{Related work}
\label{sec:relatedwork}

\firstparagraph{Categorical quantum mechanics (CQM).} Our work emerges from the CQM research programme, initiated by Abramsky and Coecke~\cite{Abramsky:2004} and developed by them and others~\cite{Abramsky:2009, Backens:2014, Coecke:2006, Coecke:2008, Coecke:2012, Coecke:2012b, Coecke:2014, Coecke:2017, Duncan:2014b, Hadzihasanovic:2015, Kissinger:2015, Selinger:2007, Vicary:2013}, which uses monoidal categories with duals to provide a high-level language for quantum programs, using in particular a graph-based language called the \textit{ZX calculus}~\cite{Coecke:2008, Coecke:2012b}. CQM verifications have been given for some programs related to those we analyze, including the Steane code~\cite{Duncan:2014}, and cluster state arguments~\cite{Coecke:2008, Duncan:2014b}. Many of the advantages of our calculus over traditional techniques---such as the power of the diagrammatic language, and its topological flavour---inherit directly from the CQM programme.\looseness=-1

The current authors have previously shown that CQM methods can be extended to a higher-categorical setting~\cite{Vicary:2012, Vicary:2012hq, Reutter:2016}, developing the work of Baez on a categorified notion of Hilbert space~\cite{Baez:1997}, and this paper develops these ideas further.

We give here some important points of distinction between traditional CQM techniques and our present work. Unlike the ZX calculus, our calculus is purely topological, and hence complex deductions can sometimes be perceived by eye in a single step.
%In common with \cite{{Vicary:2012, Vicary:2012hq}}, our methods can capture an entire quantum program, including classical control, in a single diagram; in CQM, in contrast, diagrams are sometimes decorated by variables indexing these outcomes (see for example~\cite[Section 2.1]{Abramsky:2009}, where the experimental outcome is labelled $x$), and so the formalism is not fully diagrammatic.
Also, our calculus is incomparable in strength to the ZX calculus, which is restricted (in its basic form) to Clifford quantum theory; neither calculus can simulate the other in general. As a result,  we are able to analyze many protocols that have not previously been analyzed with ZX methods, as well as discover a number of new and generalized protocols.
 
%Where similar protocols have previously been analyzed, our methods can be more efficient: for example, the CQM verification of the Steane code~\cite{Duncan:2014}, an important error correction procedure in quantum information, runs over many pages and required computer support to generate due to the extremely large size of the proof terms, in contrast to the short verifications of error correcting codes that we give in \autoref{sec:errorcorrection}.
%Finally, as we emphasize in \autoref{sec:mainresults}, we are able to discover a significant number of generalized and novel quantum programs, which have not previously been identified using CQM\ techniques.

%\paragraph{Quantum programming languages (QPL).}\DRcomm{Maybe this connection is a bit too far fetched, what do you think? Also some typos + grammar issues in this section} Recent have seen increased work on \textit{quantum programming languages}~\cite{Gay:2006, Green:2013}, which are conventional programming languages with additional features or libraries allowing the programmer to write code to control quantum hardware. Our work is complementary, providing a way to verify correctness of a program once they been formally specified; in the QPL  community, in contrast, verification has not so far been a strong focus.

\paragraph{Statistical mechanics.}  There is a rich interplay between quantum information (QI), knot theory (KT) and statistical mechanics (SM). The KT-SM and SM-QI relationships are quite well-explored in the literature, unlike the KT-QI relationship, which is our focus here.

The KT-SM relationship was first studied by Kauffman, Jones and others~\cite{Bannai:1995, Jones:1989, Kauffman:1988}, who showed how to obtain knot invariants from certain statistical mechanical models. Much of the mathematical foundations of our paper are already present in the paper~\cite{Jones:1989}, including the shaded knot notation. Work on the SM-QI relationship has focused on finding efficient quantum algorithms for approximating partition functions of statistical mechanical systems~\cite{Aharonov:2008, Aharonov:2007,vandenNest:2009,Arad:2010}, for which the best known classical algorithm is often exponential.`Chaining' these relationships allows one to obtain a statistical mechanical model from a knot, and then write down a quantum circuit approximating the model's partition function, giving overall a mapping from knots to quantum circuits, which closely matches our construction.
%The novelty here is of course the large range of new and existing quantum programs which we are able to analyze using this knot-theoretic notation.

The direct KT-QI relationship has also been emphasized by Kauffman and collaborators~\cite{Kauffman:2002, Kauffman:2009}, and also in the field of topological quantum computing~\cite{Wang:2010,Panangaden:2010}, where (as here) a strong analogy is developed between topological and quantum entanglement, although the technical details are quite different. %However, we do not yet see a direct technical connection between that work and ours.

\paragraph{Planar algebras.} The graphical notation we employ can be described formally as a \textit{shaded planar algebra}, although we do not use that terminology in this paper, preferring a more elementary presentation. The relationship between shaded planar algebras and Hadamard matrices was first suggested by Jones~\cite{Jones:1999}, and developed by the present authors~\cite{Vicary:2012, Vicary:2012hq, Reutter:2016}.

Recently, Jaffe, Liu and Wozniakowski have described a related approach to quantum information based on \textit{planar para algebras}~\cite{Jaffe:2016d, Jaffe:2016a, Jaffe:2016b,Jaffe:2017}. 
\ignore{There are several points of distinction that can be drawn between our work and theirs. Firstly, their planar para algebra setting is quite different to the mathematical structure we use. Secondly, their diagrams must be decorated with additional indices encoding measurement results, while our diagrams do not require such indices. Thirdly, there is little overlap between the quantum procedures analyzed so far in each setting.}

\paragraph{Classical verification.} There has been some work on using knot-theoretic methods for verification in linear logic~\cite{Mellies:2009, Dunn:2016} and separation logic~\cite{Wickerson:2013}. We do not see a direct technical connection to our results, although we expect this to be a  fruitful direction for further investigation.

\subsection*{Acknowledgements}

\noindent
We thank Paul-Andr\'e Mellies for suggesting the shaded tangle representation, Amar Hadzihasanovic for detailed conversations, Matty Hoban, Nathan Bowler and Niel de Beaudrap for help with  cluster states, and Arthur Jaffe, Zhengwei Liu and Alex Wozniakowski for discussions about planar para algebras.

\beginfig
\figuretopsuck
\figuretopsuck
\def\scl{0.9}%
\def\yscl{0.9}%
\begin{calign}
\nonumber
\begin{tz}[edge, yscale=1, xscale=1.3,scale=\scl,yscale=\yscl]
\node[blob, edge] at (1.5,0) {$L$};
\node[blob, edge] at (1,-1) {$M$};
\node[blob, edge] at (1.5,-2) {$N$};
\node[scale=0.8, right] at (1.5,0.5) {$A$};
\node[scale=0.8] at (1.5, 0) {$L$};
\node[scale=0.8] at (2.1, -1) {$C$};
\node[scale=0.8, left] at (0.5, -2.5) {$E$};
\node[scale=0.8, right] at (1.5,-2.5) {$F$};
\node[] at (1.5,-1) {$T$};
\node[] at (1,-2.25) {$U$};
%\node[dimension, above left] at (2.4,-1.9) {$k{:}m$};
\node[scale=1] at (0.25,-1) {$S$};
\path[blueregion] (-0.25,-3) to (0.5,-3) to [out=up, in=-135] (1,-1) to [out=up, in=-135] (1.5,0) to (1.5,1) to (-0.25,1);
\draw [redregion] (1.5,0) to [out=-45, in=45] (1.5,-2) to [out=135, in=-45] (1,-1) to [out=up, in=-135] (1.5,0);
\draw [greenregion] (0.5,-3) to [out=up, in=-135] (1,-1) to [out=-45, in=135] (1.5,-2) to (1.5,-3);
\draw [edge] (0.5,-3) to [out=up, in=-135] (1,-1) to [out=up, in=-135] node [left, scale=0.8, pos=0.65] {$B$} (1.5,0) to (1.5,1);
\draw [edge] (1,-1) to [out=-45, in=135] node [left, scale=0.8] {$D$} (1.5,-2) to (1.5,-3);
\draw [edge] (1.5,0) to [out=-45, in=45] (1.5,-2);
\end{tz}
&
\,
\sum_{t \in T} \,
\raisebox{-16pt}{$\tikz{\node at (0,0) {$\displaystyle\bigoplus_{u \in U}$};}$} \!\!\!\!
\begin{tz}[string, yscale=1, xscale=1.3,scale=\scl,yscale=\yscl]
\node[blob, edge] at (1.5,0) {$L_{st}$};
\node[blob, inner sep=0pt, edge] at (1,-1) {$M_{stu}$};
\node[blob, edge] at (1.5,-2) {$N_{ut}$};
\node[scale=0.8, right] at (1.5,0.6) {$A_s$};
\node[scale=0.8] at (1.5, 0) {$L$};
\node[scale=0.8] at (2.1, -1) {$C_t$};
\node[scale=0.8, right] at (0.5, -2.5) {$E_{su}$};
\node[scale=0.8, right] at (1.5,-2.5) {$F_u$};
%\node[dimension, above left] at (2.4,-1.9) {$k{:}m$};
\draw [edge] (0.5,-3) to [out=up, in=-135] (1,-1) to [out=up, in=-135] node [left=2pt, scale=0.8, pos=0.7] {$B_{st}$} (1.5,0) to (1.5,1);
\draw [edge] (1,-1) to [out=-45, in=135] node [left, scale=0.8, pos=0.7] {$D_{ut}$} (1.5,-2) to (1.5,-3);
\draw [edge] (1.5,0) to [out=-45, in=45] (1.5,-2);
\end{tz}
\\\nonumber
\vc{\em(a) A diagram in our\\\em generalized calculus.}
&
\vc{\em(b) Its interpretation\\\em for any $s \in S$.}
\end{calign}

\figurecaptionsuck
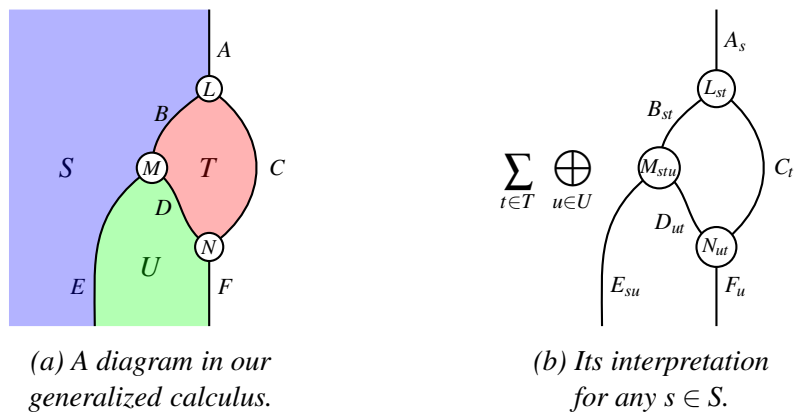
\caption{The generalized graphical calculus.}
\figurecaptionpostsuck
\figurecaptionpostsuck
\label{fig:graphicalcalculus}
\end{figure}
\section{Mathematical foundations}
\label{sec:mathematical}

\subsection{Graphical calculus}
\label{sec:graphicalcalculus}

\noindent
The graphical calculus for describing composition of multilinear maps was proposed by Penrose~\cite{Penrose:1971}, and is today widely used~\cite{Selinger:2010, Abramsky:2009, Coecke:2006, Joyal:1991, Orus:2014}. In this scheme, wires represent Hilbert spaces and vertices represent multilinear maps between them, with wiring diagrams representing composite linear maps.

In this article we use a generalized calculus that involves \emph{regions}, as well as wires and vertices; see \autoref{fig:graphicalcalculus}(a) for an example. This is an instance of the graphical calculus for 2\-categories\footnote{Here and throughout, we use the term `2-category' to refer to the weak structure, which is sometimes called `bicategory'.}~\cite{Barrett:2012, Bartlett:2014, Hummon:2012, SchommerPries:2011} applied to the 2\-category \cat{2Hilb} of finite-dimensional 2\-Hilbert spaces~\cite{Baez:1997}. The 2\-category of 2\-Hilbert spaces can be described as follows~\cite{Elgueta:2007,Vicary:2012hq}:
\begin{itemize}
  \item objects are natural numbers;
  \item 1-morphisms are matrices of finite-dimensional Hilbert spaces;
  \item 2-morphisms are matrices of linear maps.
\end{itemize}
We represent composite 2-morphisms in this 2\-category using a graphical notation involving regions, wires and vertices, which represent objects, 1\-morphisms and 2\-morphisms respectively.

\beginfig
\figuretopsuck
\figuretopsuck
{\tikzset{every picture/.style={scale=0.8}}
\begin{calign}
\nonumber
\text{\textit{(a)}}\,
\begin{tz}[scale=1,xscale=-1]
\draw [blueregion] (0,0.25) to (0,1) to [out=up, in=up, looseness=2] (-0.5,1) to [out=down, in=down, looseness=2] (-1,1) to (-1,1.75) to (0.5,1.75) to (0.5,0.25);
%\path[redregion] (0,0.25) to (0,1) to [out=up, in=up, looseness=2] (-0.5,1) to [out=down, in=down, looseness=2] (-1,1) to (-1,1.75) to (-1.5,1.75) to (-1.5, 0.25);
\draw [edge] (0,0.25) to (0,1) to [out=up, in=up, looseness=2] (-0.5,1) to [out=down, in=down, looseness=2] (-1,1) to (-1,1.75);
\end{tz}
=
\begin{tz}[scale=1,xscale=-1]
\draw [blueregion] (0,0.25) to (0,1.75) to (0.5,1.75) to (0.5,0.25);
%\draw[redregion] (0,0.25) to (0,1.75) to (-0.5,1.75) to (-0.5,0.25);
\draw [edge] (0,0.25) to (0,1.75);
\end{tz}
=
\begin{tz}[scale=1,scale=-1]
\draw [blueregion] (0,0.25) to (0,1) to [out=up, in=up, looseness=2] (-0.5,1) to [out=down, in=down, looseness=2] (-1,1) to (-1,1.75) to (0.5,1.75) to (0.5,0.25);
%\path[redregion] (0,0.25) to (0,1) to [out=up, in=up, looseness=2] (-0.5,1) to [out=down, in=down, looseness=2] (-1,1) to (-1,1.75) to (-1.5,1.75) to (-1.5, 0.25);
\draw [edge] (0,0.25) to (0,1) to [out=up, in=up, looseness=2] (-0.5,1) to [out=down, in=down, looseness=2] (-1,1) to (-1,1.75);
\end{tz}
&
\text{\textit{(b)}}\,
\begin{tz}[scale=1]
\draw [blueregion] (0,0.25) to (0,1) to [out=up, in=up, looseness=2] (-0.5,1) to [out=down, in=down, looseness=2] (-1,1) to (-1,1.75) to (0.5,1.75) to (0.5,0.25);
%\path[redregion] (0,0.25) to (0,1) to [out=up, in=up, looseness=2] (-0.5,1) to [out=down, in=down, looseness=2] (-1,1) to (-1,1.75) to (-1.5,1.75) to (-1.5, 0.25);
\draw [edge] (0,0.25) to (0,1) to [out=up, in=up, looseness=2] (-0.5,1) to [out=down, in=down, looseness=2] (-1,1) to (-1,1.75);
\end{tz}
=
\begin{tz}[scale=1]
\draw [blueregion] (0,0.25) to (0,1.75) to (0.5,1.75) to (0.5,0.25);
%\draw[redregion] (0,0.25) to (0,1.75) to (-0.5,1.75) to (-0.5,0.25);
\draw [edge] (0,0.25) to (0,1.75);
\end{tz}
=
\begin{tz}[scale=1,yscale=-1]
\draw [blueregion] (0,0.25) to (0,1) to [out=up, in=up, looseness=2] (-0.5,1) to [out=down, in=down, looseness=2] (-1,1) to (-1,1.75) to (0.5,1.75) to (0.5,0.25);
%\path[redregion] (0,0.25) to (0,1) to [out=up, in=up, looseness=2] (-0.5,1) to [out=down, in=down, looseness=2] (-1,1) to (-1,1.75) to (-1.5,1.75) to (-1.5, 0.25);
\draw [edge] (0,0.25) to (0,1) to [out=up, in=up, looseness=2] (-0.5,1) to [out=down, in=down, looseness=2] (-1,1) to (-1,1.75);
\end{tz}
&
\text{\textit{(c)}}\,
\begin{tz}[edge, scale=0.65, scale=1.17]
\path[blueregion] (-0.8,-1) rectangle (0.8,1);
\path[ fill=white,edge] (0,0) circle (0.5cm);
\end{tz}
=
\begin{tz}[edge, scale=0.65,scale=1.17]
\path[blueregion] (-0.9,-1) rectangle (0.5,1);
\end{tz}
&
\text{\textit{(d)}}\,
\begin{tz}[scale=0.75]
\draw [blueregion, edge] (0,0) circle (0.5cm);
\end{tz}
=
|S|
\end{calign}
}%
\vspace{-5pt}
{\tikzset{every picture/.style={scale=1}}%
\begin{calign}
\nonumber
\text{\textit{(e)}}\,
\begin{tz}[xscale=0.66, scale=0.45,scale=1.17]
\draw [surface] (0,0) to [out=up, in=down] (-1,1.5) to [out=up, in=down] (0,3) to (1,3) to [out=down, in=up] (2,1.5) to [out=down, in=up] (1,0);
\draw [edge] (0,0) to [out=up, in=down] (-1,1.5) to [out=up, in=down] (0,3);
\draw [edge] (1,0) to [out=up, in=down] (2,1.5) to [out=up, in=down] (1,3);
\draw [edge, fill=white] (0.5,1.5) ellipse (1 and 0.666);
\node [blob, minimum width=12pt, edge] at (-0.625,1.5) {$L$};
\end{tz}
=
\begin{tz}[xscale=0.66, scale=0.45,scale=1.17]
\draw [surface] (0,0) to [out=up, in=down] (-1,1.5) to [out=up, in=down] (0,3) to (1,3) to [out=down, in=up] (2,1.5) to [out=down, in=up] (1,0);
\draw [edge] (0,0) to [out=up, in=down] (-1,1.5) to [out=up, in=down] (0,3);
\draw [edge] (1,0) to [out=up, in=down] (2,1.5) to [out=up, in=down] (1,3);
\draw [edge, fill=white] (0.5,1.5) ellipse (1 and 0.666);
\node [blob, edge, minimum width=12pt] at (1.625,1.5) {$L$};
\end{tz}
&
\text{\textit{(f)} }
\lambda \left(
\begin{tz}[xscale=0.66, scale=0.5]
\draw [surface] (0,0) to [out=up, in=down] (-1,1.5) to [out=up, in=down] (0,3) to (1,3) to [out=down, in=up] (2,1.5) to [out=down, in=up] (1,0);
\draw [edge] (0,0) to [out=up, in=down] (-1,1.5) to [out=up, in=down] (0,3);
\draw [edge] (1,0) to [out=up, in=down] (2,1.5) to [out=up, in=down] (1,3);
\draw [edge, fill=white] (0.5,1.5) ellipse (1 and 0.666);
\end{tz}
\right)
=
\begin{tz}[xscale=0.66, scale=0.5]
\draw [surface] (0,0) to [out=up, in=down] (-1,1.5) to [out=up, in=down] (0,3) to (1,3) to [out=down, in=up] (2,1.5) to [out=down, in=up] (1,0);
\draw [edge] (0,0) to [out=up, in=down] (-1,1.5) to [out=up, in=down] (0,3);
\draw [edge] (1,0) to [out=up, in=down] (2,1.5) to [out=up, in=down] (1,3);
\draw [edge, fill=white] (0.5,1.5) ellipse (1 and 0.666);
\node [blob, minimum width=10pt, edge] at (-2,1.5) {$\lambda$};
\end{tz}
=
\begin{tz}[xscale=0.66, scale=0.5]
\draw [surface] (0,0) to [out=up, in=down] (-1,1.5) to [out=up, in=down] (0,3) to (1,3) to [out=down, in=up] (2,1.5) to [out=down, in=up] (1,0);
\draw [edge] (0,0) to [out=up, in=down] (-1,1.5) to [out=up, in=down] (0,3);
\draw [edge] (1,0) to [out=up, in=down] (2,1.5) to [out=up, in=down] (1,3);
\draw [edge, fill=white] (0.5,1.5) ellipse (1 and 0.666);
\node [blob, minimum width=10pt, edge] at (0.5,1.5) {$\lambda$};
\end{tz}
=
\begin{tz}[xscale=0.66, scale=0.5]
\draw [surface] (0,0) to [out=up, in=down] (-1,1.5) to [out=up, in=down] (0,3) to (1,3) to [out=down, in=up] (2,1.5) to [out=down, in=up] (1,0);
\draw [edge] (0,0) to [out=up, in=down] (-1,1.5) to [out=up, in=down] (0,3);
\draw [edge] (1,0) to [out=up, in=down] (2,1.5) to [out=up, in=down] (1,3);
\draw [edge, fill=white] (0.5,1.5) ellipse (1 and 0.666);
\node [blob, minimum width=10pt, edge] at (3,1.5) {$\lambda$};
\end{tz} 
\end{calign}}%

\figurecaptionsuck
\caption{Some identities in the graphical calculus.}
\figurecaptionpostsuck
\figurecaptionpostsuck
\label{fig:identities}
\end{figure}
\paragraph{Elementary description.}
While these structures are widely used in higher representation theory, they are not yet prevalent in the quantum computing community. To help the reader understand these new concepts, we also give  a direct account of the formalism  in elementary terms, that can be used without reference to the higher categorical technology (see also~\cite{Reutter:2016}).

In this direct perspective, regions are labelled by \textit{finite sets}. Wires and vertices now represent \textit{families} of Hilbert spaces and linear maps respectively, indexed by the elements of the sets labelling all adjoining regions.
% as illustrated by the following example: 
A composite surface diagram represents a family of composite linear maps, indexed by the elements of all regions open on the left or right. For regions open only at the top or bottom of the diagram, we take the direct sum over elements of the indexing set, while for closed regions, we take the vector space sum over elements of the indexing set.

We give an example in \autoref{fig:graphicalcalculus}. In the diagram on the left, regions are labelled by finite sets ($S, T, U$), with unshaded regions labelled implicitly by the 1\-element set; wires are labelled by families of finite-dimensional Hilbert spaces ($A, B, C, D, E, F$); and vertices are labelled by families of linear maps ($L, M, N$). For wires and vertices, the families are indexed by the sets associated to all neighbouring regions: for example, for $s \in S$ and $t \in T$, we have Hilbert spaces $A_s$, $B_{st}$ and $C_t$, and $L_{st}:B_{st} \otimes C_t \to A_s$ is a linear map. The single diagram on the left represents an entire family of linear maps, with the maps comprising this family  given by the right-hand diagram for different values of $s \in S$. We take the direct sum over index $u \in U$, since its region is open only at the bottom of the diagram, and the vector space sum over index $t \in T$, since its region is closed.

Given this interpretation of diagrams $D$ as families of linear maps $D_i$, we define two diagrams $D,D'$ to be \textit{equal} when all the corresponding linear maps $D_i^{},D_i'$ are equal; we define the \textit{scalar product} $\lambda D$ as the family of linear maps $\lambda D_i$; we define the \emph{adjoint} $D^\dag$ as the family of adjoint linear maps $(D_i)^\dag$; and we say that $D$ is unitary if all the maps $D_i$ are unitary. Following convention~\cite{Selinger:2010}, we depict the adjoint of a vertex by flipping it about a horizontal axis.

\paragraph{Restricted calculus.} We use a highly restricted portion of this calculus. Every shaded region we assume to be labelled by a single fixed finite set $S$. All wires bound precisely one shaded region and one unshaded region, and these wires are always labelled by a family of 1\-dimensional Hilbert spaces \C. Nonetheless, the calculus is not trivial. For example, we can build the identity on a nontrivial Hilbert space as the diagram \autoref{fig:language}(a); under the rules set out above, this is the identity map on  $\oplus _{s \in S} (\C \otimes \C) \simeq \C^{|S|}$.

Also, we add the following components to our language. In the first case there is an open region, and we use the obvious isomorphism $\C \simeq \C \otimes \C$ to build the associated families of linear maps.
{\tikzset{every picture/.style={scale=0.7}}
\begin{align}
\label{eq:s1}
\begin{tz}[yscale=1.1, yscale=-1]
\draw [edge, draw] (0,0) to [out=up, in=up, looseness=2] (1,0);
\draw [blueregion] (0,0) to (-0.5,0) to (-0.5,1) to (1.5,1) to (1.5,0) to (1,0) to [out=up, in=up, looseness=2] (0,0);
\end{tz}
& \,\,\,\,\,\,\,\,\, \forall s \in S, \C \simeq \C \otimes \C
&\ignore{
\begin{tz}[yscale=1.1]
\draw [edge, draw] (0,0) to [out=up, in=up, looseness=2] (1,0);
\draw [blueregion] (0,0) to (-0.5,0) to (-0.5,1) to (1.5,1) to (1.5,0) to (1,0) to [out=up, in=up, looseness=2] (0,0);
\end{tz}
& \forall s \in S, \C \otimes \C \simeq \C
\\
\begin{tz}[yscale=1.1]
\draw [blueregion, edge, draw] (0,0) to [out=up, in=up, looseness=2] (1,0);
\path (-0.5,0) rectangle (1.5,1);
\end{tz} 
& \textstyle \sum _{i \in S} c_i \ket i \mapsto \sum_{i \in S} c_i}
\begin{tz}[yscale=1.1, yscale=-1]
\draw [blueregion, edge, draw] (0,0) to [out=up, in=up, looseness=2] (1,0);
%\draw [redregion] (0,0) to (-0.5,0) to (-0.5,1) to (1.5,1) to (1.5,0) to (1,0) to [out=up, in=up, looseness=2] (0,0);
\path (-0.5,0) rectangle (1.5,1);
\end{tz}
& \textstyle 1 \mapsto \sum_{i \in S} \ket i 
\end{align}}%
Flipping these components about a horizontal axis denotes the adjoint of these maps, as discussed above. With these definitions the equations illustrated in \autoref{fig:identities} can be demonstrated; in that figure, the vertex $L$ and the scalar $\lambda \in \C$ are arbitrary.

\subsection{Shaded tangles}

\input{fig-Reidemeister-eptcs}
The \textit{Reidemeister moves}~\cite[Section~2.4]{Lackenby:2016} are the basic relations of classical knot theory. In this section we present an equational theory of shaded knots, which use shaded versions of the Reidemeister moves. This theory follows work of Jones~\cite{Jones:1989} on shaded tangle  invariants from statistical mechanical models.

We begin by supposing the existence of a \textit{shaded crossing}, depicted as follows:
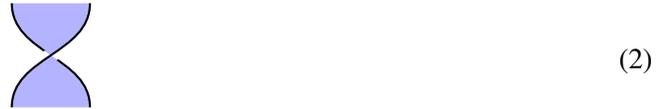
\def\scl{0.8}%
\begin{equation}
\label{eq:basicshadedcrossing}
\begin{tz}[string, scale=0.666,scale=\scl]
\path[blueregion] (1.75,0) to [out=90, in=down] (0.25,2) to (1.75,2) to [out=down, in=up] (0.25,0);
\draw[edge] (0.25,0) to [out=up, in=down] node [mask point] (1) {} (1.75,2);
\cliparoundone{1}{\draw[edge] (1.75,0) to [out=up, in=-90] (0.25,2);}
\end{tz}
\end{equation}
 We say that this crossing satisfies the \textit{basic calculus} when it satisfies the equations of \autoref{fig:reidemeister}(a)\nobreakdash--(d), and the \emph{extended calculus} when it additionally satisfies equation \autoref{fig:reidemeister}(e)\nobreakdash--(f).\footnote{In presenting this calculus, $\lambda$ is an arbitrary nonzero constant, and we implicitly use the rule described in \autoref{sec:graphicalcalculus} regarding the representation of the adjoint as a reflected diagram, which causes the crossing type to change. This calculus also defines a rotated crossing in \autoref{fig:reidemeister}(a).} A \textit{shaded tangle diagram} is a diagram constructed from the components of this calculus, the shaded cups \eqref{eq:s1}, and their adjoints.\looseness=-1

The extended calculus has the following attractive property.
\begin{theorem}[restate=thmmain, name={}] \label{thm:maintheorem}
Two shaded tangle diagrams with the same upper and lower boundaries are equal under the axioms of the extended calculus (up to overall scalar factors) just when their underlying tangles, obtained by ignoring the shading, are isotopic as classical knots.
\end{theorem}

In \cat{2Hilb}, we can classify representations of the basic calculus as follows. Note that from the discussion of \autoref{sec:graphicalcalculus}, a vertex of type \eqref{eq:basicshadedcrossing} represents in \cat{2Hilb} a linear map of type $\C^{|S|} \to \C^{|S|}$, and is therefore canonically represented by a matrix, which we assume to have matrix entries $H_{a,b}$.

\begin{theorem}[restate=thmclassification,name={}]
\label{thm:RIIclassification}
In \cat{2Hilb}, a shaded crossing yields a solution of the basic calculus just when it is equal to a self-transpose Hadamard matrix.\looseness=-1
\end{theorem}

\noindent
The following theorem identifies the additional constraint given by the extended calculus.

\begin{theorem}[restate=thmRthree,name={}]
In \cat{2Hilb}, a self-transpose Hadamard matrix satisfies the extended calculus just when:
\begin{equation} \label{eq:RIIIindex}\textstyle\sum_{r=0}^{|S|-1} \overline{H}_{ar}H_{br}H_{cr} = \sqrt{|S|}\,\overline{H}_{ab}\overline{H}_{ac}H_{bc}
\end{equation}
\end{theorem}

\noindent
A full classification of representations of this extended calculus is not known. However, it is known that solutions exist in all finite dimensions; we present this in \autoref{sec:RIIIsolutions}.

%\end{document}\end{document}\end{document}\end{document}

\subsection{Programs and specifications}
\label{sec:programsandspecifications}

\begin{figure}[]
\figuretopsuck
\figuretopsuck
\def\scl{0.8}
\def\xscl{0.85}
\def\tcdots{\scriptscriptstyle\cdots}
\def\tvdots{\tikz{\node [rotate=90, inner sep=0pt] at (0,0) {$\scriptscriptstyle\cdots$};}}
\def\tddots{\tikz{\node [rotate=-45, inner sep=0pt] at (0,0) {$\scriptscriptstyle\cdots$};}}
\def\rzero{\raisebox{2.5pt}{$0$}}
\begin{calign}
\nonumber
\begin{tz}[scale=0.5,yscale=1,scale=\scl ,xscale=\xscl]
\path[surface] (0.25,0) to [out=up, in=-135] (1,1) to [out=-45, in=up] (1.75,0);
\path[surface] (0.25,2) to [out=down, in=135] (1,1) to [out=45, in=down] (1.75,2);
\draw[edge] (0.25,0) to [out= up, in=-135] node[mask point, pos=1](1){} (1,1) to[out=45, in=down] (1.75,2);
\cliparoundone{1}{\draw[edge](1.75,0) to [out=up, in=-45] (1,1) to [out=135, in=down] (0.25,2);}
\end{tz}
\!=\!
{\scriptscriptstyle\frac 1 {\sqrt d}\!\!}
\left(
{
\small
\scriptsize
\begin{array}{@{}c@{\,}c@{\,}c@{}}
H_{11} & \tcdots & H_{1d}
\\[1pt]
\tvdots & \tddots & \tvdots
\\
H_{d1} & \tcdots & H_{dd}
\end{array}}
\right)
&
\begin{tz}[scale=0.5, yscale=1,scale=\scl,xscale=\xscl]
\path[surface,even odd rule] 
(0.25,0) to [out= up, in=-135]  (1,1) to [out= 45, in=down] (1.75,2) to (-0.75,2) to (-0.75,0) to (0.25,0)
(1.75,0) to [out= up, in=-45] (1,1) to [out= 135, in=down] (0.25,2) to (2.75,2) to (2.75,0) to (1.75,0);
\draw[edge] (0.25,0) to [out= up, in=-135] node[mask point, pos=1](1){} (1,1) to[out=45, in=down] (1.75,2);
\cliparoundone{1}{\draw[edge](1.75,0) to [out=up, in=-45] (1,1) to [out=135, in=down] (0.25,2);}
\draw[edge] (-0.75,0) to +(0,2);
\draw[edge] (2.75,0) to + (0,2);
\tidythetop
\end{tz}
\!=\!\!
\begin{aligned}
\tikz{\node [scale=0.85, inner sep=0pt] at (0,0) {$\tiny
\left(
\begin{array}{@{}c@{}c@{}c@{}c@{}c@{}c@{}}
\overline H_{11}  & \tcdots & 0 & 0 & \tcdots & 0
\\
\tvdots & \tddots &\rzero & \rzero & \tddots & \tvdots
\\
0 & 0 & \overline H_{1d} & 0 & 0 & 0
\\[1pt]
0 & 0 & 0 & \overline H_{21} & \tcdots & 0
\\
\tvdots & \tddots &   \rzero    & \rzero & \tddots & \tvdots
\\
0  & \tcdots & 0 & 0 & \tcdots & \overline H_{dd}
\end{array}
\right)$};}
\end{aligned}
&
\begin{tz}[scale=0.5,yscale=-1,scale=\scl,xscale=\xscl]
\path[surface] (0.25,0) to [out=up, in=-135] (1,1) to [out=-45, in=up] (1.75,0);
\path[surface] (0.25,2) to [out=down, in=135] (1,1) to [out=45, in=down] (1.75,2);
\draw[edge] (0.25,0) to [out= up, in=-135] node[mask point, pos=1](1){} (1,1) to[out=45, in=down] (1.75,2);
\cliparoundone{1}{\draw[edge](1.75,0) to [out=up, in=-45] (1,1) to [out=135, in=down] (0.25,2);}
\end{tz}
\!=\!
{\scriptscriptstyle\frac 1 {\sqrt d}\!\!}
\left(
{
\small
\scriptsize
\begin{array}{@{}c@{\,}c@{\,}c@{}}
\overline H_{11} & \tcdots & \overline H_{d1}
\\[1pt]
\tvdots & \tddots & \tvdots
\\
\overline H_{1d} & \tcdots & \overline H_{dd}
\end{array}}
\right)
&
\begin{tz}[scale=0.5, yscale=-1,scale=\scl,xscale=\xscl]
\path[surface,even odd rule] 
(0.25,0) to [out= up, in=-135]  (1,1) to [out= 45, in=down] (1.75,2) to (-0.75,2) to (-0.75,0) to (0.25,0)
(1.75,0) to [out= up, in=-45] (1,1) to [out= 135, in=down] (0.25,2) to (2.75,2) to (2.75,0) to (1.75,0);
\draw[edge] (0.25,0) to [out= up, in=-135] node[mask point, pos=1](1){} (1,1) to[out=45, in=down] (1.75,2);
\cliparoundone{1}{\draw[edge](1.75,0) to [out=up, in=-45] (1,1) to [out=135, in=down] (0.25,2);}
\draw[edge] (-0.75,0) to +(0,2);
\draw[edge] (2.75,0) to + (0,2);
\tidythetop
\end{tz}
\!=\!\!
\begin{aligned}
\tikz{\node [scale=0.85, inner sep=0pt] at (0,0) {$\tiny
\left(
\begin{array}{@{}c@{}c@{}c@{}c@{}c@{}c@{}}
H_{11}  & \tcdots & 0 & 0 & \tcdots & 0
\\
\tvdots & \tddots &\rzero & \rzero & \tddots & \tvdots
\\
0 & 0 & H_{1d} & 0 & 0 & 0
\\[1pt]
0 & 0 & 0 & H_{21} & \tcdots & 0
\\
\tvdots & \tddots &   \rzero    & \rzero & \tddots & \tvdots
\\
0  & \tcdots & 0 & 0 & \tcdots & H_{dd}
\end{array}
\right)$};}
\end{aligned}
\\[-1pt]\nonumber
\textit{(a) 1-qudit gate}
&
\textit{(b) 2-qudit gate}
&
\textit{(c) Adjoint 1-qudit gate}
&
\textit{(d) Adjoint 2-qudit gate}
\end{calign}

\figurecaptionsuck
\caption{Explicit expressions for the 1- and 2-qudit gates.}
\figurecaptionpostsuck
\label{fig:explicitgates}
\end{figure}
\firstparagraph{Scalar factors.} From this point onwards we drop the scalar factors appearing in the shaded tangle calculus, since they complicate the diagrams. More formally, every component we use in the remainder of the paper is proportional to an isometry, and we silently replace it with its isometric equivalent. %It is interesting that it is not possible to define everything in a way that avoids scalar factors appearing in the first place.%; similar normalization issues arise in the theory of spherical fusion categories~\cite{?}.

\paragraph{Programs.}
We write our {programs} in terms of four basic components of this shaded tangle language.
\begin{itemize}
\item \textbf{Qudits.} As mentioned above, \autoref{fig:language}(a) is interpreted as the identity map on $\C^{|S|}$, some finite-dimensional Hilbert space. This gives us our qudit.
\item \textbf{Qudit preparations.} In expression \eqref{eq:s1} above we draw a blue `cup' to indicate the state $\sum_{i \in S} \ket i \in \C^{|S|}$, which we interpret as a \textit{qudit preparation} (see \autoref{fig:language}(b).)% For qubits this is usually called the \emph{plus state}, denoted~$\ket +$.
\item \textbf{Qudit gates.} Applying our graphical calculus, given that \eqref{eq:basicshadedcrossing} corresponds to a self-transpose Hadamard with matrix entries $H_{ij}=H_{ji}$, we obtain  concrete representations for our 1- and 2\-qudit gates, given in \autoref{fig:explicitgates}.
\end{itemize}

\def\figurecircuitexample{
\begin{figure}[b]
\figuretopsuck
\figuretopsuck
\def \scl{0.8}
\def \xscl{0.9}
\begin{calign}
\nonumber
\begin{tz}[scale=0.5,yscale=1,scale=\scl,xscale=\xscl]
\path[surface] (0.25,0) to [out=up, in=-135] (1,1) to [out=-45, in=up] (1.75,0);
\path[surface] (0.25,2) to [out=down, in=135] (1,1) to [out=45, in=down] (1.75,2);
\draw[edge] (0.25,0) to [out= up, in=-135] node[mask point, pos=1](1){} (1,1) to[out=45, in=down] (1.75,2);
\cliparoundone{1}{\draw[edge](1.75,0) to [out=up, in=-45] (1,1) to [out=135, in=down] (0.25,2);}
\end{tz}
\,=\,
\frac 1 {\sqrt 2}
\left(
\begin{array}{@{}c@{\,\,\,}c@{}}
1 & 1 \\ 1 & \text{-}1
\end{array}
\right)
\,\,
\begin{tz}[scale=0.5,scale=\scl,xscale=\xscl]
\path[surface,even odd rule] 
(0.25,0) to [out= up, in=-135]  (1,1) to [out= 45, in=down] (1.75,2) to (-0.75,2) to (-0.75,0) to (0.25,0)
(1.75,0) to [out= up, in=-45] (1,1) to [out= 135, in=down] (0.25,2) to (2.75,2) to (2.75,0) to (1.75,0);
\draw[edge] (0.25,0) to [out= up, in=-135] node[mask point, pos=1](1){} (1,1) to[out=45, in=down] (1.75,2);
\cliparoundone{1}{\draw[edge](1.75,0) to [out=up, in=-45] (1,1) to [out=135, in=down] (0.25,2);}
\draw[edge] (-0.75,0) to +(0,2);
\draw[edge] (2.75,0) to +(0,2);
\tidythetop
\end{tz}
\,=\,
\left(
\footnotesize
\begin{array}{@{}c@{\,\,\,}c@{\,\,\,}c@{\,\,\,}c@{}}
1 & 0 & 0 & 0
\\ 0 & 1 & 0 & 0
\\ 0 & 0 & 1 & 0
\\ 0 & 0 & 0 & \text - 1
\end{array}
\right)
&
\begin{tz}[scale=0.5,yscale=1,scale=\scl,xscale=\xscl]
\path[surface] (0.25,0) to [out=up, in=-135] (1,1) to [out=-45, in=up] (1.75,0);
\path[surface] (0.25,2) to [out=down, in=135] (1,1) to [out=45, in=down] (1.75,2);
\draw[edge] (0.25,0) to [out= up, in=-135] node[mask point, pos=1](1){} (1,1) to[out=45, in=down] (1.75,2);
\cliparoundone{1}{\draw[edge](1.75,0) to [out=up, in=-45] (1,1) to [out=135, in=down] (0.25,2);}
\end{tz}
\,=\,
\frac{e^{\frac{\pi  i}{8}}}{\sqrt 2}
\left(\begin{array}{@{}c@{\,\,\,}c@{}}1 & \text - i \\ \text - i & 1\end{array}\right)
\,\,
\begin{tz}[scale=0.5,scale=\scl,xscale=\xscl]
\path[surface,even odd rule] 
(0.25,0) to [out= up, in=-135]  (1,1) to [out= 45, in=down] (1.75,2) to (-0.75,2) to (-0.75,0) to (0.25,0)
(1.75,0) to [out= up, in=-45] (1,1) to [out= 135, in=down] (0.25,2) to (2.75,2) to (2.75,0) to (1.75,0);
\draw[edge] (0.25,0) to [out= up, in=-135] node[mask point, pos=1](1){} (1,1) to[out=45, in=down] (1.75,2);
\cliparoundone{1}{\draw[edge](1.75,0) to [out=up, in=-45] (1,1) to [out=135, in=down] (0.25,2);}
\draw[edge] (-0.75,0) to +(0,2);
\draw[edge] (2.75,0) to + (0,2);
\tidythetop
\end{tz}
\,=\,
e^{ \text -\frac{\pi  i}{8}}
\footnotesize
\left(
\begin{array}{@{}c@{\,\,\,}c@{\,\,\,}c@{\,\,\,}c@{}}
1 & 0 & 0 & 0
\\ 0 &  i & 0 & 0
\\ 0 & 0 &  i & 0
\\ 0 & 0 & 0 & 1
\end{array}
\right)
\\[-1pt]\nonumber 
\textit{(a) A Fourier Hadamard}
& 
\textit{(b) A metaplectic Hadamard}
\end{calign}

\figurecaptionsuck
\caption{Different examples of our circuit elements.}
\figurecaptionpostsuck
\figurecaptionpostsuck
\label{fig:circuitexamples}
\end{figure}
}
\justarxiv{\figurecircuitexample}
%Note that by \autoref{fig:reidemeister}(b) and its colour-reversed version, the 1\- and 2\-qudit gates are unitary (at least up to a constant), as required for their interpretation as quantum circuit components. We also allow inverses of these 1- and 2\-qubit gates, which we denote with the opposite crossing.

\paragraph{Specifications.} We can write our specifications using the entire language, including all the cups and caps arising from \eqref{eq:s1} and their adjoints. We excluded some of these components from the program language illustrated in \autoref{fig:language} because they are not directly interpretable as circuit components. This does not prevent us using them in specifications, however, since these will not be directly executed; they exist only to define the mathematical behaviour of the overall program.

\paragraph{Examples.} We give some concrete examples of our basic circuit components. A standard Hadamard which gives a representation of the basic calculus is the \textit{qubit Fourier Hadamard}, illustrated in \autoref{fig:circuitexamples}(a). Other programs require a Hadamard representing the extended calculus; an example is the metaplectic  Hadamard illustrated in \autoref{fig:circuitexamples}(b) constructed using the methods of \autoref{sec:RIIIsolutions}. This Hadamard has been used in the cluster state literature for neighbourhood inversion on a cluster graph~\cite[Proposition~5]{Hein:2006}, an operation we verify in \autoref{sec:splicing} for a linear graph, but its strong topological properties do not seem to have been noted more generally.

\justconf{\figurecircuitexample}

\section{Entangled states}
\label{sec:entangledstates}

\noindent
In this section we describe several forms of entanglement and their representations in our calculus.

\subsection{GHZ states}
\def\drawboundingbox{\draw [red, ultra thick] (current bounding box.south west) rectangle (current bounding box.north east);}

\noindent
\emph{GHZ states} were introduced by Greenberger, Horne and Zeilinger~\cite{Greenberger:1989} to give a simplified proof of Bell's theorem. We define the unnormalized $n$\-partite qudit \GHZ~state as follows:
\begin{equation}
\ket{\GHZ_n}:=\textstyle\sum_{k=0}^{d-1} \ket{k,\cdots,k}
\end{equation}

\begin{figure}[]
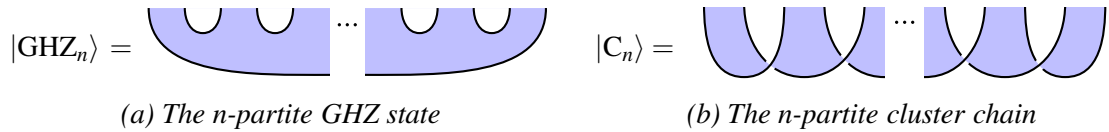

\figuretopsuck
\figuretopsuck
\def \scl{0.76}
\def \xscl{0.9}
\begin{calign}
\nonumber
\ket{\GHZ_n}\,=\,
\begin{tz}[scale=0.45,xscale=1.2,scale=\scl,xscale=\xscl]
\path [use as bounding box] (-0.1,-0.02) rectangle (11.1,-2.1);
\clip [use as bounding box] (-0.1,-0.02) rectangle (11.1,-2.1);
\begin{scope}
\clip (-0.1,-2.1) rectangle (5,0);
\draw[surface, edge, even odd rule] (0,0) to [out= down, in=180] (5,-2) to [out = 0, in =down] (10,0)
(1,0) to [out =down, in=left] (1.5,-0.75) to [out= right, in=down] (2,0)
(3,0) to [out =down, in=left] (3.5,-0.75) to [out= right, in=down] (4,0)
(6,0) to[out=down, in=left] (6.5,-0.75) to [out=right, in=down] (7,0)
(8,0) to[out=down, in=left] (8.5,-0.75) to [out=right, in=down] (9,0); 
\end{scope}
\begin{scope}[xshift=1cm]
\clip (5,-2.1) rectangle (10.1,2);
\draw[surface, edge, even odd rule] (0,0) to [out= down, in=180] (5,-2) to [out = 0, in =down] (10,0)
(1,0) to [out =down, in=left] (1.5,-0.75) to [out= right, in=down] (2,0)
(3,0) to [out =down, in=left] (3.5,-0.75) to [out= right, in=down] (4,0)
(6,0) to[out=down, in=left] (6.5,-0.75) to [out=right, in=down] (7,0)
(8,0) to[out=down, in=left] (8.5,-0.75) to [out=right, in=down] (9,0); 
\end{scope}
\node[scale=0.12,circle,fill] at (5.5,-0.5){};
\node[scale=0.12,circle,fill] at (5.72,-0.5){};
\node[scale=0.12,circle,fill] at (5.28,-0.5){};
%\drawboundingbox
\end{tz}
&
\ket{\cluster_n} =
\begin{tz}[scale=0.6,,scale=\scl,yscale=0.82,xscale=\xscl]
\begin{scope}\clip (-1,-0.05) rectangle (4,1.9);
\path[surface,even odd rule] 
(-0.5,2) to [out=down, in=left] (0.5,0) to [out= right, in=down] (1.5,2) to (0,2)
(0.5,2) to [out=down, in=left] (2,0) to [out= right, in=down] (3.5,2) 
(2.5,2) to [out=down, in=left] (4,0) to [out= right, in=down] (5.5,2)
(4.5,2) to [out=down, in=left] (6,0) to [out=right, in=down] (7.5,2) 
(6.5,2) to [out=down, in=left] (7.5,0) to[out=right, in=down] (8.5,2);
\draw[edge](-0.5,2) to [out=down, in=left] (0.5,0) to [out=right, in=down] node[mask point, pos=0.35] (1) {} (1.5,2);
\cliparoundone{1}{\draw[edge] (0.5,2) to [out=down, in=left] (2,0) to [out =right, in=down]node[mask point, pos=0.425] (1){} (3.5,2);}
\cliparoundone{1}{\draw[edge] (2.5,2) to [out=down, in=left] (4,0) to [out =right, in=down]node[mask point, pos=0.425] (1){} (5.5,2);}
\cliparoundone{1}{\draw[edge] (4.5,2) to [out=down, in=left] (6,0) to [out =right, in=down]node[mask point, pos=0.35] (1){} (7.5,2);}
\cliparoundone{1}{\draw[edge] (6.5,2) to [out=down, in=left] (7.5,0) to [out=right, in=down] (8.5,2);}
\end{scope}
\begin{scope}[xshift=1cm]
\clip (4,-0.05) rectangle (9,1.9);
\path[surface,even odd rule] 
(-0.5,2) to [out=down, in=left] (0.5,0) to [out= right, in=down] (1.5,2) to (0,2)
(0.5,2) to [out=down, in=left] (2,0) to [out= right, in=down] (3.5,2) 
(2.5,2) to [out=down, in=left] (4,0) to [out= right, in=down] (5.5,2)
(4.5,2) to [out=down, in=left] (6,0) to [out=right, in=down] (7.5,2) 
(6.5,2) to [out=down, in=left] (7.5,0) to[out=right, in=down] (8.5,2);
\draw[edge](-0.5,2) to [out=down, in=left] (0.5,0) to [out=right, in=down] node[mask point, pos=0.35] (1) {} (1.5,2);
\cliparoundone{1}{\draw[edge] (0.5,2) to [out=down, in=left] (2,0) to [out =right, in=down]node[mask point, pos=0.425] (1){} (3.5,2);}
\cliparoundone{1}{\draw[edge] (2.5,2) to [out=down, in=left] (4,0) to [out =right, in=down]node[mask point, pos=0.425] (1){} (5.5,2);}
\cliparoundone{1}{\draw[edge] (4.5,2) to [out=down, in=left] (6,0) to [out =right, in=down]node[mask point, pos=0.35] (1){} (7.5,2);}
\cliparoundone{1}{\draw[edge] (6.5,2) to [out=down, in=left] (7.5,0) to [out=right, in=down] (8.5,2);}
\end{scope}
%\node at (4.5,1.5){$\cdots$};
%\draw (-1,0) grid (9,2);
\node[scale=0.12,circle,fill] at (4.5,1.5){};
\node[scale=0.12,circle,fill] at (4.7,1.5){};
\node[scale=0.12,circle,fill] at (4.3,1.5){};
\end{tz}
\\[-1pt]\nonumber
\textit{(a) The $n$-partite GHZ state}
&
\textit{(b) The $n$-partite cluster chain}
\end{calign}

\figurecaptionsuck
\caption{Examples of entangled states.\label{fig:entangled}}
\figurecaptionpostsuck
\end{figure}

\begin{proposition}
\GHZ~states are represented as in \autoref{fig:entangled}(a).
\ignore{
\begin{equation}
\label{eq:ghzspecification}
\hspace{-0.4pt}
\ket{\GHZ_n}\,=\,
\begin{tz}[scale=0.45,xscale=1.2]
\path [use as bounding box] (-0.1,-0.02) rectangle (11.1,-2.1);
\clip [use as bounding box] (-0.1,-0.02) rectangle (11.1,-2.1);
\begin{scope}
\clip (-0.1,-2.1) rectangle (5,0);
\draw[surface, edge, even odd rule] (0,0) to [out= down, in=180] (5,-2) to [out = 0, in =down] (10,0)
(1,0) to [out =down, in=left] (1.5,-0.75) to [out= right, in=down] (2,0)
(3,0) to [out =down, in=left] (3.5,-0.75) to [out= right, in=down] (4,0)
(6,0) to[out=down, in=left] (6.5,-0.75) to [out=right, in=down] (7,0)
(8,0) to[out=down, in=left] (8.5,-0.75) to [out=right, in=down] (9,0); 
\end{scope}
\begin{scope}[xshift=1cm]
\clip (5,-2.1) rectangle (10.1,2);
\draw[surface, edge, even odd rule] (0,0) to [out= down, in=180] (5,-2) to [out = 0, in =down] (10,0)
(1,0) to [out =down, in=left] (1.5,-0.75) to [out= right, in=down] (2,0)
(3,0) to [out =down, in=left] (3.5,-0.75) to [out= right, in=down] (4,0)
(6,0) to[out=down, in=left] (6.5,-0.75) to [out=right, in=down] (7,0)
(8,0) to[out=down, in=left] (8.5,-0.75) to [out=right, in=down] (9,0); 
\end{scope}
\node[scale=0.12,circle,fill] at (5.5,-0.5){};
\node[scale=0.12,circle,fill] at (5.72,-0.5){};
\node[scale=0.12,circle,fill] at (5.28,-0.5){};
%\drawboundingbox
\end{tz}
\end{equation}}%
\end{proposition}
\noindent This arises directly from the representation of GHZ states in the CQM programme~\cite{Coecke:2017}. Important special cases for qubits are the {$\ket{+}$ state} $\ket{\GHZ_1} = \ket 0 + \ket 1$, and the Bell state $\ket{\GHZ_2} = \ket {00} + \ket {11}$.
\ignore{
\begin{calign}\begin{tz}[scale=0.75]\path[surface,edge] (0,2) to [out=down, in=left] (0.5,1) to [out=right, in=down] (1,2);
\end{tz}
&
\begin{tz}[scale=0.5]\path[surface,edge, even odd rule] (0,2) to [out=down, in=left] (1.5,0.5) to[out=right, in=down] (3,2)
(1,2) to [out=down, in=left] (1.5,1.5) to [out=right, in=down] (2,2);
\end{tz}\\*\nonumber \ket{+} =\ket{\GHZ_1}= \ket{0} + \ket{1}& \ket{\GHZ_2}=\ket{00} + \ket{11}\end{calign}
}

\subsection{Cluster chains}

\noindent
Another important class of entangled states are the \emph{cluster states} or \textit{graph states}~\cite{Briegel:2001,Schlingemann:2001,Hein:2006} and their qudit generalizations associated to Hadamard matrices~\cite{Cui:2015}. Cluster states have numerous applications, most prominently in the theory of measurement based quantum computation~\cite{Raussendorf:2001, Raussendorf:2009} and quantum error correction~\cite{Schlingemann:2001}.
Here we will focus on qudit \emph{cluster chains}, cluster states entangled along a chain.

Given a self-transpose $d$\-dimensional Hadamard matrix $H$, the $n$\-partite qudit cluster chain associated to $H$ is the following, where we conjugate the matrix due to our conventions:
\begin{equation}
\label{eq:clusterchainindex}
%\hspace{-0.2cm}
\ket{\cluster_n} := \textstyle \sum_{a_1,\ldots, a_n=0}^{d-1} \overline{H}_{a_1,a_2}\cdots \overline{H}_{a_{n{-}1},a_n} \ket{a_1\cdots a_n}
\end{equation}

\begin{proposition} \label{prop:clusterchain}Cluster chains are represented as in \autoref{fig:entangled}(b).
\end{proposition}

\noindent
This is essentially the same as the representation of cluster states in the CQM programme~\cite{Coecke:2008, Duncan:2014b}. If all Hadamard matrices in \eqref{eq:clusterchainindex} are the Fourier Hadamard, then this recovers conventional cluster chains~\cite{Briegel:2001}.

\subsection{Tangle gates and tangle states}

\begin{figure}[b]
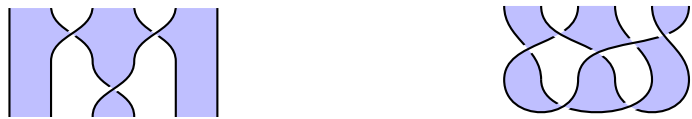

\figuretopsuck
\figuretopsuck
\def\scl{0.9}
\begin{calign}
\nonumber
\def\maskscl{1}
\begin{tz}[scale=0.45,scale=0.7,scale=\scl]
\clip (1,2) rectangle (10,6);
%\draw (0,0) grid (10,6);
\path[surface, even odd rule] 
(-0.75, 6) to (-0.75,0) to [out=right, in=down] (1.5, 2) to (1.5,6) 
(0,6) to (0,2) to [out=down, in=left] (2.25, 0) to [out=right, in=down] (4.5,2) to [out=up, in=down] (6,4) to [out=up, in=down] (7.5,6)
(4.5,6) to [out=down, in=up] (3,4) to (3,2) to [out=down, in=left] (5.25,0) to [out=right, in=down] (7.5,2) to (7.5,4) to[out=up, in=down] (6,6)
 (3,6) to [out=down, in=up] (4.5,4) to[out=down, in=up] (6,2)to[out=down, in=left] (8.25,0) to [out=right, in=down] (10.5,2) to (10.5,6)
(9,6) to(9,2) to  [out=down, in=left] (11.25,0) to (11.25,6);
\draw[edge] (-0.75,0) to [out=right, in=down] node[mask point, scale=\maskscl,pos=0.47] (1){} (1.5,2) to (1.5,6);
\cliparoundone{1}{\draw[edge] (0,6) to (0,2) to  [out=down, in=left] (2.25,0) to [out=right, in=down] node[mask point, scale=\maskscl, pos=0.48] (3){} (4.5,2) to [out=up, in=down] node[mask point, scale=\maskscl, pos=0.5] (4){}(6,4) to[out=up, in=down]node[mask point, scale=\maskscl, pos=0.5] (5){} (7.5,6);}
\cliparoundtwo{3}{5}{\draw[edge] (4.5,6) to [out=down, in=up] node[mask point, scale=\maskscl, pos=0.5] (2){} (3,4) to (3,2) to [out=down, in=left] (5.25,0) to [out=right, in=down]node[mask point, scale=\maskscl, pos=0.48] (6){} (7.5,2) to (7.5,4) to [out=up, in=down](6,6);}
\cliparoundthree{2}{4}{6}{\draw[edge] (3,6) to [out=down, in=up] (4.5,4) to [out=down, in=up] (6,2) to [out=down, in=left] (8.25,0) to [out=right, in=down]node[mask point, scale=\maskscl, pos=0.5] (7){} (10.5,2) to (10.5,6);}
\cliparoundone{7}{\draw[edge] (9,6) to (9,2) to [out=down,in=left] (11.24,0);}
\draw [decorate, decoration=brace] (-1.5,-0.2) to node [xshift=-1pt] {\circlenumber 1} +(0,2);
\draw [decorate, decoration=brace] (-1.5,2.1) to node [xshift=-1pt] {\circlenumber 2} +(0,3.7);
\tidythetop
\end{tz}
&
\begin{tz}[scale=0.42,scale=\scl]
%\path [use as bounding box] (1.9,1) rectangle +(7.2,3);
%\draw (0,0) grid (5,5);
\path [use as bounding box] (1.96,0) rectangle (7.04,3);
\path [surface, even odd rule]
    (6,3) to [out=down, in=up] node [mask point, pos=0.38] (19) {} (7,1) to [out=down, in=down, looseness=1.5] (5,1) to [out=up, in=down] node [mask point, pos=0.67] (14) {} (3,3)
    (5,3) to [out=down, in=up] (6,1) to [out=down, in=down, looseness=1] node [mask point, pos=0.29] (17) {} (3,1) to [out=up, in=down] (2,3)
    (7,3) to [out=down, in=up] node [mask point, pos=0.5] (18) {} node [mask point, pos=0.74] (15) {} (4,1) to [out=down, in=down, looseness=1.5] node [mask point, pos=0.28] (16) {} (2,1) to [out=up, in=down] node [mask point, pos=0.38] (13) {} (4,3);
\draw [thick, white] (5,3) to +(1,0);
\draw [thick, white] (3,3) to +(1,0);
\cliparoundtwo{15}{17}{\draw [edge] (6,3) to [out=down, in=up] (7,1) to [out=down, in=down, looseness=1.5] (5,1) to [out=up, in=down] (3,3);}
\cliparoundtwo{18}{16}{\draw [edge] (5,3) to [out=down, in=up] (6,1) to [out=down, in=down, looseness=1] (3,1);}
\cliparoundone{13}{\draw [edge] (3,1) to [out=up, in=down] (2,3);}
\cliparoundtwo{19}{14}{\draw [edge] (7,3) to [out=down, in=up] (4,1) to [out=down, in=down, looseness=1.5] (2,1) to [out=up, in=down] (4,3);}
%\fixboundingbox
\end{tz}
\end{calign}

\figurecaptionsuck
\caption{A tangle gate and a tangle state.\label{fig:tanglegate}}
\figurecaptionpostsuck
\end{figure}

\noindent
More generally, a \textit{tangle gate} is any circuit built from 1\- and 2\-qudit gates and their adjoints, and a \textit{tangle state} is a tangle with no inputs built from qudit preparations and tangle gates (see \autoref{fig:tanglegate}.) Such tangle states and gates can be arbitrarily complex, and have all the algebraic richness of knot topology. If a Hadamard represents the extended calculus, then two tangle states or gates are equal just when the corresponding tangles are isotopic, as established by \autoref{thm:maintheorem}.

\section{Manipulating quantum states}
\label{sec:entanglement}

\noindent
In this section we verify a wide variety of programs for creating and manipulating entangled states, including a new program for robust state transfer within a cluster chain--based quantum computer.

\subsection{Constructing GHZ states (\autoref{fig:exampletangles})}
\label{sec:creatingghz}

\firstparagraph{Overview.} We use our formalism to design and verify a program for constructing $n$\-partite GHZ states.

\paragraph{Program \autoref{fig:exampletangles}(a).} Begin by preparing $n$ qudits, then apply a sequence of 2\- and 1\-qudit gates as indicated in \autoref{fig:exampletangles}(a) for $n=3$ qudits.

\paragraph{Specification \autoref{fig:exampletangles}(b).} This is the tangle state corresponding to a 3\-qudit GHZ-state (\autoref{fig:entangled}(a)).

\paragraph{Verification.} Immediate by isotopy: the middle and rightmost qudit preparations in \autoref{fig:exampletangles}(a) move up and left, underneath the diagonal strand, producing \autoref{fig:exampletangles}(b). This only requires the basic calculus.

%\requiresRII
\paragraph{Novelty.} The GHZ version is known for the qubit Fourier Hadamard and was described very recently~\cite{Uchida:2015} for the \textit{qudit Fourier matrices} $H_{ab} =\textstyle{\frac{1}{\sqrt{d}} e^{ \frac{2\pi i}{d} ab}}$. For the self-transpose qudit Hadamard case covered here, the procedure seems new.

\ignore{
\subsection{Creating cluster chains}
\label{sec:creatingcluster}

\firstparagraph{Overview.} Analogously to \autoref{sec:creatingghz}, we can consider the design of a program to create a cluster chain.

\paragraph{Program.} We illustrate this in \autoref{fig:entangled}(b): we begin with $n$ qudit preparations, then perform 2\-qudit gates a total of $n-1$ times.

\paragraph{Specification.} We also take expression \autoref{fig:entangled}(b) to be the specification, \autoref{prop:clusterchain} shows that this recovers the standard algebraic definition~\eqref{eq:clusterchainindex}.

\paragraph{Verification.} Trivial,  the program and specification are equal.

%\requiresRII

\paragraph{Novelty.} This procedure is well known for conventional and generalized cluster states~\cite{Raussendorf:2001, Cui:2015}. Our treatment is not fundamentally different to the CQM\ analysis of Coecke, Duncan and Perdrix~\cite{Coecke:2008, Duncan:2014b}.

}

\def\figuresequivalence{%
\begin{figure}[]
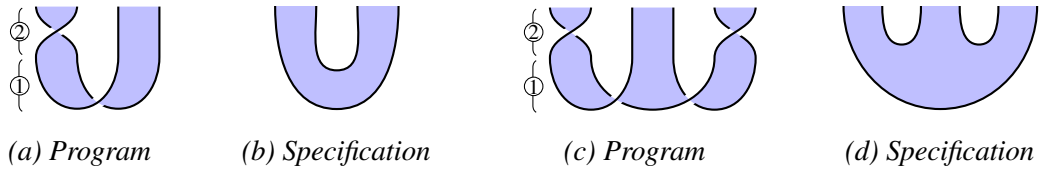

\figuretopsuck
\figuretopsuck
\vspace{-5pt}
\def\lscl{0.33}
\def\maskscale{1.6}
\def\maskscl{1.6}
\def\yscl{0.8}
\begin{calign}
\nonumber
\begin{tz}[xscale=0.85,scale=\lscl, yscale=\yscl]
%\clip (-5,3.97) rectangle (4.6, -0.1);
\path[surface,even odd rule] 
(1.5,4) to [out=down, in=up](0,2) to [out=down, in=left] (1.5,0) to [out=right, in=down] (3,2)  to (3,4)
(0,4) to [out=down, in=up](1.5,2) to [out=down, in=left] (3,0) to [out=right, in=down] (4.5,2) to (4.5,4);
\draw[edge] (1.5,4) to [out=down, in=up] node[mask point,scale=\maskscale,pos=0.5] (1){} (0,2) to [out=down, in=left] (1.5,0) to [out=right, in=down]node[mask point,scale=\maskscale, pos=0.3] (2){} (3,2) to (3,4);
\cliparoundtwo{1}{2}{\draw[edge] (0,4) to [out=down, in=up](1.5,2) to [out=down, in=left] (3,0) to [out=right, in=down] (4.5,2) to (4.5,4);}
\draw [decorate, decoration=brace] (-0.5,-0.1) to node [xshift=-1pt] {\circlenumber 1} +(0,2);
\draw [decorate, decoration=brace] (-0.5,2.1) to node [xshift=-1pt] {\circlenumber 2} +(0,1.8);
\tidythetop
\end{tz}
&
\begin{tz}[xscale=0.85,scale=\lscl, yscale=\yscl]
\clip (-0.1,3.97) rectangle (4.6, -0.1);
\path[surface,edge,even odd rule] 
(0,4) to [out=down, in=left] (2.25,0) to [out=right, in=down] (4.5,4)
(1.5,4) to [out=down, in=left] (2.25,1.5) to [out=right, in=down] (3,4);
\end{tz}
&
\begin{tz}[xscale=0.85,scale=\lscl, yscale=\yscl]
%\clip (-5,3.97) rectangle (7.6, -0.1);
\path[surface, even odd rule] (0,4) to [out=down, in=up] (1.5,2) to [out=down, in=left] (3.75,0) to [out=right, in=down] (6,2) to [out=up, in=down] (7.5,4)
(1.5,4) to [out=down,in=up] (0,2) to [out=down, in=left] (1.5,0) to [out=right, in=down] (3,2) to (3,4)
(4.5,4) to (4.5,2) to [out=down, in=left] (6,0) to [out=right, in=down] (7.5,2) to [out=up, in=down] (6,4);
\draw[edge] (1.5,4) to [out=down, in=up] node[mask point, scale=\maskscl,,pos=0.5] (1){}(0,2) to [out=down, in=left] (1.5,0) to [out=right, in=down] node[mask point, scale=\maskscl,pos=0.38] (2){}(3,2) to (3,4);
\cliparoundtwo{1}{2}{\draw[edge] (0,4) to [out=down, in=up] (1.5,2) to [out= down, in=left] (3.75,0) to [out= right, in=down]node[mask point, scale=\maskscl,, pos= 0.41](1){} (6,2)  to [out=up, in=down]node[mask point, scale=\maskscl,, pos=0.5](2){} (7.5,4); }
\cliparoundtwo{1}{2}{\draw[edge] (4.5,4) to (4.5,2) to [out=down, in=left] (6,0) to [out=right, in=down] (7.5,2) to [out=up, in=down] (6,4);}
\draw [decorate, decoration=brace] (-0.5,-0.1) to node [xshift=-1pt] {\circlenumber 1} +(0,2);
\draw [decorate, decoration=brace] (-0.5,2.1) to node [xshift=-1pt] {\circlenumber 2} +(0,1.8);
\tidythetop
\end{tz}
&
\begin{tz}[xscale=0.8,scale=\lscl, yscale=\yscl]
\clip (-0.1,3.97) rectangle (7.6, -0.1);
\path[surface,even odd rule, edge]
(0,4) to [out=down, in=left] (3.75,0) to [out=right, in=down] (7.5,4)
(1.5,4) to [out=down, in=left] (2.25,2.5) to [out=right, in=down] (3,4)
(4.5,4) to[out=down, in=left] (5.25,2.5) to [out=right, in=down] (6,4);
\end{tz}
\\*\nonumber
\textit{(a) Program} & \textit{(b) Specification} & \textit{(c) Program} & \textit{(d) Specification}
\end{calign}

\figurecaptionsuck
\caption{Converting 2- and 3-party cluster states into GHZ states.\label{fig:equivalent2}}
\figurecaptionpostsuck
\end{figure}

}%
\justconf{\figuresequivalence}
\subsection{Local unitary equivalence (\autoref{fig:equivalent2})}
\label{sec:localequivalence}

\firstparagraph{Overview.} In the case of 2 or 3 parties, cluster chains can be converted into GHZ states by applying 1\-qudit gates on certain sites. This means that, in a strong sense, they are equivalent computational resources. The reverse process, converting GHZ states to cluster chains, could be just as easily described.

\paragraph{Program.} For 2 and 3 parties, we illustrate the programs in \autoref{fig:equivalent2}(a) and (c), respectively. {\circlenumber 1}~Construct a cluster state. \circlenumber 2 Perform a 1\-qubit gate at certain sites.

\justarxiv{\figuresequivalence}

\paragraph{Specification.} Illustrated in \autoref{fig:equivalent2}(b) and (d), these are instances of the general GHZ specification~\autoref{fig:entangled}(a).

\paragraph{Verification.} Immediate by isotopy.
In both cases, loops of string in the corners of \autoref{fig:equivalent2}(a) and (c) contract to the top, giving \autoref{fig:equivalent2}(b) and (d), respectively. This only requires the basic calculus.

%\requiresRII

\paragraph{Novelty.} This is known both for conventional~\cite{Briegel:2001} and generalized~\cite{Cui:2015} cluster chains. For more than 3 parties it is known to be false, and indeed our method fails in these instances.

\begin{figure}[b]
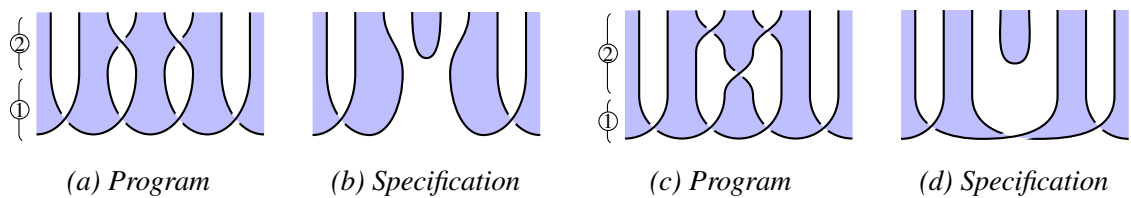

\figuretopsuck
\figuretopsuck
\def\maskscl{1.7}
\def\scl {0.7}
\def\xscl{0.88}
\begin{calign}
\nonumber
\hspace{-5pt}
\begin{tz}[scale=0.45,xscale=0.7,scale=\scl,xscale=\xscl]
%\clip (-0.85,3.97) rectangle (11.35,-0.1);
\path[surface, even odd rule] 
(-0.75, 4) to (-0.75,0) to [out=right, in=down] (1.5, 2) to (1.5,4) 
(0,4) to (0,2) to [out=down, in=left] (2.25, 0) to [out=right, in=down] (4.5,2) to [out=up, in=down] (3,4)
(4.5,4) to [out=down, in=up] (3,2) to [out=down, in=left] (5.25,0) to [out=right, in=down] (7.5,2) to [out=up, in=down] (6,4)
(7.5,4) to [out=down, in=up] (6,2) to[out=down, in=left] (8.25,0) to [out=right, in=down] (10.5,2) to (10.5,4)
(9,4) to(9,2) to  [out=down, in=left] (11.25,0) to (11.25,4);
\ignore{
\begin{scope}
\clip  (3,4) to [out=down, in=up] (4.5,2) to (6,2) to [out=up, in=down] (7.5,4);
\path[fill=white, postaction={classical}] (4.5,4) to [out=down, in=up] (3,2) to (7.5,2) to[out=up, in=down] (6,4);
\end{scope}
}
\draw[edge] (-0.75,0) to [out=right, in=down] node[mask point, scale=\maskscl,pos=0.47] (1){} (1.5,2) to (1.5,4);
\cliparoundone{1}{\draw[edge] (0,4) to (0,2) to  [out=down, in=left] (2.25,0) to [out=right, in=down] node[mask point, scale=\maskscl, pos=0.48] (1){} (4.5,2) to [out=up, in=down] node[mask point, scale=\maskscl, pos=0.5] (2){}(3,4);}
\cliparoundtwo{1}{2}{\draw[edge] (4.5,4) to [out=down, in=up] (3,2) to [out=down, in=left] (5.25,0) to [out=right, in=down]node[mask point, scale=\maskscl, pos=0.48] (1){} (7.5,2) to[out=up, in=down]node[mask point, scale=\maskscl, pos=0.5] (2){} (6,4);}
\cliparoundtwo{1}{2}{\draw[edge] (7.5,4) to [out=down, in=up] (6,2) to [out=down, in=left] (8.25,0) to [out=right, in=down]node[mask point, scale=\maskscl, pos=0.5] (1){} (10.5,2) to (10.5,4);}
\cliparoundone{1}{\draw[edge] (9,4) to (9,2) to [out=down,in=left] (11.24,0);}
\draw [decorate, decoration=brace] (-1.5,-0.2) to node [xshift=-1pt] {\circlenumber 1} +(0,2);
\draw [decorate, decoration=brace] (-1.5,2.1) to node [xshift=-1pt] {\circlenumber 2} +(0,1.7);
\tidythetop
\end{tz}
&
\begin{tz}[scale=0.45,xscale=0.7,scale=\scl,xscale=\xscl]
\clip (-0.85,3.97) rectangle (11.35,-0.1);
\path[surface,even odd rule]
(-0.75,4) to (-0.75,0) to [out=right, in=down] (1.5,2) to (1.5,4)
(0,4) to (0,2) to [out=down, in=left] (2.25,0) to [out=right, in=down] (4.,2) to [out=up, in=down] (3,4)
(7.5,4) to [out=down, in=up] (6.5,2) to [out=down, in=left] (8.25,0) to [out=right, in=down] (10.5,2) to (10.5,4)
(9,4) to (9,2) to[out=down, in=left] (11.25,0) to (11.25,4);
\path[surface] (4.5,4) to [out=down, in=left] (5.25,2.5) to [out=right, in=down] (6,4);
\draw[edge] (-0.75,0) to [out=right, in=down] node[mask point, scale=\maskscl,pos=0.47] (1){} (1.5,2) to (1.5,4);
\cliparoundone{1}{\draw[edge] (0,4) to (0,2) to  [out=down, in=left] (2.25,0) to [out=right, in=down] (4.,2) to [out=up, in=down](3,4);}
\draw[edge](4.5,4) to [out=down, in=left] (5.25,2.5) to [out=right, in=down] (6,4);
\draw[edge] (7.5,4) to [out=down, in=up] (6.5,2) to [out=down, in=left] (8.25,0) to [out=right, in=down] node[mask point, scale=\maskscl, pos=0.5](1){}(10.5,2) to (10.5,4);
\cliparoundone{1}{\draw[edge] (9,4) to (9,2) to [out=down, in=left] (11.25,0);}
\end{tz}
&
\begin{tz}[scale=0.45,scale=0.7,scale=\scl,xscale=\xscl]
%\clip (-0.85,5.97) rectangle (11.35,-0.1);
\path[surface, even odd rule] 
(-0.75, 6) to (-0.75,0) to [out=right, in=down] (1.5, 2) to (1.5,6) 
(0,6) to (0,2) to [out=down, in=left] (2.25, 0) to [out=right, in=down] (4.5,2) to [out=up, in=down] (6,4) to [out=up, in=down] (7.5,6)
(4.5,6) to [out=down, in=up] (3,4) to (3,2) to [out=down, in=left] (5.25,0) to [out=right, in=down] (7.5,2) to (7.5,4) to[out=up, in=down] (6,6)
 (3,6) to [out=down, in=up] (4.5,4) to[out=down, in=up] (6,2)to[out=down, in=left] (8.25,0) to [out=right, in=down] (10.5,2) to (10.5,6)
(9,6) to(9,2) to  [out=down, in=left] (11.25,0) to (11.25,6);
\draw[edge] (-0.75,0) to [out=right, in=down] node[mask point, scale=\maskscl,pos=0.47] (1){} (1.5,2) to (1.5,6);
\cliparoundone{1}{\draw[edge] (0,6) to (0,2) to  [out=down, in=left] (2.25,0) to [out=right, in=down] node[mask point, scale=\maskscl, pos=0.48] (3){} (4.5,2) to [out=up, in=down] node[mask point, scale=\maskscl, pos=0.5] (4){}(6,4) to[out=up, in=down]node[mask point, scale=\maskscl, pos=0.5] (5){} (7.5,6);}
\cliparoundtwo{3}{5}{\draw[edge] (4.5,6) to [out=down, in=up] node[mask point, scale=\maskscl, pos=0.5] (2){} (3,4) to (3,2) to [out=down, in=left] (5.25,0) to [out=right, in=down]node[mask point, scale=\maskscl, pos=0.48] (6){} (7.5,2) to (7.5,4) to [out=up, in=down](6,6);}
\cliparoundthree{2}{4}{6}{\draw[edge] (3,6) to [out=down, in=up] (4.5,4) to [out=down, in=up] (6,2) to [out=down, in=left] (8.25,0) to [out=right, in=down]node[mask point, scale=\maskscl, pos=0.5] (7){} (10.5,2) to (10.5,6);}
\cliparoundone{7}{\draw[edge] (9,6) to (9,2) to [out=down,in=left] (11.24,0);}
\draw [decorate, decoration=brace] (-1.5,-0.2) to node [xshift=-1pt] {\circlenumber 1} +(0,2);
\draw [decorate, decoration=brace] (-1.5,2.1) to node [xshift=-1pt] {\circlenumber 2} +(0,3.7);
\tidythetop
\end{tz}
&
\begin{tz}[scale=0.45,scale=0.7,scale=\scl,xscale=\xscl]
\clip (-0.85,5.97) rectangle (11.35,-0.1);
\path[surface,even odd rule]
(-0.75,6) to (-0.75,0) to [out=right, in=down] (1.5,2) to (1.5,6)
(0,6) to (0,2) to [out=down, in=left] (3.5,0) to [out=right, in=down] (7.5,2) to (7.5,6)
(3,6) to (3,2) to [out=down, in=left] (6.5,0) to [out=right, in=down] (10.5,2) to (10.5,6)
(9,6) to (9,2) to[out=down, in=left] (11.25,0) to (11.25,6);
\path[surface] (4.5,6) to [out=down, in=left] (5.25,3.5) to [out=right, in=down] (6,6);
\draw[edge] (-0.75,0) to [out=right, in=down]node[mask point, scale=\maskscl,pos=0.48] (1){} (1.5,2) to (1.5,6);
\cliparoundone{1}{\draw[edge] (0,6) to (0,2) to [out=down, in=left] (3.5,0) to [out=right, in=down]node[mask point, scale=\maskscl, pos= 0.34,minimum width=35pt] (2){} (7.5,2) to (7.5,6);}
\cliparoundone{2}{\draw[edge] (3,6) to (3,2) to [out=down, in=left] (6.5,0) to [out=right, in=down]node[mask point, scale=\maskscl, pos= 0.6] (3){} (10.5,2) to (10.5,6);}
\cliparoundone{3}{\draw[edge] (9,6) to (9,2) to [out=down, in=left] (11.25,0);}
\draw[edge] (4.5,6) to [out=down, in=left] (5.25,3.5) to [out=right, in=down] (6,6);
\end{tz}
\\*\nonumber
\textit{(a) Program} & \textit{(b) Specification}
&
\textit{(c) Program} & \textit{(d) Specification}
\end{calign}

\figurecaptionsuck
\caption{Cutting and splicing a cluster chain.\label{fig:cutting}}
\figurecaptionpostsuck
\end{figure}

\subsection{Cutting cluster chains (\autoref{fig:cutting}(a) and (b))}
\label{sec:cutting}

\firstparagraph{Overview.} Given a cluster chain of length $n$ we can \textit{cut} a target node from the chain, yielding two chains of total length $n{-}1$ and the target node in the $\ket +$ state.\footnote{In some variants the target node is instead destroyed by a projective measurement, and controlled operations performed on the adjacent qudits~\cite[Section 3]{Hein:2006}; the mathematical structure is identical to the version we analyze. A similar comment applies to the splicing procedure of \autoref{sec:splicing}.}

\paragraph{Program \autoref{fig:cutting}(a).} \circlenumber 1 Prepare a cluster state, of which only a central part is shown. \circlenumber 2 Perform two adjoint 2\-qudit gates both involving a central target qudit.

\paragraph{Specification \autoref{fig:cutting}(b).} Prepare two separate cluster chains, and separately prepare the target node in the $\ket +$ state.

\paragraph{Verification.} Immediate by isotopy: starting with  \autoref{fig:cutting}(a), we cancel the inverse pairs of crossings on the left and right of the central qudit, yielding \autoref{fig:cutting}(b). This only requires the basic calculus.

%\requiresRII

\paragraph{Novelty.} For the qubit Fourier Hadamard matrix this is well known; see~\cite{vandenNest:2004} and~\cite[Section 3]{Hein:2006}. Here, and in the next subsection, our verification provides new insight into how these chain manipulation programs work.

% For conventional graph states built from the Fourier matrix \eqref{eq:Fourier}, the neighboring qubits usually need to be corrected by a phase - and indeed, for the protocol to succeed the Fourier matrix needs to be transformed into a matrix fulfilling RIII.\DRcomm{Clear?} In other words, Reidemeister III essentially explains the mathematics behind local complementations of graph states. \DRcomm{Maybe nicer way to say this, it's all a bit sketchy} This protocol is completely new for qudit cluster states associated to more general Hadamard matrices. \textcolor{red}{Maybe reference back to \eqref{eq:Ising} and the interpretation as a quarter turn around the X-axis - this is what is used to locally complement cluster states.}

\subsection{Splicing cluster chains (\autoref{fig:cutting}(c) and (d))}
\label{sec:splicing}
\firstparagraph{Overview.} Given a cluster chain of length $n$ we can \textit{splice} a target node from the chain, yielding a single chain of length $n{-}1$, and the target node in the $\ket +$ state.

\paragraph{Program \autoref{fig:cutting}(c).} \hspace{-1pt}\circlenumber 1 Prepare a cluster state (only a central part is shown). \hspace{-1pt}\circlenumber 2 Perform a 1\-qudit gate on the target qudit, and then two 2\-qudit gates involving the target qudit and each of its adjacent qudits.

\paragraph{Specification \autoref{fig:cutting}(d).} Prepare a cluster chain of length $n{-}1$, and separately prepare the target node in the $\ket +$ state.

\paragraph{Verification.} By isotopy, although harder to see by eye than previous examples. Looking closely, one can see that the target qudit in \autoref{fig:cutting}(c) is unlinked from the other strings, and so the entire shaded tangle can be deformed to give \autoref{fig:cutting}(d). This requires the extended calculus.

%\requiresRIII
\paragraph{Novelty.} It is well-known that certain local operations on cluster chains splice the chain (\textit{neighbourhood inversion} on graph states, see~\cite{vandenNest:2004} and~\cite[Prop. 5]{Hein:2006}); our analysis is more general since it applies for any qudit Hadamard satisfying the extended calculus. The standard procedures use cluster chains based on the qubit Fourier Hadamard, and require additional phase corrections, which effectively serve to convert the Hadamard into one representing the extended calculus. We avoid this by building the cluster chain itself from a Hadamard representing the extended calculus.

\section{Quantum error correction}
\label{sec:errorcorrection}

\noindent
We now apply our calculus to the theory of quantum error correction. We give a graphical verification of the phase and Shor codes, and a substantial new generalization of both based on unitary error bases.

\def\E{\mathcal E}
\def\P{\mathcal P}
\paragraph{Basic definitions.} We begin by establishing notation.
For $n,k,p,d \in \N$, an $[[n,k,p]]^\E_d$ code uses $n$ physical qudits to encode $k$ logical qudits, in a way which is robust against errors occurring on at most $\lfloor{(p-1)/2}\rfloor$ physical qudits, such that each error is drawn from the subgroup $\E \subseteq U(d)$. We will be concerned with two types of errors: \textit{full qudit errors}, for which $\E = U(d)$, and \textit{phase errors}, for which $\E = \P \subset U(d)$, the subgroup of diagonal unitary matrices. The Knill-Laflamme theorem~\cite{Knill:1997} gives a way to identify these codes.
\begin{definition} An operator $e: (\C^d)^n \to (\C^d)^n$ is \textit{$(p,\E)$\-local} when it is of the form $e=U_1 \otimes \cdots \otimes U_n$, such that $U_i \in \E$ for all $1 \leq i \leq n$, and such that at most $p-1$ of the operators $U_i$ are not the identity.
\end{definition}
\begin{theorem}[Knill-Laflamme~\cite{Knill:1997}]
\label{thm:knilllaflamme}
An isometry
$i: (\C^d)^k \to (\C^d)^n$
gives an $[[n,k,p]]_d ^\E$ code just when, for any $(p,\E)$-local operator \mbox{$e:(\C^d)^n\to(\C^d)^n$}, the following composite is proportional to the identity:\vspace{-5pt}
\begin{equation}
(\C^d)^k \stackrel {\textstyle i} \to (\C^d)^n \stackrel {\textstyle e} \to (\C^d)^n \stackrel {\textstyle i^\dag} \to (\C^d)^k
\end{equation}
\end{theorem}

\noindent
Informally, the Knill-Laflamme theorem says that we have a $[[n,k,p]]^\E_d$ code just when, if we perform the encoding map, then perform a $(p,\E)$-local error, then perform the adjoint of the encoding map, the result is proportional to our initial state. To be clear, any proportionality factor is allowed, even~0.

\paragraph{Representing errors.} Following the general rules of our graphical calculus presented in \autoref{sec:graphicalcalculus}, we represent arbitrary qudit phases and qudit gates as follows, respectively:
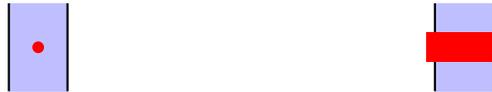
\def\scl{0.6}%
\vspace{-5pt}
\begin{calign}
\nonumber
\begin{tz}[yscale=1,scale=\scl]
\draw [surface] (0,0.25) rectangle (1,1.75);
\draw [fill=red, draw=none] (0.5,1) circle (0.1 and 0.1);
\draw [edge] (0,0.25) to +(0,1.5);
\draw [edge] (1,0.25) to +(0,1.5);
\end{tz}
&
\begin{tz}[scale=\scl]
\draw [surface] (0,0.25) rectangle (1,1.75);
\draw [edge] (0,0.25) to +(0,1.5);
\draw [edge] (1,0.25) to +(0,1.5);
\draw [fill=red, draw=none] (-0.15,0.75) rectangle +(1.3,0.5);
\end{tz}
\end{calign}
We draw them in red as they are interpreted here as errors.

\subsection{The phase code}
\label{sec:phasecode}

\begin{figure}[]
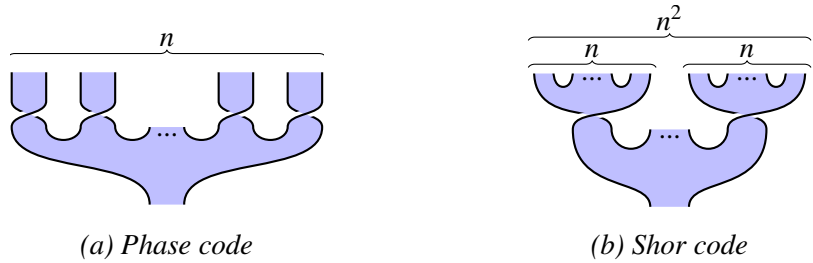

\figuretopsuck
\figuretopsuck
\def\scl{0.8}%
\def \yscl{0.9}%
\begin{calign}
\nonumber%
\def\maskscl{1.2}%
\def \rd{1.215}
\begin{tz}[scale=0.49,yscale=0.75,xscale=0.9,scale=\scl,yscale=\yscl]
%\draw (0,0) grid (9,7);
\clip (-0.04,0) rectangle (9.04,7.6);
\path[surface,even odd rule](4,0) to [out=up, in=down] (0,3) to [out=up, in=down] (1,4) to (1,4+\rd) to (8,4+\rd) to (8,4) to [out=down, in=up] (9,3) to [out=down, in=up] (5,0)
(0,4+\rd) to (0,4) to [out=down, in=up] (1,3) to[out=down, in=left] (1.5,2.5) to [out=right, in=down] (2,3) to [out=up, in=down] (3,4) to (3,4+\rd)
(2,4+\rd) to (2,4) to [out=down, in=up] (3,3) to [out=down, in=left] (3.5,2.5) to [out=right, in=down] (4,3) to(5,3) to [out=down, in=left] (5.5,2.5) to [out=right, in=down] (6,3) to [out=up, in=down] (7,4) to (7,4+\rd)
(6,4+\rd) to (6,4) to [out=down, in=up] (7,3) to [out=down, in=left] (7.5,2.5) to [out=right, in=down] (8,3) to [out=up, in=down] (9,4) to (9,4+\rd);
\draw[edge] (4,0) to [out=up, in=down] (0,3) to [out=up, in=down]node[mask point, scale=\maskscl, pos=0.5] (1){} (1,4) to (1,4+\rd);
\cliparoundone{1}{\draw[edge]  (0,4+\rd) to (0,4) to [out=down, in=up] (1,3) to [out=down, in=left] (1.5,2.5) to [out=right, in=down] (2,3) to [out=up,in=down] node[mask point, scale=\maskscl, pos=0.5] (2){} (3,4) to (3,4+\rd);}
\cliparoundone{2}{\draw[edge] (2,4+\rd) to (2,4) to [out=down, in=up] (3,3) to [out=down, in=left] (3.5,2.5) to [out=right, in=down] (4,3);}
\draw[edge] (5,3) to [out=down, in=left] (5.5,2.5) to [out=right, in=down] (6,3) to [out=up, in=down]node[mask point, scale=\maskscl, pos=0.5] (3){} (7,4) to (7,4+\rd);
\cliparoundone{3}{\draw[edge] (6,4+\rd) to (6,4) to [out=down, in=up] (7,3) to [out=down, in=left] (7.5,2.5) to [out=right, in=down] (8,3) to [out=up, in=down]node[mask point, scale=\maskscl, pos=0.5] (4){}  (9,4) to (9,4+\rd);}
\cliparoundone{4}{\draw[edge] (5,0) to [out=up, in=down] (9,3) to [out=up, in=down] (8,4) to (8,4+\rd);}
%\draw (0,0) grid (9,4);
\draw [decorate, decoration=brace] (-0.1,4.5+\rd) to node [above] {$n$} +(9.2,0);
\node at (4.5,2.7) {...};
\draw [white, ultra thick] (-0.1,5.2) to +(9.2,0);
\end{tz}
&
\def\maskscl{0.8}%
\def\lone{1.5}%
\def\ltwo{0.7}%
\begin{tz}[scale=0.55,xscale=0.9,scale=\scl,yscale=\yscl]
\clip (-0.42, 0) rectangle (6.92, 5.2);
%\draw (-1,0) grid (9,7);
\begin{scope}
%\clip (-0.02, -0.5)rectangle (6.57, 4.48);
\path[surface,even odd rule]
    (2.75,3.5) to [out=down, in=up, out looseness=\lone, in looseness=\ltwo] (0.75,2) to [out=down, in=up, out looseness=\lone, in looseness=\ltwo] (2.75,0) to (3.75,0) to [out=up, in=down, out looseness=\ltwo, in looseness=\lone] (5.75,2) to [out=up, in=down, out looseness=\ltwo, in looseness=\lone] (3.75,3.5)
    (-0.25,3.5) to [out=down, in=up, out looseness=\lone, in looseness=\ltwo] (1.75,2) to [out=down, in=left] (2.25,1.5) to [out=right, in=down] (2.75,2) to (3.75,2 ) to [out=down, in=left] (4.25,1.5) to [out=right, in=down] (4.75,2) to [out=up, in=down, out looseness=\ltwo, in looseness=\lone] (6.75,3.5)
    (0.25,3.5) to [out=down, in=left] +(0.25,-0.4) to [out=right, in=down] +(0.25,0.4)
    (1.75,3.5) to [out=down, in=left] (2,3.1) to [out=right, in=down] (2.25,3.5)
    (4.25,3.5) to [out=down, in=left] (4.5,3.1) to [out=right, in=down] (4.75,3.5)
    (5.75,3.5) to [out=down, in=left] (6,3.1) to [out=right, in=down] (6.25,3.5);
\end{scope}
\draw[edge] (2.75,3.5) to [out=down, in=up, out looseness=\lone, in looseness=\ltwo] node [mask point, scale=\maskscl, pos=0.7, minimum width=30pt] (1) {} (0.75,2) to [out=down, in=up, out looseness=\lone, in looseness=\ltwo] (2.75,0);
\cliparoundone{1}{\draw[edge] (-0.25,3.5) to [out=down, in=up, out looseness=\lone, in looseness=\ltwo] (1.75,2) to [out=down, in=left] (2.25,1.5) to [out=right, in=down] (2.75,2);}
\draw [edge] (3.75,2) to [out=down, in=left] (4.25,1.5) to [out=right, in=down] (4.75,2) to [out=up, in=down, out looseness=\ltwo, in looseness=\lone] node [mask point, scale=\maskscl, pos=0.3, minimum width=30pt] (2) {} (6.75,3.5);
\cliparoundone{2}{\draw[edge] (3.75,0) to [out=up, in=down, out looseness=\ltwo, in looseness=\lone] (5.75,2) to [out=up, in=down, out looseness=\ltwo, in looseness=\lone] (3.75,3.5);}
\draw[edge] (0.25,3.5) to [out=down, in=left] (0.5,3.1) to [out=right, in=down] (0.75,3.5);
\draw[edge] (1.75,3.5) to [out=down, in=left] (2,3.1) to [out=right, in=down] (2.25,3.5);
\draw[edge] (4.25,3.5) to [out=down, in=left] (4.5,3.1) to [out=right, in=down] (4.75,3.5);
\draw[edge] (5.75,3.5) to [out=down, in=left] (6,3.1) to [out=right, in=down] (6.25,3.5);
\ignore{
\node[scale=0.15,circle,fill] at (3.25,1.7){};
\node[scale=0.15,circle,fill] at (3.5,1.7){};
\node[scale=0.15,circle,fill] at (3,1.7){};
\node[scale=0.1,circle,fill] at (1.25,4.3){};
\node[scale=0.1,circle,fill] at (1.35,4.3){};
\node[scale=0.1,circle,fill] at (1.15,4.3){};
\node[scale=0.1,circle,fill] at (5.25,4.3){};
\node[scale=0.1,circle,fill] at (5.35,4.3){};
\node[scale=0.1,circle,fill] at (5.15,4.3){};
}%
\draw [decorate, decoration=brace] (-0.4,3.6) to node [above] {$n$} +(3.3,0);
\draw [decorate, decoration=brace] (3.6,3.6) to node [above] {$n$} +(3.3,0);
\draw [decorate, decoration=brace] (-0.4,4.3) to node [above] {$n^2$} +(7.3,0);
\node at (3.25,1.8) {...};
\node at (1.25,3.3) {...};
\node at (5.25,3.3) {...};
\draw [white, ultra thick] (-0.3,3.5) to +(9.1,0);
\end{tz}
\\*\nonumber 
\textit{(a) Phase code}
&
\textit{(b) Shor code}
\end{calign}

\figurecaptionsuck
\caption{The encoding maps $i$ for the phase and Shor codes.
\label{fig:phaseshor}}
\figurecaptionpostsuck
\end{figure}

\firstparagraph{Overview.} We present a $[[n,1,n]]^\P_d$ code: that is, a code which uses $n$ physical qudits to encode 1 logical qudit in a way that corrects $\lfloor (n{-}1)/2 \rfloor$ phase errors on the physical qudits. The data is a family of $n$ $d$\-dimensional Hadamard matrices.

\paragraph{Program.} The encoding map $i$ is depicted in \autoref{fig:phaseshor}(a).

\paragraph{Specification.} Satisfaction of the conditions of \autoref{thm:knilllaflamme}.

\paragraph{Verification.} In \autoref{fig:phaseerror} we illustrate the $n=3$ version of the code. We must show that the composite $i^\dag \circ e \circ i$, for any 3-local phase error in which 2 qudits are corrupted by arbitrary phases, is proportional to the identity. Given the symmetry of the encoding map, there are two cases: the errors can occur on adjacent or nonadjacent qudits. We analyze the case of adjacent errors here; the verification for nonadjacent errors is analogous. In the first image of \autoref{fig:phaseerror} we represent the composite $i^\dag \circ e \circ i$, using some artistic licence to draw the closed curves as circles. We apply moves to cause the errors to become `captured' by bubbles floating in unshaded regions, which therefore (see \autoref{fig:identities}(f)) give rise to overall scalar factors. This only requires the basic calculus.

%\requiresRII

\paragraph{Novelty.} A major novel feature is the visceral sense of how the protocol works that \autoref{fig:phaseerror} conveys: the phase errors are `captured by bubbles' and turned into scalar factors. We believe this intuition has not been described elsewhere. In terms of the mathematics, for the qubit Fourier Hadamard, this code is well known~\cite{Shor:1995, Nielsen:2009}. The generalization to arbitrary qudit Hadamard follows from work of Ke~\cite{Ke:2010}. Our treatment reveals a further generalization: each of the $n$ Hadamards used to build the encoding map $i$ may be distinct, since throughout the verification, we only apply the basic calculus moves to a Hadamard and its own adjoint. Our usual requirement for the Hadamards to be self-transpose is not necessary here, since we never rotate the crossings.

\justarxiv{\renewcommand{\quad}{\hskip0.15em\relax}}%
\ifarxiv
        \begin{figure*}[b]
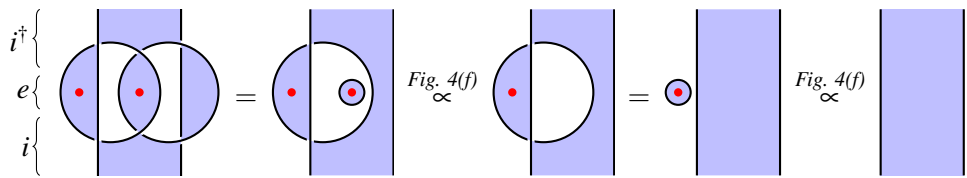

\else
        \begin{figure*}
\fi%
\figuretopsuck
\def\maskscl{0.3}
\def\scl{1.1}
\justarxiv{\def\scl{0.85}}
\def\yscl{1}
\def\bottom{0.5}
\def\top{2.5}
\def\height{2}
\[\justarxiv{\hspace{-0.8cm}}\begin{tz}[scale=\scl, yscale=\yscl]
\clip (0.,\bottom) rectangle (2.98,\top);
\path[surface] (1.5,0) to (1.5,3) to (2.5,3) to (2.5,0)
(1.65,1.5) circle (0.6)
(2.35,1.5) circle (0.6);
\draw[edge] (1.5,0) to node[mask point, scale=\maskscl, pos=0.31] (1){} node[mask point, scale=\maskscl, pos=0.69] (2){} +(0,3) ;
\cliparoundtwo{1}{2}{\draw[edge] (1.65,1.5) circle (0.6);}
\node[mask point, scale=\maskscl, rotate=-35](3) at (2,1.985){};
\node[mask point, scale=\maskscl, rotate=35](4) at (2,1.015){};
\cliparoundtwo{3}{4}{\draw[edge] (2.35,1.5) circle (0.6);}
\node[mask point, scale=\maskscl, rotate=-15](5) at (2.5,2.08){};
\node[mask point, scale=\maskscl, rotate=15](6) at (2.5,0.92){};
\cliparoundtwo{5}{6}{\draw[edge] (2.5,0) to + (0,3);}
\path[fill,red] (1.275, 1.5) circle (0.05);
\path[fill,red] (2, 1.5) circle (0.05);
\draw[decorate,decoration=brace] (0.8,\bottom) to node[left]{$i$} (0.8,1.2);
\draw[decorate,decoration=brace] (0.8,1.3) to node[left] {$e$} (0.8,1.7);
\draw[decorate,decoration=brace] (0.8,1.8) to node[left]{$i^\dagger$} (0.8,\top);
\end{tz}
\quad
\stackrel {} =
\quad
\begin{tz}[scale=\scl, yscale=\yscl]
\clip (1.03,\bottom) rectangle (2.5,\top);
\path[surface] (1.5,\bottom) to (1.5,\top) to (2.5,\top) to (2.5,\bottom)
(1.65,1.5) circle (0.6)
(2,1.5) circle (0.15);

\draw[edge] (1.5,0.5) to node[mask point, scale=\maskscl, pos=0.22] (1){} node[mask point, scale=\maskscl, pos=0.78] (2){} +(0,2) ;
\cliparoundtwo{1}{2}{\draw[edge] (1.65,1.5) circle (0.6);}

\draw[edge] (2,1.5) circle (0.15);

\draw[edge] (2.5,\bottom) to (2.5,\top);
\path[fill,red] (1.275, 1.5) circle (0.05);
\path[fill,red] (2, 1.5) circle (0.05);

\end{tz}
\quad
\stackrel{\text{\em Fig. \ref{fig:identities}(f)}}{\propto}
\quad
\begin{tz}[scale=\scl, yscale=\yscl]
\clip (1.03,\bottom) rectangle (2.5,\top);
\path[surface] (1.5,\bottom) to (1.5,\top) to (2.5,\top) to (2.5,\bottom)
(1.65,1.5) circle (0.6)
%(2.725,1.5) circle (0.15)
;

\draw[edge] (1.5,0.5) to node[mask point, scale=\maskscl, pos=0.22] (1){} node[mask point, scale=\maskscl, pos=0.78] (2){} +(0,2) ;
\cliparoundtwo{1}{2}{\draw[edge] (1.65,1.5) circle (0.6);}

%\draw[edge] (2.725,1.5) circle (0.15);

\draw[edge] (2.5,\bottom) to +(0,\height);
\path[fill,red] (1.275, 1.5) circle (0.05);
%\path[fill,red] (2.725, 1.5) circle (0.05);
\end{tz}
\quad
\stackrel {} =
\quad
\begin{tz}[scale=\scl, yscale=\yscl]
\path[surface] (1.5,\bottom) to (1.5,\top) to (2.5,\top) to (2.5,\bottom)
(1.275,1.5) circle (0.15)
%(2.725,1.5) circle (0.15)
;
\draw[edge] (1.5,\bottom) to  +(0,\height);
\draw[edge] (1.275,1.5) circle (0.15);
%\draw[edge] (2.725,1.5) circle (0.15);
\draw[edge] (2.5,\bottom) to + (0,\height);
\path[fill,red] (1.275, 1.5) circle (0.05);
%\path[fill,red] (2.725, 1.5) circle (0.05);
\end{tz}
\quad
\stackrel{\text{\em Fig. \ref{fig:identities}(f)}}{\propto}
\quad
\begin{tz}[scale=\scl, yscale=\yscl]
\path[surface] (1.5,\bottom) to (1.5,\top) to (2.5,\top) to (2.5,\bottom);
\draw[edge] (1.5,\bottom) to  +(0,\height);
\draw[edge] (2.5,\bottom) to + (0,\height);
\end{tz}
\]

\figurecaptionsuck
\caption{Verification of the $[[3,1,3]]^\P_d$ phase code.}\label{fig:phaseerror}
\figurecaptionpostsuck
\end{figure*}
\justarxiv{\renewcommand{\quad}{\hskip1em\relax}}%

\subsection{The Shor code}
\label{sec:shorcode}

\firstparagraph{Overview.} We present a $[[n^2,1,n]]^{U(d)}_d$ code: that is, a code which uses $n^2$ physical qudits to encode 1 logical qudit in a way that corrects $\lfloor (n{-}1)/2\rfloor$ arbitrary physical qudit errors. The data is a family of $n$ $d$\-dimensional Hadamard matrices.

\paragraph{Program.}
We choose the encoding map from \autoref{fig:phaseshor}(b).

\paragraph{Specification.} Satisfaction of the conditions of \autoref{thm:knilllaflamme}.

\paragraph{Verification.} In \autoref{fig:shorerror} we illustrate one error configuration for the $n=3$ case, where $e$ encodes two full qudit errors. All other cases work similarly. The general principle is the same as for \autoref{sec:phasecode}. This only requires the basic calculus.

\begin{figure*}
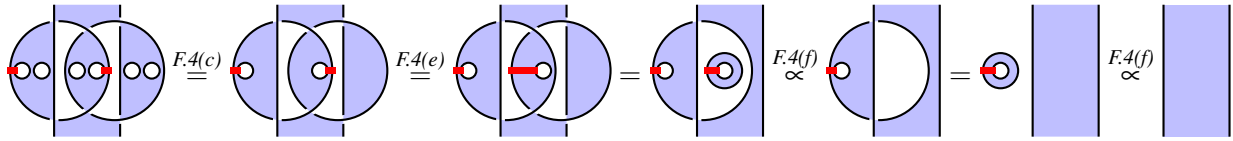

\def \l {0.25}
\def \rad {0.18}
\def\w{0.07}
\def\h{0.15}
\def \maskscl{0.7}
\def\scl{0.45}
\def\outer{0.4}
\begin{equation}
\nonumber
\hspace{-10pt}
\begin{tz}[scale=\scl]
\path[surface, even odd rule] (1,-1.5) to +(0,3) to (2+2*\l,1.5) to (2+2*\l,-1.5)
 (1+0.5*\l, 0) circle (1+0.5*\l)
 (2+1.5*\l, 0) circle (1+0.5*\l)
 (0.33-0.33*\rad,0) circle (\rad)
 (0.66+0.33*\rad,0) circle (\rad)
(1.33+\l-0.33*\rad,0) circle (\rad)
 (1.66+\l+0.33*\rad,0) circle (\rad)
(2.33+2*\l-0.33*\rad,0) circle (\rad)
 (2.66+2*\l+0.33*\rad,0) circle (\rad);
\draw[edge] (1,-1.5) to node[mask point, scale=\maskscl, pos=0.13] (1){} node[mask point, scale=\maskscl, pos=0.87] (2){} +(0,3);
\node[mask point, scale=\maskscl,rotate=7] (3) at (2+2*\l,-1.11){};
\node[mask point, scale=\maskscl,rotate=-7] (4) at (2+2*\l,1.11){};
\cliparoundtwo{3}{4}{\draw[edge] (2+2*\l,-1.5) to +(0,3);}
\cliparoundtwo{1}{2}{\draw[edge] (1+0.5*\l, 0) circle (1+0.5*\l);}
\node[mask point, scale=\maskscl,rotate=35] (4) at (1.5+\l,-0.93){};
\node[mask point, scale=\maskscl,rotate=-35] (5) at (1.5+\l,0.93){};
\cliparoundtwo{4}{5}{\draw[edge] (2+1.5*\l, 0) circle (1+0.5*\l);}
\draw[edge] (0.33-0.33*\rad,0) circle (\rad);
\draw[edge] (0.66+0.33*\rad,0) circle (\rad);
\draw[edge] (1.33+\l-0.33*\rad,0) circle (\rad);
\draw[edge]  (1.66+\l+0.33*\rad,0) circle (\rad);
\draw[edge] (2.33+2*\l-0.33*\rad,0) circle (\rad);
\draw[edge] (2.66+2*\l+0.33*\rad,0) circle (\rad);
\path[fill=red] (-\w,-0.5*\h) rectangle (0.33-1.33*\rad+\w,0.5*\h);
\path[fill=red] (1.66+\l+1.33*\rad-\w,-0.5*\h) rectangle (2+\l+\w,0.5*\h);
\ignore{
\draw[decorate,decoration=brace] (-0.2,-1.5) to node[left]{$i$} (-0.2,-0.2);
\draw[decorate,decoration=brace] (-0.2,-0.2) to node[left] {$e$} (-0.2,0.2);
\draw[decorate,decoration=brace] (-0.2,0.2) to node[left]{$i^\dagger$} (-0.2,1.5);
}
\end{tz}
\stackrel{\text{\em F.\ref{fig:identities}(c)}}{=}
\begin{tz}[scale=\scl]
\path[surface, even odd rule] (1,-1.5) to +(0,3) to (2+2*\l,1.5) to (2+2*\l,-1.5)
 (1+0.5*\l, 0) circle (1+0.5*\l)
 (2+1.5*\l, 0) circle (1+0.5*\l)
 (0.33-0.33*\rad,0) circle (\rad)
% (0.66+0.33*\rad,0) circle (\rad)
%(1.25+\l,0) circle (\rad)
 (1.66+\l+0.33*\rad,0) circle (\rad)
%(2.33+2*\l-0.33*\rad,0) circle (\rad)
% (2.66+2*\l+0.33*\rad,0) circle (\rad)
;
\draw[edge] (1,-1.5) to node[mask point, scale=\maskscl, pos=0.13] (1){} node[mask point, scale=\maskscl, pos=0.87] (2){} +(0,3);
\node[mask point, scale=\maskscl,rotate=7] (3) at (2+2*\l,-1.11){};
\node[mask point, scale=\maskscl,rotate=-7] (4) at (2+2*\l,1.11){};
\cliparoundtwo{3}{4}{\draw[edge] (2+2*\l,-1.5) to +(0,3);}
\cliparoundtwo{1}{2}{\draw[edge] (1+0.5*\l, 0) circle (1+0.5*\l);}
\node[mask point, scale=\maskscl,rotate=35] (4) at (1.5+\l,-0.93){};
\node[mask point, scale=\maskscl,rotate=-35] (5) at (1.5+\l,0.93){};
\cliparoundtwo{4}{5}{\draw[edge] (2+1.5*\l, 0) circle (1+0.5*\l);}
\draw[edge] (0.33-0.33*\rad,0) circle (\rad);
%\draw[edge] (0.66+0.33*\rad,0) circle (\rad);
%\draw[edge] (1.25+\l,0) circle (\rad);
\draw[edge] (1.66+\l+0.33*\rad,0) circle (\rad);
%\draw[edge] (2.33+2*\l-0.33*\rad,0) circle (\rad);
%\draw[edge] (2.66+2*\l+0.33*\rad,0) circle (\rad);
\path[fill=red] (-\w,-0.5*\h) rectangle (0.33-1.33*\rad+\w,0.5*\h);
\path[fill=red] (1.66+\l+1.33*\rad-\w,-0.5*\h) rectangle (2+\l+\w,0.5*\h);
%\path[fill=red] (1.33+\l+0.66*\rad-\w,-0.5*\h) rectangle (1.66+\l-0.66*\rad+\w,0.5*\h);
\end{tz}
\stackrel{\text{\em F.\ref{fig:identities}(e)}}{=}
\begin{tz}[scale=\scl]
\path[surface, even odd rule] (1,-1.5) to +(0,3) to (2+2*\l,1.5) to (2+2*\l,-1.5)
 (1+0.5*\l, 0) circle (1+0.5*\l)
 (2+1.5*\l, 0) circle (1+0.5*\l)
 (0.33-0.33*\rad,0) circle (\rad)
% (0.66+0.33*\rad,0) circle (\rad)
%(1.25+\l,0) circle (\rad)
 (1.66+\l+0.33*\rad,0) circle (\rad)
%(2.33+2*\l-0.33*\rad,0) circle (\rad)
% (2.66+2*\l+0.33*\rad,0) circle (\rad)
;
\draw[edge] (1,-1.5) to node[mask point, scale=\maskscl, pos=0.13] (1){} node[mask point, scale=\maskscl, pos=0.87] (2){} +(0,3);
\node[mask point, scale=\maskscl,rotate=7] (3) at (2+2*\l,-1.11){};
\node[mask point, scale=\maskscl,rotate=-7] (4) at (2+2*\l,1.11){};
\cliparoundtwo{3}{4}{\draw[edge] (2+2*\l,-1.5) to +(0,3);}
\cliparoundtwo{1}{2}{\draw[edge] (1+0.5*\l, 0) circle (1+0.5*\l);}
\node[mask point, scale=\maskscl,rotate=35] (4) at (1.5+\l,-0.93){};
\node[mask point, scale=\maskscl,rotate=-35] (5) at (1.5+\l,0.93){};
\cliparoundtwo{4}{5}{\draw[edge] (2+1.5*\l, 0) circle (1+0.5*\l);}
\draw[edge] (0.33-0.33*\rad,0) circle (\rad);
%\draw[edge] (0.66+0.33*\rad,0) circle (\rad);
%\draw[edge] (1.25+\l,0) circle (\rad);
\draw[edge]  (1.66+\l+0.33*\rad,0) circle (\rad);
%\draw[edge] (2.33+2*\l-0.33*\rad,0) circle (\rad);
%\draw[edge] (2.66+2*\l+0.33*\rad,0) circle (\rad);
\path[fill=red] (-\w,-0.5*\h) rectangle (0.33-1.33*\rad+\w,0.5*\h);
%\path[fill=red] (1.66+\l+1.33*\rad-\w,-0.5*\h) rectangle (2+\l+\w,0.5*\h);
\path[fill=red] (1.+\l-\w,-0.5*\h) rectangle (1.66+\l-0.66*\rad+\w,0.5*\h);
\end{tz}
\stackrel {} =
\begin{tz}[scale=\scl]
\path[surface, even odd rule] (1,-1.5) to +(0,3) to (2+2*\l,1.5) to (2+2*\l,-1.5)
 (1+0.5*\l, 0) circle (1+0.5*\l)
 (0.33-0.33*\rad,0) circle (\rad)
(1.5+0.5*\l, 0) circle (\outer)
 (1.5+0.5*\l,0) circle (\rad)
;
\draw[edge] (1.5+0.5*\l, 0) circle (\outer);
\draw[edge] (1.5+0.5*\l,0) circle (\rad);
\draw[edge] (1,-1.5) to node[mask point, scale=\maskscl, pos=0.13] (1){} node[mask point, scale=\maskscl, pos=0.87] (2){} +(0,3);
\draw[edge] (2+2*\l,-1.5) to +(0,3);
\cliparoundtwo{1}{2}{\draw[edge] (1+0.5*\l, 0) circle (1+0.5*\l);}
\node[mask point, scale=\maskscl,rotate=35] (4) at (1.5+\l,-0.93){};
\node[mask point, scale=\maskscl,rotate=-35] (5) at (1.5+\l,0.93){};
\draw[edge] (0.33-0.33*\rad,0) circle (\rad);
\path[fill=red] (-\w,-0.5*\h) rectangle (0.33-1.33*\rad+\w,0.5*\h);
\path[fill=red] (1.5-\outer+0.5*\l-\w,-0.5*\h) rectangle (1.5-\rad+0.5*\l+\w,0.5*\h);
\end{tz}
%\justarxiv{\\ \nonumber}
\stackrel{\text{\em F.\ref{fig:identities}(f)}}{\propto}
\begin{tz}[scale=\scl]
\path[surface, even odd rule] (1,-1.5) to +(0,3) to (2+2*\l,1.5) to (2+2*\l,-1.5)
 (1+0.5*\l, 0) circle (1+0.5*\l)
 (0.33-0.33*\rad,0) circle (\rad)
;
\draw[edge] (1,-1.5) to node[mask point, scale=\maskscl, pos=0.13] (1){} node[mask point, scale=\maskscl, pos=0.87] (2){} +(0,3);
\draw[edge] (2+2*\l,-1.5) to +(0,3);
\cliparoundtwo{1}{2}{\draw[edge] (1+0.5*\l, 0) circle (1+0.5*\l);}
\node[mask point, scale=\maskscl,rotate=35] (4) at (1.5+\l,-0.93){};
\node[mask point, scale=\maskscl,rotate=-35] (5) at (1.5+\l,0.93){};
\draw[edge] (0.33-0.33*\rad,0) circle (\rad);
\path[fill=red] (-\w,-0.5*\h) rectangle (0.33-1.33*\rad+\w,0.5*\h);
\end{tz}
\stackrel {} =
\begin{tz}[scale=\scl]
\path[surface, even odd rule] (1,-1.5) to +(0,3) to (2+2*\l,1.5) to (2+2*\l,-1.5)
 (0.33-0.33*\rad,0) circle (\rad)
 (0.33-0.33*\rad,0) circle (\outer)
;
\draw[edge] (1,-1.5) to  +(0,3);
\draw[edge] (2+2*\l,-1.5) to +(0,3);
\draw[edge] (0.33-0.33*\rad,0) circle (\rad);
\draw[edge] (0.33-0.33*\rad,0) circle (\outer);
\path[fill=red] (0.33-0.33*\rad-\outer-\w,-0.5*\h) rectangle (0.33-1.33*\rad+\w,0.5*\h);
\end{tz}
\stackrel{\text{\em F.\ref{fig:identities}(f)}}{\propto}
\begin{tz}[scale=\scl]
\path[surface, even odd rule] (1,-1.5) to +(0,3) to (2+2*\l,1.5) to (2+2*\l,-1.5)
;
\draw[edge] (1,-1.5) to  +(0,3);
\draw[edge] (2+2*\l,-1.5) to +(0,3);
\end{tz}
\end{equation}

\figurecaptionsuck
\caption{Verification of the $[[9,1,3]]^{U(d)}_d$ Shor code.}\label{fig:shorerror}
%\figurecaptionpostsuck
\end{figure*} 

%\requiresRII

\paragraph{Novelty.} For $d=2$, $n=3$ and the qubit Fourier Hadamard, this is exactly Shor's 9-qubit code~\cite{Shor:1995}. The qudit generalization for a general Hadamard is discussed in~\cite{Ke:2010}. As with the phase code, our version is more general still, since each of the Hadamards can be different.

\subsection{Unitary error basis codes}
\label{sec:uebcodes}

\firstparagraph{Overview.} We show that the phase and Shor codes described above still work correctly when the Hadamards are replaced by \textit{unitary error bases} (UEBs). These new codes have the same types $[[n,1,n]]^\P_{d^2}$, $[[n^2,1,n]]^{U(d)}_{d^2}$ as the phase and Shor codes, except with the additional restriction that the systems are of square dimension, since unitary error bases always have a square number of elements.

\paragraph{Unitary error bases (UEBs).}
UEBs are fundamental structures in quantum information which play a central role in quantum teleportation and dense coding~\cite{Werner:2001}, and also in error correction when they satisfy the additional axioms of a \textit{nice error basis}~\cite{Knill:1996-2}. However, the new UEB codes we present here are seemingly unrelated, and do \textit{not} require the additional nice error basis axioms.

\begin{definition}
On a finite-dimensional Hilbert space $H$, a \textit{unitary error basis} is a basis of unitary operators $U_i : H \to H$ such that $\Tr(U_i^{\dag} U_j^{\phantom\dag})=\delta_{ij}\dim(H)$.
\end{definition}
\ignore{
\begin{example}The most common unitary error basis in quantum computation is given by the Pauli matrices~\cite{Nielsen:2009}.
\begin{calign}
\label{eq:Paulimatrices}
\begin{pmatrix} 1&0\\0&1\end{pmatrix}
&
\begin{pmatrix}0&1\\1&0\end{pmatrix}
&
\begin{pmatrix} 0&\text{-}i\\i&0\end{pmatrix}
&
\begin{pmatrix}1&0\\0&\text{-}1\end{pmatrix}
\end{calign}
\end{example}
}

\noindent
UEBs have an elegant presentation in terms of our graphical calculus~\cite{Vicary:2012, Reutter:2016}.
\begin{theorem}[See~\cite{Reutter:2016}]
Unitary error bases correspond to vertices of the following type, satisfying equations analogous to \autoref{fig:reidemeister}(b) and (c):
\begin{equation}
\label{eq:uebvertex}
\begin{tz}[xscale=1,scale=0.5]
\path[surface](0.25,2) to (1.75,2) to [out=down, in=45] (1,1) to [out=135, in=down] (0.25,2);
\draw[edge] (0.25,0) to [out= up, in=-135] node [mask point,pos=1] (1) {}(1,1)
to [out= 45, in=down] (1.75,2); 
\cliparoundone{1}{\draw[edge] (0.25,2) to[out=down, in=135] (1,1) to [out=
-45, in=up] (1.75,0);}
\end{tz}
\end{equation}
\end{theorem}

\noindent
For a precise description of the necesasry equations see \cite[Proposition 9]{Reutter:2016}. The wires with unshaded regions on both sides represent the Hilbert space $\C^d$, and the shaded region is labelled by a set of cardinality~$d^2$.

\paragraph{Program.}
We choose the encoding maps of \autoref{fig:uebcodes}(a) and (b) to generalize the phase and Shor codes.

\paragraph{Specification.} Satisfaction of the conditions of \autoref{thm:knilllaflamme}.

\paragraph{Verification.} The procedure is identical to the phase and Shor code verifications, the only difference being that some regions are differently shaded. To make this clear, in \autoref{fig:uebcodes}(c) and (d) we give the graphical representations of the Knill-Laflamme $i^\dag \circ e \circ i$ composites for these new codes; compare these images to the first graphics in \Autoref{fig:phaseerror,fig:shorerror}.

\begin{figure}[b]
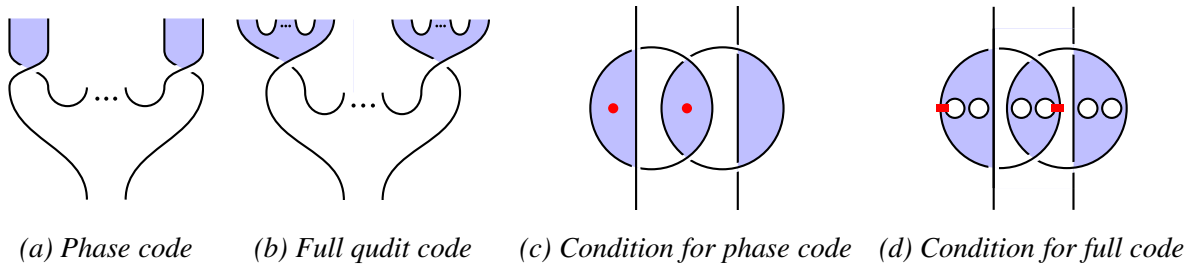

\figuretopsuck
\figuretopsuck
\begin{calign}
\nonumber
\def\maskscl{1.2}
\def \rd{0.885}
\def\scl{0.8}
\def\xscl{0.9}
\begin{tz}[scale=0.55,yscale=0.87,xscale=1,scale=\scl,xscale=\xscl]
\clip (1.98,0) rectangle (7.02,4.885);
\begin{scope}
\clip (0,5) to (0,3) to [out=up, in=down] (1,4) to (2,4) to [out=down, in=up] (3,3) to (3.1,3.1) to (5.9,3.1) to (6,3) to [out=up, in=down] (7,4) to (8,4) to [out=down, in=up] (9,3) to (10,3) to (10,5) to (0,5);
\path[surface,even odd rule](4,0) to [out=up, in=down] (2,3) to [out=up, in=down] (3,4) to (3,4+\rd) to (6,4+\rd) to  (6,4) to [out=down, in=up] (7,3) to [out=down, in=up] (5,0)
(2,4+\rd) to (2,4) to [out=down, in=up] (3,3) to [out=down, in=left] (3.5,2.5) to [out=right, in=down] (4,3) to(5,3) to [out=down, in=left] (5.5,2.5) to [out=right, in=down] (6,3) to [out=up, in=down] (7,4) to (7,4+\rd)
;
\end{scope}
\draw[edge] (4,0) to [out=up, in=down] (2,3) to [out=up, in=down]node[mask point, scale=\maskscl, pos=0.5] (1){} (3,4) to (3,4+\rd);
\cliparoundone{1}{\draw[edge] (2,4+\rd) to (2,4) to [out=down, in=up] (3,3) to [out=down, in=left] (3.5,2.5) to [out=right, in=down] (4,3);}
\draw[edge] (5,3) to [out=down, in=left] (5.5,2.5) to [out=right, in=down] (6,3) to [out=up, in=down]node[mask point, scale=\maskscl, pos=0.5] (2){} (7,4) to (7,4+\rd);
\cliparoundone{2}{\draw[edge] (6,4+\rd) to (6,4) to [out=down, in=up] (7,3) to [out=down, in=up] (5,0);}
%\draw (0,0) grid (9,4);
\node[scale=0.15,circle,fill] at (4.5,2.7){};
\node[scale=0.15,circle,fill] at (4.27,2.7){};
\node[scale=0.15,circle,fill] at (4.73,2.7){};
\draw [white, ultra thick] (-0.1,4.9) to +(9.2,0);
\end{tz}
&
\def\maskscl{1.2}
\def \rd{0.885}
\def\scl{0.8}
\def\xscl{0.9}
%i \quad:=\quad
\begin{tz}[scale=0.55,yscale=1,scale=\scl,xscale=\xscl]
\clip (-0.02,-0.48) rectangle (6.52,3.748);
%\draw (0,0) grid (7,8);
%\draw (0,-2) grid (7,5);
\begin{scope}
\clip (2.5,3.75) to [out=down, in=up] (0.75,2) to (-0.02,2) to (-0.02, 3.75) to (2.5,3.75)
(6.5, 3.75) to [out=down, in=up] (4.75,2) to (3,2) to (3,3.75) to (6.5,3.75);
\path[surface,even odd rule] (0,3.75) to [out=down, in=up] (1.75,2) to (3,2) to (3,3.75) to (0,3.75)
(4,3.75) to [out=down, in=up] (5.75,2) to (6.5,2) to (6.5,3.75) to (4,3.75)
(0.5,3.75) to [out=down, in=left] (0.75,3.35) to [out=right, in=down] (1,3.75)
(1.5,3.75) to [out=down, in=left] (1.75,3.35) to [out=right, in=down] (2,3.75)
 (4.5,3.75) to [out=down, in=left] (4.75,3.35) to [out=right, in=down] (5,3.75)
(5.5, 3.75) to [out=down, in=left] (5.75,3.35) to [out=right, in=down] (6,3.75);
\end{scope}
\draw[edge] (2.5,3.75) to  [out=down, in=up]node[mask point, scale=\maskscl, pos=0.65] (1){} (0.75,2) to [out=down, in=up] (2.75,-0.5);
\cliparoundone{1}{\draw[edge] (0,3.75) to [out=down, in=up] (1.75,2) to [out=down, in=left] (2.25,1.5) to [out=right, in=down] (2.75,2);}
\draw[edge] (3.75,2 ) to [out=down, in=left] (4.25,1.5) to [out=right, in=down] (4.75,2) to [out=up, in=down] node[mask point, scale=\maskscl, pos=0.35] (2){}(6.5,3.75) ;
\cliparoundone{2}{\draw[edge] (3.75,-0.5) to [out=up, in=down] (5.75,2) to [out=up, in=down]  (4,3.75);}
\draw[edge] (0.5,3.75) to [out=down, in=left] (0.75,3.35) to [out=right, in=down] (1,3.75);
\draw[edge] (1.5,3.75) to [out=down, in=left] (1.75,3.35) to [out=right, in=down] (2,3.75);
\draw[edge] (4.5,3.75) to [out=down, in=left] (4.75,3.35) to [out=right, in=down] (5,3.75);
\draw[edge] (5.5, 3.75) to [out=down, in=left] (5.75,3.35) to [out=right, in=down] (6,3.75);
\node[scale=0.15,circle,fill] at (3.25,1.7){};
\node[scale=0.15,circle,fill] at (3.5,1.7){};
\node[scale=0.15,circle,fill] at (3,1.7){};
\node[scale=0.1,circle,fill] at (1.25,3.55){};
\node[scale=0.1,circle,fill] at (1.35,3.55){};
\node[scale=0.1,circle,fill] at (1.15,3.55){};
\node[scale=0.1,circle,fill] at (5.25,3.55){};
\node[scale=0.1,circle,fill] at (5.35,3.55){};
\node[scale=0.1,circle,fill] at (5.15,3.55){};
\begin{scope}
\clip (2.75,-1.5) to [out=up, in=down] (3.75,-0.5) to (3.75,-1.5);
\path[surface] (3.75,-1.5) to [out=up, in=down] (2.75,-0.5) to (2.75,-1.5);
\end{scope}
\draw[edge] (2.75,-1.5) to [out=up, in=down] node[mask point, scale=\maskscl] (3){}(3.75,-0.5);
\cliparoundone{3}{\draw[edge] (3.75,-1.5) to [out=up, in=down] (2.75,-0.5);}
%\draw [decorate, decoration=brace] (-0.1,4.6) to node [above] {$n$} +(2.7,0);
%\draw [decorate, decoration=brace] (3.9,4.6) to node [above] {$n$} +(2.7,0);
%\draw [decorate, decoration=brace] (-0.2,5.1) to node [above] {$n^2$} +(6.9,0);
%\draw (0,-1) grid(7,7);
\draw [white, ultra thick] (-0.1,3.75) to +(9,0);
\end{tz}
&
\def\maskscl{0.3}
\def\scl{1.3}
\def\yscl{1}
\def\bottom{0.5}
\def\top{2.5}
\def\height{2}
\begin{tz}[scale=\scl*0.8, yscale=\yscl]
\clip (1.,\bottom) rectangle (2.98,\top);
\path[surface] (1.5,0) to (1.5,3) to (2.5,3) to (2.5,0)
(1.65,1.5) circle (0.6)
(2.35,1.5) circle (0.6);
\begin{scope}
\clip (1.5,0) to (1.5,3) to (2.5,3) to (2.5,0)
(1.65,1.5) circle (0.6);
\clip (1.5,0) to (1.5,3) to (2.5,3) to (2.5,0)
(2.35,1.5) circle (0.6);
\path[fill=white]  (1.5,0) to (1.5,3) to (2.5,3) to (2.5,0);
\end{scope}
\draw[edge] (1.5,0) to node[mask point, scale=\maskscl, pos=0.31] (1){} node[mask point, scale=\maskscl, pos=0.69] (2){} +(0,3) ;
\cliparoundtwo{1}{2}{\draw[edge] (1.65,1.5) circle (0.6);}
\node[mask point, scale=\maskscl, rotate=-35](3) at (2,1.985){};
\node[mask point, scale=\maskscl, rotate=35](4) at (2,1.015){};
\cliparoundtwo{3}{4}{\draw[edge] (2.35,1.5) circle (0.6);}
\node[mask point, scale=\maskscl, rotate=-15](5) at (2.5,2.08){};
\node[mask point, scale=\maskscl, rotate=15](6) at (2.5,0.92){};
\cliparoundtwo{5}{6}{\draw[edge] (2.5,0) to + (0,3);}
\path[fill,red] (1.275, 1.5) circle (0.05);
\path[fill,red] (2, 1.5) circle (0.05);
\end{tz}
&
\def \l {0.25}
\def \rad {0.18}
\def\w{0.07}
\def\h{0.15}
\def \maskscl{0.7}
\def\scl{0.68}
\def\outer{0.4}
\begin{tz}[scale=\scl*0.8]
\path[surface, even odd rule] (1,-1.5) to +(0,3) to (2+2*\l,1.5) to (2+2*\l,-1.5)
 (1+0.5*\l, 0) circle (1+0.5*\l)
 (2+1.5*\l, 0) circle (1+0.5*\l)
 (0.33-0.33*\rad,0) circle (\rad)
 (0.66+0.33*\rad,0) circle (\rad)
(1.33+\l-0.33*\rad,0) circle (\rad)
 (1.66+\l+0.33*\rad,0) circle (\rad)
(2.33+2*\l-0.33*\rad,0) circle (\rad)
 (2.66+2*\l+0.33*\rad,0) circle (\rad);
 \begin{scope}
\clip (1,-1.5) to +(0,3) to (2+2*\l,1.5) to (2+2*\l,-1.5)
 (1+0.5*\l, 0) circle (1+0.5*\l);
 \clip  (1,-1.5) to +(0,3) to (2+2*\l,1.5) to (2+2*\l,-1.5)
  (2+1.5*\l, 0) circle (1+0.5*\l);
  \path[fill=white](1,-1.5) to +(0,3) to (2+2*\l,1.5) to (2+2*\l,-1.5);
\end{scope}
\draw[edge] (1,-1.5) to node[mask point, scale=\maskscl, pos=0.13] (1){} node[mask point, scale=\maskscl, pos=0.87] (2){} +(0,3);
\node[mask point, scale=\maskscl,rotate=7] (3) at (2+2*\l,-1.11){};
\node[mask point, scale=\maskscl,rotate=-7] (4) at (2+2*\l,1.11){};
\cliparoundtwo{3}{4}{\draw[edge] (2+2*\l,-1.5) to +(0,3);}
\cliparoundtwo{1}{2}{\draw[edge] (1+0.5*\l, 0) circle (1+0.5*\l);}
\node[mask point, scale=\maskscl,rotate=35] (4) at (1.5+\l,-0.93){};
\node[mask point, scale=\maskscl,rotate=-35] (5) at (1.5+\l,0.93){};
\cliparoundtwo{4}{5}{\draw[edge] (2+1.5*\l, 0) circle (1+0.5*\l);}
\draw[edge] (0.33-0.33*\rad,0) circle (\rad);
\draw[edge] (0.66+0.33*\rad,0) circle (\rad);
\draw[edge] (1.33+\l-0.33*\rad,0) circle (\rad);
\draw[edge]  (1.66+\l+0.33*\rad,0) circle (\rad);
\draw[edge] (2.33+2*\l-0.33*\rad,0) circle (\rad);
\draw[edge] (2.66+2*\l+0.33*\rad,0) circle (\rad);
\path[fill=red] (-\w,-0.5*\h) rectangle (0.33-1.33*\rad+\w,0.5*\h);
\path[fill=red] (1.66+\l+1.33*\rad-\w,-0.5*\h) rectangle (2+\l+\w,0.5*\h);
\draw[edge] (1,-1.91) to (1,1.91) ;
\draw[edge] (2+2*\l,-1.91) to (2+2*\l,-1.5) ;
\draw[edge] (2+2*\l,1.5) to (2+2*\l,1.91) ;
\end{tz}
\\\nonumber
\textit{(a) Phase code} & \textit{(b) Full qudit code}
&
\textit{(c) Condition for phase code}
& 
\textit{(d) Condition for full code}
\end{calign}

\figurecaptionsuck
\caption{The UEB codes and their Knill-Laflamme conditions.\label{fig:uebcodes}}
\figurecaptionpostsuck
\figurecaptionpostsuck
\end{figure}

\paragraph{Novelty.}
As error correcting codes, these have the same strength as the traditional phase and Shor codes. However, they are constructed from completely different data\footnote{Although some UEBs can be constructed from Hadamards, they do not all arise in that way~\cite{Musto:2015, Reutter:2016}.}, and therefore push the theory of quantum error correcting codes in a new direction. This showcases the power of our approach to uncover new paradigms in quantum information.

\newpage
\bibliographystyle{eptcs}
\bibliography{references-eptcs}

\newpage
\ifarxiv
\appendix
\else
\appendices
\fi

\section{Reidemeister III Hadamard matrices}
\label{sec:RIIIsolutions}
The additional RIII condition \eqref{eq:RIIIindex} induces substantial constraints on a self-transpose Hadamard matrix. Here, we show that these equations have solutions in all finite dimensions. We consider two different families of solutions: \emph{Potts-Hadamard matrices}, and \emph{metaplectic invariants}. Almost everything in this section follows directly from results of Jones~\cite{Jones:1989} on building link invariants from statistical mechanical models.

\paragraph{Potts-Hadamard matrices.}
A \textit{Potts-Hadamard matrix} is a self-transpose Hadamard matrix of the following form, that satisfies \eqref{eq:RIIIindex}:
\begin{equation}
\label{eq:Potts-Hadamard}
\begin{tz}[string, scale=0.5]
\path[blueregion] (1.75,0) to [out=90, in=down] (0.25,2) to (1.75,2) to [out=down, in=up] (0.25,0);
\draw[edge] (0.25,0) to [out=up, in=down] node [mask point] (1) {} (1.75,2);
\cliparoundone{1}{\draw[edge] (1.75,0) to [out=up, in=-90] (0.25,2);}
\end{tz}
=
\lambda\,
\begin{tz}[string, edge,xscale=1.5]
\path[blueregion] (0.25,0) rectangle + (0.5,1);
\draw[edge] (0.25,0) to + (0,1);
\draw[edge] (0.75,0) to +(0,1);
\end{tz}
+\mu\,
\begin{tz}[string,xscale=1.5]\path[blueregion,draw,edge]  (0.25,1) to [out=down, in=left] (0.5,0.6) to [out=right, in=down] (0.75,1);
\path[blueregion,draw,edge] (0.25,0) to [out=up, in=left] (0.5,0.4) to [out=right, in=up] (0.75,0);
\end{tz}
\end{equation}

\noindent
In tensor notation, this means that $H_{a,b} = \lambda\, \delta_{a,b} \, + \, \mu$. We can classify Potts-Hadamard matrices exactly.
\begin{theorem}[restate=thmpotts,name={}]
\label{thm:pottshadamard}
Every $d$\-dimensional Potts-Hadamard matrix has $\mu = \frac{1}{\sqrt{d}}\, \overline{\lambda}$ with  \begin{equation}\label{eq:conditionsonlambda}\lambda \in U(1)\hspace{0.2cm} \text{ and }\hspace{0.2cm}\lambda^2+\overline{\lambda}^2 = -\sqrt{d}\end{equation} where $d$ is the dimension of the Hadamard matrix. 
This has the following solutions:
\begin{itemize}

\item $d=2$ and $\lambda \in \{ e^{\frac{3\pi i}{8}}, e^{-\frac{3 \pi i}{8}}, e^{-\frac{ 5\pi i}{8}}, e^{\frac{5\pi i}{8}} \}$;
\item $d=3$ and $\lambda \in \{ e^{\frac{5\pi i}{12}}, e^{-\frac{5 \pi i}{12}}, e^{-\frac{7 \pi i}{12}}, e^{\frac{7 \pi i}{12}} \}$;

\item $d=4$ and $\lambda \in \{i, -i\}$.
\end{itemize}
\end{theorem}
\noindent
The $d=2$ Potts-Hadamard matrices have the following form:
 \begin{calign}
\nonumber
\frac{e^{\text -\frac{\pi  i}{8}}}{\sqrt{2}} \left(\begin{array}{@{}c@{\,\,\,}c@{}}1 & i \\ i & 1\end{array}\right)
&
\frac{e^{\frac{\pi i}{8}}}{\sqrt{2}} \left(\begin{array}{@{}c@{\,\,\,}c@{}}1 & \text{-}i \\ \text{-}i & 1\end{array}\right)
&
\frac{e^{\frac{7\pi i}{8}}}{\sqrt{2}} \left(\begin{array}{@{}c@{\,\,\,}c@{}}1 & i \\ i & 1\end{array}\right)
&
\frac{e^{\text -\frac{7\pi i}{8}}}{\sqrt{2}} \left(\begin{array}{@{}c@{\,\,\,}c@{}}1 & \text{-}i \\ \text{-}i & 1\end{array}\right)
\end{calign}
In fact, it can be shown by direct calculation that these are the only two dimensional self-transpose Hadamard matrices fulfilling \eqref{eq:RIIIindex}. 

Equation \eqref{eq:Potts-Hadamard} (together with \eqref{eq:conditionsonlambda}) is a rescaled version of the defining relation of Kauffman's bracket polynomial~\cite{Kauffman:1987}; evaluating one of these matrices on a closed link diagram therefore yields (after suitable renormalization) the Jones polynomial of the link at certain roots of unity. 

\paragraph{Metaplectic invariants.} Following Jones and others~\cite{Jones:1989, Goldschmidt:1989,Jaffe:2016d}, given $d \in \N$ with $d>0$, we make the following definitions:
\begin{align}\label{eq:omega}
\xi &:= -e^{\frac{\pi i}{d}}
\ignore{
 \left\{
\begin{array}{rl}
- e^{\frac{\pi i}{d}}
& \text{$d$ odd}
\\
e^{\frac{\pi i}{d}}
& \text{$d$ even}
\end{array}
\right.
}%
&
\omega &:= \textstyle\frac{1}{\sqrt{d}}\sum_{k=0}^{d-1} \xi ^{k^2}
\end{align}
Let $\lambda$ be a square root of $\omega$, and for $0\leq a,b\leq d-1$, define $H_{a,b}$ as follows:
\begin{equation}\label{eq:defmeta} H_{a,b} = \frac{\overline{\lambda}}{\sqrt{d}} \,\xi^{(a-b)^2}
\end{equation}
Then we have the following.
\begin{theorem}[restate=thmmetaplectic,name={}]
\label{prop:metaplectic}
The coefficients $H_{a,b}$ define a self-transpose Hadamard matrix satisfying \eqref{eq:RIIIindex}.
\end{theorem}
\noindent
This establishes that solutions to our graphical equations can be found in all finite dimensions.

\section{Omitted proofs}
\thmmain*
\begin{proof}It is well known that two tangle diagrams are isotopic just when they can be transformed into each other using local Reidemeister moves. All Reidemeister moves can be obtained from arbitrary rotations and reflections of the moves depicted in \autoref{fig:reidemeisterunshaded}.
\begin{figure}
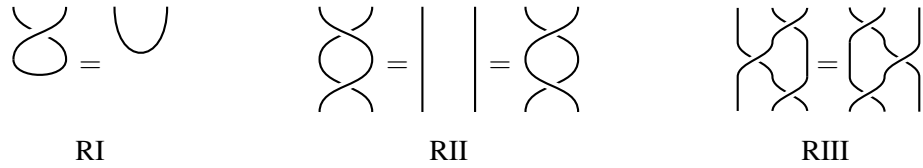

\figuretopsuck
\figuretopsuck
\tikzset{every picture/.style={scale=0.7}}
\begin{calign}
\nonumber
%\text{(a) }
\begin{tz}[xscale=-1, scale=1]
\path [use as bounding box] (0,-1) rectangle +(1,2);
\draw [edge] (0,1) to [out=down, in=up] node (1) [mask point] {} (1,0);
\cliparoundone{1}{\draw [edge] (1,0) to [out=down, in=down] (0,0) to [out=up, in=down] (1,1);}
\end{tz}
\,=\,
\begin{tz}
\path [use as bounding box] (0,-2) rectangle +(1,2);
\draw [edge] (0,0) to [out=down, in=down, looseness=3] (1,0);
\end{tz}
&
%\text{(b) }
\def\maskscl{0.65}%
\begin{tz}
%\path [use as bounding box, red, ultra thick] (0,0) rectangle (3,2);
\draw [edge] (0,0) to [out=up, in=down] node [mask point,scale=\maskscl] (1) {} (1,1) to [out=up, in=down] node [mask point,scale=\maskscl] (2) {} (0,2);
\cliparoundtwo{1}{2}{\draw [edge] (1,2) to [out=down, in=up] (0,1) to [out=down, in=up] (1,0);}
\end{tz}
\,=\,
\begin{tz}
\draw [edge] (0,0) to (0,2);
\draw [edge] (1,0) to (1,2);
\end{tz}
\,=\,
\begin{tz}[xscale=-1]
%\path [use as bounding box, red, ultra thick] (0,0) rectangle (3,2);
\draw [edge] (0,0) to [out=up, in=down] node [mask point,scale=\maskscl] (1) {} (1,1) to [out=up, in=down] node [mask point,scale=\maskscl] (2) {} (0,2);
\cliparoundtwo{1}{2}{\draw [edge] (1,2) to [out=down, in=up] (0,1) to [out=down, in=up] (1,0);}
\end{tz}
&
\def\maskscl{1.35}%
%\text{(c) }
\begin{tz}[scale=0.33, xscale=2, yscale=-1]
\path [use as bounding box] (0.8,0) rectangle (3,6);
\draw [edge] (3,0) to [out=up, in=down] node [mask point,scale=\maskscl] (1) {} (2,2) to [out=up, in=down] node [mask point, scale=\maskscl] (2) {} (1,4) to (1,6);
\draw [edge] (3,4) to [out=up, in=down] node [mask point, scale=\maskscl] (3) {} (2,6);
\cliparoundone{1}{\draw [edge] (2,0) to [out=up, in=down] (3,2) to (3,4);}
\cliparoundtwo{2}{3}{\draw [edge] (1,0) to (1,2) to [out=up, in=down] (2,4) to [out=up, in=down] (3,6);}
%\tidythetop
\draw [white] (0,6) to (4,6) to (4,0) to (0,0);
\end{tz}
=
\begin{tz}[scale=0.33, xscale=2]
%\path [use as bounding box] (0,0) rectangle (3,6);
\draw [edge] (1,0) to [out=up, in=down] node [mask point,scale=\maskscl] (1) {} (2,2) to [out=up, in=down] node [mask point,scale=\maskscl] (2) {} (3,4) to (3,6);
\draw [edge] (1,4) to [out=up, in=down] node [mask point,scale=\maskscl] (3) {} (2,6);
\cliparoundone{1}{\draw [edge] (2,0) to [out=up, in=down] (1,2) to (1,4);}
\cliparoundtwo{2}{3}{\draw [edge] (3,0) to (3,2) to [out=up, in=down] (2,4) to [out=up, in=down] (1,6);}
\end{tz}
\\\nonumber
\text{RI} & \text{RII} & \text {RIII}
\end{calign}

\figurecaptionsuck
\caption{The unshaded Reidemeister moves up to rotations and reflections.\label{fig:reidemeisterunshaded}}
\figurecaptionpostsuck
\end{figure}
 Since our tangles are shaded, they transform under \textit{shaded Reidemeister moves} - ordinary Reideimeister moves with a choice of checkerboard shading. Thus, up to rotations and reflections, there are 2 shaded versions of RI, 4 shaded versions of RII and 2 shaded versions of RIII. To prove \autoref{thm:maintheorem}, we therefore have to show that (up to scalar factors) all these shaded Reidemeister moves are implied by the basic axioms of the extended calculus presented in~\autoref{fig:reidemeister}. Using the shaded RII equations, it can be shown that the two shaded RIII equations are equivalent. Using shaded RII and RIII, it can be shown that the two shaded RI equations are equivalent. Therefore, two shaded tangles are isotopic if and only if they can be transformed into each other using all four shaded RII equations and one shaded RI and RIII equation, respectively.\end{proof}

\thmclassification*
\vspace{-3pt}
\begin{proof} Solutions to the shaded Reimeister II equations in \cat{2Hilb} \autoref{fig:reidemeister}(b) and (c) were classified in terms of Hadamard matrices in \cite[Proposition 7]{Reutter:2016}. The additional equation \autoref{fig:reidemeister}(d) implies that the corresponding Hadamard matrix is self-transpose. An equivalent classification using slightly different terminology can be found in \cite{Jones:1989}.
\end{proof}

\thmRthree*
 \begin{proof}
 Translating the Reimeister III equation \autoref{fig:reidemeister}(f) into the corresponding family of tensor diagrams (as described in \autoref{fig:graphicalcalculus}) yields the following:
 \[
 \def\maskscl{1}%
 \def\scl{0.8}%
 \begin{tz}[scale=0.45, xscale=1.5, yscale=-1,scale=\scl]
\path [use as bounding box] (0,0) rectangle (3,6);
\path [surface, even odd rule] (4,0) to (4,6) to (3,6) to [out=down, in=up] (2,4) to [out=down, in=up] (1,2) to (1,0) to (2,0) to [out=up, in=down] (3,2) to (3,4) to [out=up, in=down] (2,6) to (1,6) to (1,4) to [out=down, in=up] (2,2) to [out=down, in=up] (3,0)
 (0,0) to (0,6) to (4,6) to (4,0);
\draw [edge] (3,0) to [out=up, in=down] node [mask point,scale=\maskscl] (1) {} (2,2) to [out=up, in=down] node [mask point, scale=\maskscl] (2) {} (1,4) to (1,6);
\draw [edge] (3,4) to [out=up, in=down] node [mask point, scale=\maskscl] (3) {} (2,6);
\cliparoundone{1}{\draw [edge] (2,0) to [out=up, in=down] (3,2) to (3,4);}
\cliparoundtwo{2}{3}{\draw [edge] (1,0) to (1,2) to [out=up, in=down] (2,4) to [out=up, in=down] (3,6);}
%\tidythetop
\draw [white] (0,6) to (4,6) to (4,0) to (0,0);
\end{tz}
=\sqrt{|S|}\,
\begin{tz}[scale=0.45, xscale=1.5,scale=\scl]
\path [use as bounding box] (0,0) rectangle (3,6);
\path [surface, even odd rule] (0,0) to (0,6) to (1,6) to [out=down, in=up] (2,4) to [out=down, in=up] (3,2) to (3,0) to (2,0) to [out=up, in=down] (1,2) to (1,4) to [out=up, in=down] (2,6) to (3,6) to (3,4) to [out=down, in=up] (2,2) to [out=down, in=up] (1,0);
\draw [edge] (1,0) to [out=up, in=down] node [mask point,scale=\maskscl] (1) {} (2,2) to [out=up, in=down] node [mask point,scale=\maskscl] (2) {} (3,4) to (3,6);
\draw [edge] (1,4) to [out=up, in=down] node [mask point,scale=\maskscl] (3) {} (2,6);
\cliparoundone{1}{\draw [edge] (2,0) to [out=up, in=down] (1,2) to (1,4);}
\cliparoundtwo{2}{3}{\draw [edge] (3,0) to (3,2) to [out=up, in=down] (2,4) to [out=up, in=down] (1,6);}
\end{tz}
\hspace{0.5cm}\leftrightsquigarrow\hspace{0.5cm}
\bigforall_{a,b,c=0}^{d-1} \left( \sum_{x=0}^{d-1} \,
\begin{tz}[scale=0.45, xscale=1.5, yscale=-1,scale=\scl]
\path [use as bounding box] (0,0) rectangle (3,6);
\path [surface, even odd rule] (4,0) to (4,6) to (3,6) to [out=down, in=up] (2,4) to [out=down, in=up] (1,2) to (1,0) to (2,0) to [out=up, in=down] (3,2) to (3,4) to [out=up, in=down] (2,6) to (1,6) to (1,4) to [out=down, in=up] (2,2) to [out=down, in=up] (3,0)
 (0,0) to (0,6) to (4,6) to (4,0);
\draw [edge] (3,0) to [out=up, in=down] node [mask point,scale=\maskscl] (1) {} (2,2) to [out=up, in=down] node [mask point, scale=\maskscl] (2) {} (1,4) to (1,6);
\draw [edge] (3,4) to [out=up, in=down] node [mask point, scale=\maskscl] (3) {} (2,6);
\cliparoundone{1}{\draw [edge] (2,0) to [out=up, in=down] (3,2) to (3,4);}
\cliparoundtwo{2}{3}{\draw [edge] (1,0) to (1,2) to [out=up, in=down] (2,4) to [out=up, in=down] (3,6);}
%\tidythetop
\draw [white] (0,6) to (4,6) to (4,0) to (0,0);
\node[blob,scale=0.75] at (2.5,5) {$H_{c,x}$};
\node[blob,scale=0.75] at (2.5,1) {$H_{b,x}$};
\node[blob,scale=0.75] at (1.5,3) {$\overline{H}_{a,x}$};
\node[dimension] at (0.5,3) {$a$};
\node[dimension] at (2.5,3) {$x$};
\node[dimension] at (2.5,5.79) {$c$};
\node[dimension] at (2.5,0.21) {$b$};
\end{tz}
=\sqrt{|S|}\,
\begin{tz}[scale=0.45, xscale=1.5,scale=\scl]
\path [use as bounding box] (0,0) rectangle (3,6);
\path [surface, even odd rule] (0,0) to (0,6) to (1,6) to [out=down, in=up] (2,4) to [out=down, in=up] (3,2) to (3,0) to (2,0) to [out=up, in=down] (1,2) to (1,4) to [out=up, in=down] (2,6) to (3,6) to (3,4) to [out=down, in=up] (2,2) to [out=down, in=up] (1,0);
\draw [edge] (1,0) to [out=up, in=down] node [mask point,scale=\maskscl] (1) {} (2,2) to [out=up, in=down] node [mask point,scale=\maskscl] (2) {} (3,4) to (3,6);
\draw [edge] (1,4) to [out=up, in=down] node [mask point,scale=\maskscl] (3) {} (2,6);
\cliparoundone{1}{\draw [edge] (2,0) to [out=up, in=down] (1,2) to (1,4);}
\cliparoundtwo{2}{3}{\draw [edge] (3,0) to (3,2) to [out=up, in=down] (2,4) to [out=up, in=down] (1,6);}
\node[blob,scale=0.75] at (1.5,5) {$\overline{H}_{a,b}$};
\node[blob, scale=0.75] at ( 1.5,1) {$\overline{H}_{a,c}$};
\node[blob,scale=0.75] at (2.5,3) {$H_{b,c}$};
\node[dimension] at (0.5,3) {$a$};
\node[dimension] at (2.5,5.) {$b$};
\node[dimension] at (2.5,1) {$c$};
\end{tz}
\right)
\]
Here $a,b,$ and $c$ label the left, top right, and bottom right shaded region, respectively. The central shaded region is labelled by $x$ and summed over. Note that the Hadamard matrix $H$ is self-transpose. Thus, this results in equation \eqref{eq:RIIIindex}. Similarly, the Reidemeister I equation \autoref{fig:reidemeister}(e) translates into the following equation which is a direct algebraic consequence of \eqref{eq:RIIIindex} for $a=b$ (with $\lambda=\sqrt{|S|}\, \overline{H}_{a,a}$): $\textstyle\sum_{r=0}^{|S|-1} H_{c,r} = \lambda$.

 An equivalent classification using slightly different terminology can be found in \cite{Jones:1989}.
\end{proof}

\thmpotts*
\vspace{-3pt}
\begin{proof} For a shaded crossing of the form \eqref{eq:Potts-Hadamard} the two Reidemeister II equations look as follows:
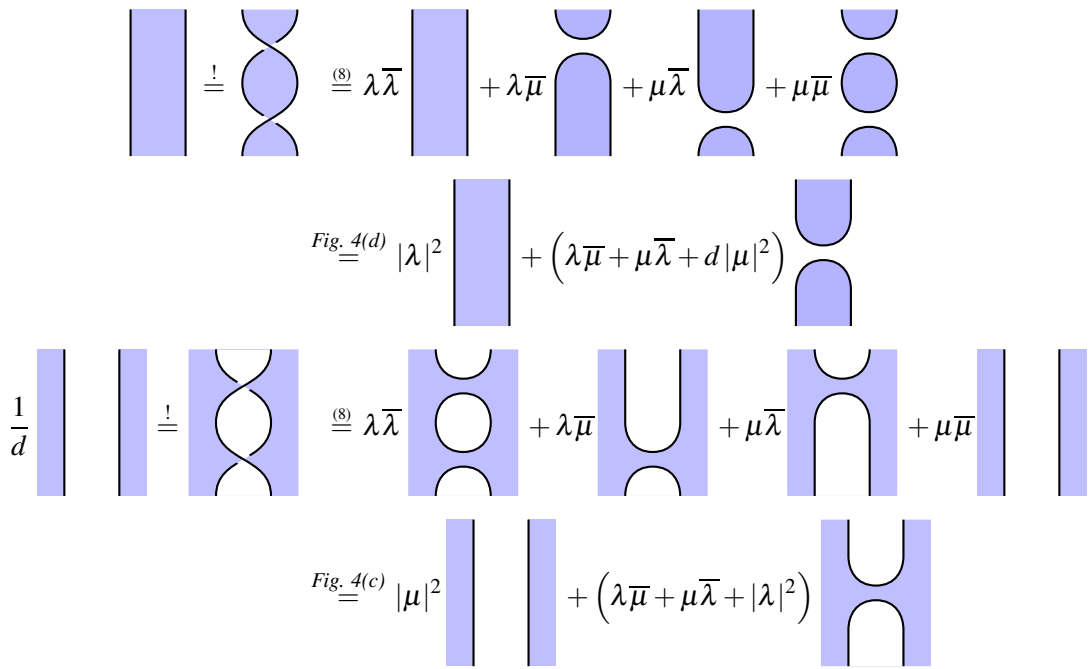
\def\maskscl{0.65}%
\def\scl{0.75}
\begin{align}\nonumber
\begin{tz}[string,scale=\scl,xscale=0.75]
\draw [surface] (0,0) rectangle (1,2);
\draw [edge] (0,0) to (0,2);
\draw [edge] (1,0) to (1,2);
\end{tz}
\,\, \stackrel{!}{=}\, \,
\begin{tz}[string,scale=\scl,xscale=0.75]
%\path [use as bounding box, red, ultra thick] (0,0) rectangle (3,2);
\draw [surface] (0,0) to [out=up, in=down] (1,1) to [out=up, in=down] (0,2) to (1,2) to [out=down, in=up] (0,1) to [out=down, in=up] (1,0);
\draw [edge] (0,0) to [out=up, in=down] node [mask point,scale=\maskscl] (1) {} (1,1) to [out=up, in=down] node [mask point,scale=\maskscl] (2) {} (0,2);
\cliparoundtwo{1}{2}{\draw [edge] (1,2) to [out=down, in=up] (0,1) to [out=down, in=up] (1,0);}
\end{tz}\,
&\,\,\,\,\,\,\superequals{eq:Potts-Hadamard}\, \lambda \overline{\lambda}\,\,
\begin{tz}[string,scale=\scl,xscale=0.75]
\draw [surface] (0,0) rectangle (1,2);
\draw [edge] (0,0) to (0,2);
\draw [edge] (1,0) to (1,2);
\end{tz}\, +\lambda\overline{\mu} \,\,
\begin{tz}[string,xscale=1.5,scale=\scl]\path[blueregion,draw,edge]  (0.25,2) to [out=down, in=left] (0.5,1.6) to [out=right, in=down] (0.75,2);
\path[blueregion,draw,edge] (0.25,0) to (0.25,1) to [out=up, in=left] (0.5,1.4) to [out=right, in=up] (0.75,1) to (0.75,0);
\end{tz}
\,+ \mu \overline{\lambda} \,\, 
\begin{tz}[string,xscale=1.5,scale=\scl]\path[blueregion,draw,edge] (0.25,2) to (0.25,1) to [out=down, in=left] (0.5,0.6) to [out=right, in=down] (0.75,1) to (0.75,2);
\path[blueregion,draw,edge]  (0.25,0) to [out=up, in=left] (0.5,0.4) to [out=right, in=up] (0.75,0) ;
\end{tz}
\,+\mu \overline{\mu}\,\,
\begin{tz}[string,xscale=1.5,scale=\scl]
\path[blueregion,draw,edge]  (0.25,2) to [out=down, in=left] (0.5,1.6) to [out=right, in=down] (0.75,2);
\path[blueregion,draw,edge]  (0.25,1) to [out=up, in=left] (0.5,1.4) to [out=right, in=up] (0.75,1) to [out=down, in=right] (0.5,0.6) to [out=left, in=down] (0.25,1);
\path[blueregion,draw,edge]  (0.25,0) to [out=up, in=left] (0.5,0.4) to [out=right, in=up] (0.75,0) ;
\end{tz}\\\nonumber
& \stackrel{\text{\em Fig. \ref{fig:identities}(d)}}{=} \,
|\lambda|^2 \,\,
\begin{tz}[string,scale=\scl,xscale=0.75]
\draw [surface] (0,0) rectangle (1,2);
\draw [edge] (0,0) to (0,2);
\draw [edge] (1,0) to (1,2);
\end{tz}
\, + \left( \lambda \overline{\mu} + \mu \overline{\lambda} + d\, |\mu|^2\right)
\begin{tz}[string,xscale=1.5,scale=\scl]\path[blueregion,draw,edge]  (0.25,2) to (0.25,1.5) to [out=down, in=left] (0.5,1.1) to [out=right, in=down] (0.75,1.5) to (0.75,2);
\path[blueregion,draw,edge] (0.25,0) to (0.25,0.5) to [out=up, in=left] (0.5,0.9) to [out=right, in=up] (0.75,0.5) to (0.75,0);
\end{tz}
\\
\nonumber
\frac{1}{d}\,
\begin{tz}[string,scale=\scl,xscale=0.75]
\draw[surface] (-0.5,0) rectangle (0,2);
\draw[surface](1,0) rectangle (1.5,2);
\draw [edge] (0,0) to (0,2);
\draw [edge] (1,0) to (1,2);
\end{tz}
\stackrel{!}{=}
\begin{tz}[string,scale=\scl,xscale=-0.75]
%\path [use as bounding box, red, ultra thick] (0,0) rectangle (3,2);
\draw [surface,even odd rule] (0,0) to [out=up, in=down] (1,1) to [out=up, in=down] (0,2) to (1,2) to [out=down, in=up] (0,1) to [out=down, in=up] (1,0)
(-0.5,0) rectangle (1.5,2);
\draw [edge] (0,0) to [out=up, in=down] node [mask point,scale=\maskscl] (1) {} (1,1) to [out=up, in=down] node [mask point,scale=\maskscl] (2) {} (0,2);
\cliparoundtwo{1}{2}{\draw [edge] (1,2) to [out=down, in=up] (0,1) to [out=down, in=up] (1,0);}
\end{tz}\,
&\,\,\,\,\,\,\superequals{eq:Potts-Hadamard}\, \lambda \overline{\lambda}\,
\begin{tz}[string,xscale=1.5,scale=\scl]
\path[surface,even odd rule] (0.25,2) to [out=down, in=left] (0.5,1.6) to [out=right, in=down] (0.75,2)
 (0.25,1) to [out=up, in=left] (0.5,1.4) to [out=right, in=up] (0.75,1) to [out=down, in=right] (0.5,0.6) to [out=left, in=down] (0.25,1)
 (0.25,0) to [out=up, in=left] (0.5,0.4) to [out=right, in=up] (0.75,0) 
 (0,0) rectangle (1,2);
\path[draw,edge]  (0.25,2) to [out=down, in=left] (0.5,1.6) to [out=right, in=down] (0.75,2);
\path[draw,edge]  (0.25,1) to [out=up, in=left] (0.5,1.4) to [out=right, in=up] (0.75,1) to [out=down, in=right] (0.5,0.6) to [out=left, in=down] (0.25,1);
\path[draw,edge]  (0.25,0) to [out=up, in=left] (0.5,0.4) to [out=right, in=up] (0.75,0) ;
\end{tz}
\, +\lambda\overline{\mu} 
\begin{tz}[string,xscale=1.5,scale=\scl]
\path[surface,even odd rule](0.25,2) to (0.25,1) to [out=down, in=left] (0.5,0.6) to [out=right, in=down] (0.75,1) to (0.75,2)
(0.25,0) to [out=up, in=left] (0.5,0.4) to [out=right, in=up] (0.75,0) 
(0,0) rectangle (1,2);
\path[draw,edge] (0.25,2) to (0.25,1) to [out=down, in=left] (0.5,0.6) to [out=right, in=down] (0.75,1) to (0.75,2);
\path[draw,edge]  (0.25,0) to [out=up, in=left] (0.5,0.4) to [out=right, in=up] (0.75,0) ;
\end{tz}
\,+ \mu \overline{\lambda} 
\begin{tz}[string,xscale=1.5,scale=\scl]
\path[surface,even odd rule]
 (0.25,2) to [out=down, in=left] (0.5,1.6) to [out=right, in=down] (0.75,2)
 (0.25,0) to (0.25,1) to [out=up, in=left] (0.5,1.4) to [out=right, in=up] (0.75,1) to (0.75,0)
 (0,0) rectangle (1,2);
\path[draw,edge]  (0.25,2) to [out=down, in=left] (0.5,1.6) to [out=right, in=down] (0.75,2);
\path[draw,edge] (0.25,0) to (0.25,1) to [out=up, in=left] (0.5,1.4) to [out=right, in=up] (0.75,1) to (0.75,0);
\end{tz}
\,+\mu \overline{\mu}
\begin{tz}[string,scale=\scl,xscale=0.75]
\draw[surface] (-0.5,0) rectangle (0,2);
\draw[surface](1,0) rectangle (1.5,2);
\draw [edge] (0,0) to (0,2);
\draw [edge] (1,0) to (1,2);
\end{tz}\\ \nonumber
&\stackrel{\text{\em Fig. \ref{fig:identities}(c)}}{=} \,
|\mu|^2 
\begin{tz}[string,scale=\scl,xscale=0.75]
\draw[surface] (-0.5,0) rectangle (0,2);
\draw[surface](1,0) rectangle (1.5,2);
\draw [edge] (0,0) to (0,2);
\draw [edge] (1,0) to (1,2);
\end{tz}
\, + \left( \lambda \overline{\mu} + \mu \overline{\lambda} +  |\lambda|^2\right)
\begin{tz}[string,xscale=1.5,scale=\scl]
\path[surface,even odd rule] 
(0.25,2) to (0.25,1.5) to [out=down, in=left] (0.5,1.1) to [out=right, in=down] (0.75,1.5) to (0.75,2)
(0.25,0) to (0.25,0.5) to [out=up, in=left] (0.5,0.9) to [out=right, in=up] (0.75,0.5) to (0.75,0)
(0,0) rectangle (1,2);
\path[draw,edge]  (0.25,2) to (0.25,1.5) to [out=down, in=left] (0.5,1.1) to [out=right, in=down] (0.75,1.5) to (0.75,2);
\path[draw,edge] (0.25,0) to (0.25,0.5) to [out=up, in=left] (0.5,0.9) to [out=right, in=up] (0.75,0.5) to (0.75,0);
\end{tz}
\end{align}
In other words, $|\lambda|=1,\,\,|\mu| = \frac{1}{\sqrt{d}}$, and $\lambda \overline{\mu} + \mu \overline{\lambda} = -1$. Reidemeister III yields the following:
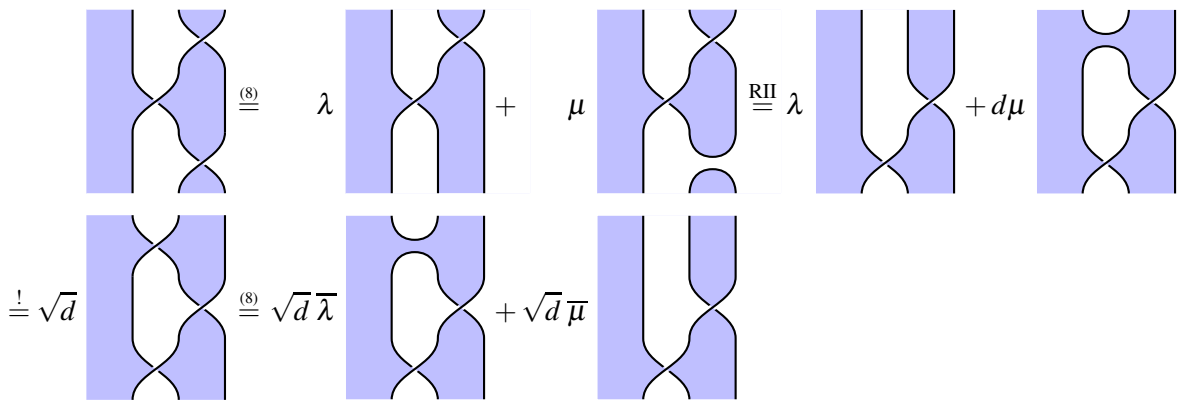
\def\scl{0.7}
\begin{align}\nonumber 
&\begin{tz}[scale=0.45, xscale=1.5, yscale=-1,scale=\scl]
\path [use as bounding box] (0,0) rectangle (3,6);
\path [surface, even odd rule] (4,0) to (4,6) to (3,6) to [out=down, in=up] (2,4) to [out=down, in=up] (1,2) to (1,0) to (2,0) to [out=up, in=down] (3,2) to (3,4) to [out=up, in=down] (2,6) to (1,6) to (1,4) to [out=down, in=up] (2,2) to [out=down, in=up] (3,0)
 (0,0) to (0,6) to (4,6) to (4,0);
\draw [edge] (3,0) to [out=up, in=down] node [mask point,scale=\maskscl] (1) {} (2,2) to [out=up, in=down] node [mask point, scale=\maskscl] (2) {} (1,4) to (1,6);
\draw [edge] (3,4) to [out=up, in=down] node [mask point, scale=\maskscl] (3) {} (2,6);
\cliparoundone{1}{\draw [edge] (2,0) to [out=up, in=down] (3,2) to (3,4);}
\cliparoundtwo{2}{3}{\draw [edge] (1,0) to (1,2) to [out=up, in=down] (2,4) to [out=up, in=down] (3,6);}
%\tidythetop
\draw [white] (0,6) to (4,6) to (4,0) to (0,0);
\end{tz}
\,\superequals{eq:Potts-Hadamard}\,\hphantom{\sqrt{d}\,\,}
\lambda \,\,
\begin{tz}[scale=0.45, xscale=1.5, yscale=-1,scale=\scl]
\path [use as bounding box] (0,0) rectangle (3,6);
\path [surface, even odd rule] 
(4,0) to (3,0) to [out=up, in=down] (2,2) to [out=up, in=down] (1,4) to (1,6) to (4,6)
(1,0) to (1,2) to [out=up, in=down] (2,4) to (2,6) to (3,6) to (3,2) to [out=down, in=up] (2,0)
 (0,0) to (0,6) to (4,6) to (4,0);
\draw [edge] (3,0) to [out=up, in=down] node [mask point,scale=\maskscl] (1) {} (2,2) to [out=up, in=down] node [mask point, scale=\maskscl] (2) {} (1,4) to (1,6);
\cliparoundone{1}{\draw [edge] (2,0) to [out=up, in=down] (3,2) to (3,6);}
\cliparoundone{2}{\draw [edge] (1,0) to (1,2) to [out=up, in=down] (2,4) to (2,6);}
%\tidythetop
\draw [white] (0,6) to (4,6) to (4,0) to (0,0);
\end{tz} \,+ \hphantom{\sqrt{d}\,\,}\mu \,\,
\begin{tz}[scale=0.45, xscale=1.5, yscale=-1,scale=\scl]
\path [use as bounding box] (0,0) rectangle (3,6);
\path [surface, even odd rule] 
(4,0) to (3,0) to [out=up, in=down] (2,2) to [out=up, in=down] (1,4) to (1,6) to (4,6)
(1,0) to (1,2) to [out=up, in=down] (2,4) to [out=up, in=left] (2.5,4.8) to [out=right, in=up] (3,4) to (3,2) to [out=down, in=up] (2,0) 
(2,6) to [out=down, in=left] (2.5,5.2) to [out=right, in=down] (3,6)
 (0,0) to (0,6) to (4,6) to (4,0);
\draw [edge] (3,0) to [out=up, in=down] node [mask point,scale=\maskscl] (1) {} (2,2) to [out=up, in=down] node [mask point, scale=\maskscl] (2) {} (1,4) to (1,6);
\cliparoundtwo{1}{2}{\draw [edge] (2,0) to [out=up, in=down] (3,2) to (3,4) to [out=up, in=right] (2.5,4.8) to [out=left, in=up] (2,4) to [out=down, in=up] (1,2) to (1,0);}
\draw [edge] (2,6) to [out=down, in=left] (2.5,5.2) to [out=right, in=down] (3,6);
%\tidythetop
\draw [white] (0,6) to (4,6) to (4,0) to (0,0);
\end{tz}
\,\stackrel{\text{RII}}{=}\, \lambda \,\,
\begin{tz}[scale=0.45, xscale=-1.5, yscale=1,scale=\scl]
\path [use as bounding box] (1,0) rectangle (4,6);
\path [surface, even odd rule] 
(4,0) to (3,0) to [out=up, in=down] (2,2) to [out=up, in=down] (1,4) to (1,6) to (4,6)
(1,0) to (1,2) to [out=up, in=down] (2,4) to (2,6) to (3,6) to (3,2) to [out=down, in=up] (2,0);
\draw [edge] (3,0) to [out=up, in=down] node [mask point,scale=\maskscl] (1) {} (2,2) to [out=up, in=down] node [mask point, scale=\maskscl] (2) {} (1,4) to (1,6);
\cliparoundone{1}{\draw [edge] (2,0) to [out=up, in=down] (3,2) to (3,6);}
\cliparoundone{2}{\draw [edge] (1,0) to (1,2) to [out=up, in=down] (2,4) to (2,6);}
%\tidythetop
\draw [white] (0,6) to (4,6) to (4,0) to (0,0);
\end{tz}
\,+ 
d \mu \,\,
\begin{tz}[scale=0.45, xscale=-1.5, yscale=1,scale=\scl]
\path [use as bounding box] (1,0) rectangle (4,6);
\path [surface, even odd rule] 
(4,0) to (3,0) to [out=up, in=down] (2,2) to [out=up, in=down] (1,4) to (1,6) to (4,6)
(1,0) to (1,2) to [out=up, in=down] (2,4) to [out=up, in=left] (2.5,4.8) to [out=right, in=up] (3,4) to (3,2) to [out=down, in=up] (2,0) 
(2,6) to [out=down, in=left] (2.5,5.2) to [out=right, in=down] (3,6);
\draw [edge] (3,0) to [out=up, in=down] node [mask point,scale=\maskscl] (1) {} (2,2) to [out=up, in=down] node [mask point, scale=\maskscl] (2) {} (1,4) to (1,6);
\cliparoundtwo{1}{2}{\draw [edge] (2,0) to [out=up, in=down] (3,2) to (3,4) to [out=up, in=right] (2.5,4.8) to [out=left, in=up] (2,4) to [out=down, in=up] (1,2) to (1,0);}
\draw [edge] (2,6) to [out=down, in=left] (2.5,5.2) to [out=right, in=down] (3,6);
%\tidythetop
\draw [white] (0,6) to (4,6) to (4,0) to (0,0);
\end{tz} 
\\
\nonumber
\stackrel{!}{=}\sqrt{d}\,\,&
\begin{tz}[scale=0.45, xscale=1.5,scale=\scl]
\path [use as bounding box] (0,0) rectangle (3,6);
\path [surface, even odd rule] (0,0) to (0,6) to (1,6) to [out=down, in=up] (2,4) to [out=down, in=up] (3,2) to (3,0) to (2,0) to [out=up, in=down] (1,2) to (1,4) to [out=up, in=down] (2,6) to (3,6) to (3,4) to [out=down, in=up] (2,2) to [out=down, in=up] (1,0);
\draw [edge] (1,0) to [out=up, in=down] node [mask point,scale=\maskscl] (1) {} (2,2) to [out=up, in=down] node [mask point,scale=\maskscl] (2) {} (3,4) to (3,6);
\draw [edge] (1,4) to [out=up, in=down] node [mask point,scale=\maskscl] (3) {} (2,6);
\cliparoundone{1}{\draw [edge] (2,0) to [out=up, in=down] (1,2) to (1,4);}
\cliparoundtwo{2}{3}{\draw [edge] (3,0) to (3,2) to [out=up, in=down] (2,4) to [out=up, in=down] (1,6);}
\end{tz}
\,\superequals{eq:Potts-Hadamard} \,\sqrt{d}\,\,\overline{\lambda}\,\,
\begin{tz}[scale=0.45, xscale=-1.5, yscale=1,scale=\scl]
\path [use as bounding box] (1,0) rectangle (4,6);
\path [surface, even odd rule] 
(4,0) to (3,0) to [out=up, in=down] (2,2) to [out=up, in=down] (1,4) to (1,6) to (4,6)
(1,0) to (1,2) to [out=up, in=down] (2,4) to [out=up, in=left] (2.5,4.8) to [out=right, in=up] (3,4) to (3,2) to [out=down, in=up] (2,0) 
(2,6) to [out=down, in=left] (2.5,5.2) to [out=right, in=down] (3,6);
\draw [edge] (3,0) to [out=up, in=down] node [mask point,scale=\maskscl] (1) {} (2,2) to [out=up, in=down] node [mask point, scale=\maskscl] (2) {} (1,4) to (1,6);
\cliparoundtwo{1}{2}{\draw [edge] (2,0) to [out=up, in=down] (3,2) to (3,4) to [out=up, in=right] (2.5,4.8) to [out=left, in=up] (2,4) to [out=down, in=up] (1,2) to (1,0);}
\draw [edge] (2,6) to [out=down, in=left] (2.5,5.2) to [out=right, in=down] (3,6);
%\tidythetop
\draw [white] (0,6) to (4,6) to (4,0) to (0,0);
\end{tz} 
\,+ \sqrt{d}\,\,\overline{\mu}\,\,
\begin{tz}[scale=0.45, xscale=-1.5, yscale=1,scale=\scl]
\path [use as bounding box] (1,0) rectangle (4,6);
\path [surface, even odd rule] 
(4,0) to (3,0) to [out=up, in=down] (2,2) to [out=up, in=down] (1,4) to (1,6) to (4,6)
(1,0) to (1,2) to [out=up, in=down] (2,4) to (2,6) to (3,6) to (3,2) to [out=down, in=up] (2,0);
\draw [edge] (3,0) to [out=up, in=down] node [mask point,scale=\maskscl] (1) {} (2,2) to [out=up, in=down] node [mask point, scale=\maskscl] (2) {} (1,4) to (1,6);
\cliparoundone{1}{\draw [edge] (2,0) to [out=up, in=down] (3,2) to (3,6);}
\cliparoundone{2}{\draw [edge] (1,0) to (1,2) to [out=up, in=down] (2,4) to (2,6);}
%\tidythetop
\draw [white] (0,6) to (4,6) to (4,0) to (0,0);
\end{tz}
\end{align}
In short, $\mu = \frac{\overline{\lambda}}{\sqrt{d}}$. Together with the constraints from RII this proves the theorem.
\end{proof}
\thmmetaplectic*
\vspace{-3pt}
\begin{proof}
Note that $\omega$ (defined in \eqref{eq:omega}) and its square root $\lambda$ have modulus one. A proof of this fact using the discrete Fourier transform can be found in \cite[Proposition 2.15]{Jaffe:2016d}. Therefore, $|H_{a,b}| = \frac{1}{\sqrt{d}}$. The matrix $H$ is unitary, since 
\[  \sum_{c=0}^{d-1} H_{a,c} \overline{H}_{b,c} \superequals{eq:defmeta} \frac{1}{d}\sum_{c=0}^{d-1} \xi^{(a-c)^2-(b-c)^2} = \frac{1}{d} \xi^{a^2-b^2}\sum_{c=0}^{d-1} \xi^{2bc-2ac}\superequals{eq:omega}\frac{1}{d} \xi^{a^2-b^2} \sum_{c=0}^{d-1} e^{\frac{2 \pi i}{d} (b-a)c} = \delta_{a,b}.\]
It satisfies \eqref{eq:RIIIindex}, since 
\begin{align}\nonumber \sum_{r=0}^{d-1} \overline{H}_{a,r} H_{b,r} H_{c,r} &\superequals{eq:defmeta} \frac{\overline{\lambda}}{d^{\frac{3}{2}}} \sum_{r=0}^{d-1} \xi^{-(a-r)^2 + (b-r)^2 + (c-r)^2} = \frac{\overline{\lambda}}{d^{\frac{3}{2}}} \xi^{b^2+c^2-a^2}\sum_{r=0}^{d-1} \xi^{r^2+2r(a-b-c)} \\ \label{eq:metaplecticstep}&=\frac{\overline{\lambda}}{d^{\frac{3}{2}}} \xi^{b^2+c^2-a^2 - (a-b-c)^2}\sum_{r=0}^{d-1} \xi^{(r + (a-b-c))^2} =\frac{\overline{\lambda}}{d^{\frac{3}{2}}} \xi^{b^2+c^2-a^2 - (a-b-c)^2}\sum_{r=0}^{d-1} \xi^{r ^2} \\\nonumber &\superequals{eq:omega} \frac{\lambda}{d} \xi^{-2a^2+2ab+2ac -2bc}.   \end{align}
\noindent
The second equality in \eqref{eq:metaplecticstep} holds since $\xi^{d^2} =1$. On the other hand,
\[\sqrt{d} \,\overline{H}_{a,b} \overline{H}_{a,c} H_{b,c} = \frac{\lambda}{d} \xi^{-(a-b)^2-(a-c)^2+(b-c)^2} = \frac{\lambda}{d} \xi^{-2a^2+2ab+2ac-2bc}.\]
Thus, $H$ is a self-transpose Hadamard matrix fulfilling \eqref{eq:RIIIindex}.

\end{proof}
\end{document}

%% file: header-eptcs.tex
\usepackage{amsmath,amsthm,amssymb}
\usepackage{graphicx}    % to include graphics
\usepackage{calc}
\usepackage{docmute}
\usepackage{array}
\usepackage{xspace}
\usepackage[numbers,sort,compress]{natbib}
\usepackage[font={it}]{caption}
\usepackage{bbm}            % for \mathbbm
\usepackage{enumerate}  % for (i) in enumerate
\usepackage{etoolbox}      % programming facilities
\usepackage{hyperref}       % hyperlinks
\usepackage{xstring} % for EqIf
\usepackage{mathtools}
\usepackage{thm-restate}
\usepackage{docmute}
\def\subparagraph{}
\usepackage[compact]{titlesec}
\usepackage{enumitem}
\setlist[itemize]{topsep=0mm, leftmargin=6mm, itemsep=-1.25pt}
\allowdisplaybreaks[1]

\newtheorem{theorem}{Theorem}[section]
\newtheorem{proposition}[theorem]{Proposition}

\theoremstyle{definition}
\newtheorem{definition}[theorem]{Definition}
\newtheorem{varremark}[theorem]{Remark}
\newtheorem{example}[theorem]{Example}

\makeatletter
\renewcommand\Autoref[1]{\@first@ref#1,@}
\def\@throw@dot#1.#2@{#1}% discard everything after the dot
\def\@set@refname#1{%    % set \@refname to autoefname+s using \getrefbykeydefault
    \edef\@tmp{\getrefbykeydefault{#1}{anchor}{}}%
    \def\@refname{\@nameuse{\expandafter\@throw@dot\@tmp.@autorefname}s}%
}
\def\@first@ref#1,#2{%
  \ifx#2@\autoref{#1}\let\@nextref\@gobble% only one ref, revert to normal \autoref
  \else%
    \@set@refname{#1}%  set \@refname to autoref name
    \@refname~\ref{#1}% add autoefname and first reference
    \let\@nextref\@next@ref% push processing to \@next@ref
  \fi%
  \@nextref#2%
}
\def\@next@ref#1,#2{%
   \ifx#2@ and~\ref{#1}\let\@nextref\@gobble% at end: print and+\ref and stop
   \else, \ref{#1}% print  ,+\ref and continue
   \fi%
   \@nextref#2%
}
\makeatother

\usepackage{tikz}

\usetikzlibrary{calc, decorations.markings, decorations.pathmorphing, intersections, shapes, fit, shapes, decorations.pathreplacing}
\pgfdeclarelayer{back}
\pgfdeclarelayer{front}
\pgfsetlayers{back,main,front}
\makeatletter
\pgfkeys{%
  /tikz/on layer/.code={\pgfonlayer{#1}\begingroup\aftergroup\endpgfonlayer\aftergroup\endgroup},
  /tikz/node on layer/.code={\gdef\node@@on@layer{\setbox\tikz@tempbox=\hbox\bgroup\pgfonlayer{#1}\unhbox\tikz@tempbox\endpgfonlayer\egroup}\aftergroup\node@on@layer},
  /tikz/end node on layer/.code={\endpgfonlayer\endgroup\endgroup}
}
\def\node@on@layer{\aftergroup\node@@on@layer}

\newenvironment{tz}
{\begin{aligned}\begin{tikzpicture}}
{\end{tikzpicture}\end{aligned}\,}
\tikzset{blob/.style={draw, circle, fill=white, inner sep=1pt, minimum width=5pt, font=\scriptsize, node on layer=front, thick}}
\tikzset{greenregion/.style={fill=green, fill opacity=0.3, draw=none}}
\tikzset{redregion/.style={fill=red, fill opacity=0.3, draw=none}}
\tikzset{blueregion/.style={fill=blue, fill opacity=0.3, draw=none}}
\tikzset{yellowregion/.style={fill=yellow, fill opacity=0.5, draw=none}}
\tikzset{cyanregion/.style={fill=cyan, fill opacity=0.3, draw=none}}
\tikzset{orangeregion/.style={fill=orange, fill opacity=0.6, draw=none}}
\tikzset{solidgreenregion/.style={fill=green!30, fill opacity=1, draw=none}}
\tikzset{solidredregion/.style={fill=red!30, fill opacity=1, draw=none}}
\tikzset{solidblueregion/.style={fill=blue!30, fill opacity=1, draw=none}}
\tikzset{solidyellowregion/.style={fill=yellow!30, fill opacity=1, draw=none}}
\tikzset{string/.style={line width=0.7pt}}
\tikzset{zig/.style={decoration={zigzag,segment length=3, amplitude=0.5}}}
\tikzset{bnd/.style={draw,string}}   %alternative: draw, string 
\tikzset{projector/.style={circle, draw, font=\scriptsize, inner sep=-5pt, minimum width=0.35cm, string, fill=white}}
\tikzset{dimension/.style={font=\scriptsize, inner sep=1pt}}
\tikzset{edge/.style={draw,thick}}
\tikzset{surface/.style={draw=none, fill=blue!50, fill opacity=0.5}}
\tikzset{classical/.style={draw=none, fill=red!50, fill opacity=0.5}}
\tikzset{background/.style={fill=gray!40, fill opacity=0.4}}
\tikzset{partition/.style={}}
\tikzset{hor/.style={draw, dashed, ultra thin ,opacity=1}}

\makeatletter
\def\calign@preamble{&\hfil\strut@\setboxz@h{\@lign$\m@th\displaystyle{##}$}\ifmeasuring@\savefieldlength@\fi\set@field\hfil\tabskip\alignsep@}
\let\cmeasure@\measure@
\patchcmd\cmeasure@{\divide\@tempcntb\tw@}{}{}{}
\patchcmd\cmeasure@{\divide\@tempcntb\tw@}{}{}{}
\patchcmd\cmeasure@{\ifodd\maxfields@\global\advance\maxfields@\@ne\fi}{}{}{}
\newenvironment{calign}{\let\align@preamble\calign@preamble\let\measure@\cmeasure@\align}{\endalign}
\makeatother

\makeatletter
\let\oldmarginpar\marginpar
\renewcommand*{\marginpar}[1]{%
   \begingroup%
     \if@firstcolumn\else\reversemarginpar\fi
   \oldmarginpar{#1}%
   \endgroup%
}
\makeatother
\newcounter{jvcommcounter}
\newcounter{drcommcounter}
\renewcommand{\-}[0]{\nobreakdash-\hspace{0pt}}
\newcommand{\Tr}{\mathrm{Tr}}

\newcommand\cat[1]{\ensuremath{\mathbf{#1}}}

\newcommand\C{\ensuremath{\mathbb{C}}\xspace}

\newcommand\superequals[1]{\stackrel {\makebox[0pt]{\tiny\eqref{#1}}} =}

\newcommand{\ket}[1]{\left|#1\right\rangle}
\newcommand{\bra}[1]{\left\langle#1\right|}

\newcommand\N{\mathbb{N}}
\newcommand\vc[1]{\begin{tabular}{ccccccc}#1\end{tabular}}

\allowdisplaybreaks[1]
\newcommand\ignore[1]{}

\def\maskpointheight{8pt}
\tikzset{mask point/.style={transform shape, sloped, 
minimum width=15pt, minimum height=\maskpointheight, inner sep=0pt, ultra thin, font=\tiny}}
\def\fixboundingbox{\path [use as bounding box] (current bounding box.south west) rectangle (current bounding box.north east);}
\tikzset{
clip even odd rule/.code={\pgfseteorule},
invclip/.style={clip,insert path=[clip even odd rule]{
   [reset cm](-\maxdimen,-\maxdimen)rectangle(\maxdimen,\maxdimen)
    }}} 
\newcommand\clipintersection[1]{(#1.north west) -- (#1.north east) -- (#1.south east) -- (#1.south west) -- (#1.north west)}

\newcommand\cliparoundthree[4]{\begin{pgfscope}\begin{scope}[overlay]
    \path [invclip] \clipintersection{#1} -- \clipintersection{#2}--\clipintersection{#3} -- (#1.north west);#4
\end{scope}\end{pgfscope}}  
\newcommand\cliparoundone[2]{\begin{pgfscope}\begin{scope}[overlay]
\path[invclip] \clipintersection{#1};#2
\end{scope}\end{pgfscope}}
\newcommand\cliparoundtwo[3]{\begin{pgfscope}\begin{scope}[overlay]
    \path [invclip] \clipintersection{#1}  \clipintersection{#2};#3
\end{scope}\end{pgfscope}}  
\newcommand\cliparoundfour[5]{\begin{pgfscope}\begin{scope}[overlay]
    \path [invclip] \clipintersection{#1} \clipintersection{#2} \clipintersection{#3} \clipintersection{#4};#5
\end{scope}\end{pgfscope}}   
\newcommand\cliparoundfive[6]{\begin{pgfscope}\begin{scope}[overlay]
    \path [invclip] \clipintersection{#1} \clipintersection{#2} \clipintersection{#3} \clipintersection{#4}\clipintersection{#5};#6
\end{scope}\end{pgfscope}}  

\usepackage[justification=centering]{caption}
\newcommand \GHZ{\ensuremath{\mathrm{GHZ}}}
\newcommand \cluster{\ensuremath{\mathrm{C}}}
\renewcommand\paragraph[1]{
  
\vspace{2pt}\noindent
\textbf{#1}}
\newcommand\firstparagraph[1]{\vspace{-2pt}\noindent
\textbf{#1}}
\newcommand\axiom{Reidemeister moves.}
\newcommand\requiresRIII{\paragraph{\axiom} Requires the extended calculus.}
\newcommand\requiresRII{\paragraph{\axiom} Requires the basic calculus only.}

\newcommand\circlenumber[1]{\ensuremath{\tikz{\node [circle, draw, inner sep=-1pt, minimum width=7pt, font=\scriptsize, fill=white] at (0,0) {#1};}}}

\allowdisplaybreaks 
\def\figuretopsuck{\vspace{-14pt}}
\def\figurecaptionsuck{\vspace{-7pt}}
\newcommand\bigforall{\mbox{\huge $\mathsurround0pt\forall$}}

\makeatletter
\def\slashedarrowfill@#1#2#3#4#5{%
  $\m@th\thickmuskip0mu\medmuskip\thickmuskip\thinmuskip\thickmuskip
\relax#5#1\mkern-7mu%
\cleaders\hbox{$#5\mkern-2mu#2\mkern-2mu$}\hfill
\mathclap{#3}\mathclap{#2}%
\cleaders\hbox{$#5\mkern-2mu#2\mkern-2mu$}\hfill
\mkern-7mu#4$%
}
\def\rightslashedarrowfill@{%
  \slashedarrowfill@\relbar\relbar\mapstochar\rightarrow}
\newcommand\xslashedrightarrow[2][]{%
  \ext@arrow 0055{\rightslashedarrowfill@}{#1}{#2}}
\def\Rightslashedarrowfill@{%
  \slashedarrowfill@\relbar\relbar\mapstochar\Rightarrow}
\newcommand\xslashedRightarrow[2][]{%
  \ext@arrow 0055{\Rightslashedarrowfill@}{#1}{#2}}
\makeatother

\renewcommand{\to}[1][]{\ensuremath{\xrightarrow{#1}}}

\tikzset{programlabel/.style={anchor=west, align=left, scale=0.75}}
\tikzset{protocoldiagram/.style={yscale=0.7, xscale=0.7}}
\allowdisplaybreaks

\def\tidythetop{\path [use as bounding box] (current bounding box.north west) rectangle (current bounding box.south east);
\draw [white, ultra thick] ([xshift=-5pt] current bounding box.north west) to ([xshift=5pt] current bounding box.north east);
}

%% file: fig-Reidemeister-eptcs.tex
% 
%\documentclass[10pt,conference]{IEEEtran}

%\input{header}
%\begin{document}

%aergaerg aergaerg aergaerg eargaerg aergaerg aergaerg aeraerg aergaerg aeaeg aergaerg aergaerg aergaerg aergaerg aergaerg aergaer aeraer aeraerg aeraerg aergaerg aer

\beginfig
\figuretopsuck
\figuretopsuck
\tikzset{every picture/.style={scale=0.7}}
\begin{align}
\nonumber
&
\def\lscl{0.7}
\def\lxscl{0.8}
\text{\textit{(a)} }
\begin{tz}[string, scale=1, scale=\lscl, xscale=-\lxscl]
\path [use as bounding box] (-0.5,0) rectangle +(3,2.);
\path[blueregion] (1.75,0) to [out=90, in=down] (0.25,2) to (-0.5,2) to (-0.5,0) to (0.25,0) to [out=up, in=down] (1.75,2) to (2.5,2) to (2.5,0);
\draw[edge] (0.25,0) to [out=up, in=down] node [mask point] (1) {} (1.75,2);
\cliparoundone{1}{\draw[edge] (1.75,0) to [out=up, in=-90] (0.25,2);}
\end{tz}
=
\def\dx{0.1}
\def\dy{-0.2}
\def\l{0.77}
\def\maskscl{1.4}
\begin{tz}[scale=0.666, scale=\lscl, xscale=\lxscl]
%\draw (0,0) grid (5,3);
\draw [surface] (1,3) to [out=down, in=left, in looseness=\l] (1.5+\dx,0.5-\dy) to [out=right, in=left] (3.5-\dx,2.5+\dy) to [out=right, in=up, out looseness=\l] (4,0) to (5,0) to (5,3) to (2,3) to [out=down, in=up] (3,0) to (0,0) to (0,3);
\draw [edge] (1,3) to [out=down, in=left, in looseness=\l] (1.5+\dx,0.5-\dy) to [out=right, in=left] node [mask point,scale=\maskscl] (1) {} (3.5-\dx,2.5+\dy) to [out=right, in=up, out looseness=\l] (4,0);
\cliparoundone{1}{\draw [edge] (2,3) to [out=down, in=up] (3,0);}
\end{tz}
&&
\textit{(b) }
\def\maskscl{0.65}%
\begin{tz}
%\path [use as bounding box, red, ultra thick] (0,0) rectangle (3,2);
\draw [surface] (0,0) to [out=up, in=down] (1,1) to [out=up, in=down] (0,2) to (1,2) to [out=down, in=up] (0,1) to [out=down, in=up] (1,0);
\draw [edge] (0,0) to [out=up, in=down] node [mask point,scale=\maskscl] (1) {} (1,1) to [out=up, in=down] node [mask point,scale=\maskscl] (2) {} (0,2);
\cliparoundtwo{1}{2}{\draw [edge] (1,2) to [out=down, in=up] (0,1) to [out=down, in=up] (1,0);}
\end{tz}
=
\begin{tz}
\draw [surface] (0,0) rectangle (1,2);
\draw [edge] (0,0) to (0,2);
\draw [edge] (1,0) to (1,2);
\end{tz}
=
\begin{tz}[xscale=-1]
%\path [use as bounding box, red, ultra thick] (0,0) rectangle (3,2);
\draw [surface] (0,0) to [out=up, in=down] (1,1) to [out=up, in=down] (0,2) to (1,2) to [out=down, in=up] (0,1) to [out=down, in=up] (1,0);
\draw [edge] (0,0) to [out=up, in=down] node [mask point,scale=\maskscl] (1) {} (1,1) to [out=up, in=down] node [mask point,scale=\maskscl] (2) {} (0,2);
\cliparoundtwo{1}{2}{\draw [edge] (1,2) to [out=down, in=up] (0,1) to [out=down, in=up] (1,0);}
\end{tz}
&&
\textit{(c) }
\def\lscl{0.7}
\def\lxscl{0.6}
\def\maskscl{1.2}
\begin{tz}[scale=\lscl, xscale=\lxscl]
\path [use as bounding box, red, ultra thick] (0,0) rectangle (3,2);
\draw [surface]
    (0,0) to [out=up, in=up, looseness=1.7] (3,0)
    (0,2) to [out=down, in=down, looseness=1.7] (3,2);
\draw [edge] (0,0) to [out=up, in=up, looseness=1.7] node [mask point,scale=\maskscl, pos=00.21] (1) {} node [mask point, pos=0.79,scale=\maskscl] (2) {} (3,0);
\cliparoundtwo{1}{2}{\draw [edge] (0,2) to [out=down, in=down, looseness=1.7] (3,2);}
\end{tz}
=
\displaystyle\frac 1 {|S|}
\begin{tz}[scale=\lscl, xscale=\lxscl]
\path [use as bounding box, red, ultra thick] (0,0) rectangle (3,2);
\draw [surface]
    (0,0) to [out=up, in=up, looseness=0.9] (3,0)
    (0,2) to [out=down, in=down, looseness=0.9] (3,2);
\draw [edge] (0,0) to [out=up, in=up, looseness=0.9] (3,0);
\draw [edge] (0,2) to [out=down, in=down, looseness=0.9] (3,2);
\end{tz}
=
\begin{tz}[scale=\lscl, xscale=\lxscl]
\path [use as bounding box, red, ultra thick] (0,0) rectangle (3,2);
\draw [surface]
    (0,0) to [out=up, in=up, looseness=1.7] (3,0)
    (0,2) to [out=down, in=down, looseness=1.7] (3,2);
\draw [edge] (0,2) to [out=down, in=down, looseness=1.7] node [mask point, pos=00.21,scale=\maskscl] (1) {} node [mask point, pos=0.79,scale=\maskscl] (2) {} (3,2);
\cliparoundtwo{1}{2}{\draw [edge] (0,0) to [out=up, in=up, looseness=1.7] (3,0);}
\end{tz}
\\[-5pt]\nonumber
&
\textit{(d) }
\def\dx{0.1}
\def\dy{-0.2}
\def\l{0.77}
\def\lxscl{0.7}
\def\lyscl{0.8}
\def\maskscl{1.3}
\begin{tz}[scale=0.66, xscale=\lxscl, yscale=\lyscl]
%\draw (0,0) grid (5,3);
\draw [surface] (1,3) to [out=down, in=left, in looseness=\l] (1.5+\dx,0.5-\dy) to [out=right, in=left] (3.5-\dx,2.5+\dy) to [out=right, in=up, out looseness=\l] (4,0) to (5,0) to (5,3) to (2,3) to [out=down, in=up] (3,0) to (0,0) to (0,3);
\draw [edge] (1,3) to [out=down, in=left, in looseness=\l] (1.5+\dx,0.5-\dy) to [out=right, in=left] node [mask point,scale=\maskscl] (1) {} (3.5-\dx,2.5+\dy) to [out=right, in=up, out looseness=\l] (4,0);
\cliparoundone{1}{\draw [edge] (2,3) to [out=down, in=up] (3,0);}
\end{tz}
=
\begin{tz}[scale=0.66, yscale=-1, xscale=\lxscl, yscale=\lyscl]
%\draw (0,0) grid (5,3);
\draw [surface] (1,3) to [out=down, in=left, in looseness=\l] (1.5+\dx,0.5-\dy) to [out=right, in=left] (3.5-\dx,2.5+\dy) to [out=right, in=up, out looseness=\l] (4,0) to (5,0) to (5,3) to (2,3) to [out=down, in=up] (3,0) to (0,0) to (0,3);
\draw [edge] (2,3) to [out=down, in=up] node [mask point,scale=\maskscl] (1) {} (3,0);
\cliparoundone{1}{\draw [edge] (1,3) to [out=down, in=left, in looseness=\l] (1.5+\dx,0.5-\dy) to [out=right, in=left] (3.5-\dx,2.5+\dy) to [out=right, in=up, out looseness=\l] (4,0);}
\end{tz}
\def\maskscl{0.5}%
&&
\textit{(e) }
\begin{tz}[xscale=-1, scale=1]
\path [use as bounding box] (0,1.5) rectangle +(1,-1.9);
\draw [surface] (0,1) to [out=down, in=up] (1,0) to [out=down, in=down] (0,0) to [out=up, in=down] (1,1);
\draw [edge] (0,1) to [out=down, in=up] node (1) [mask point] {} (1,0);
\cliparoundone{1}{\draw [edge] (1,0) to [out=down, in=down] (0,0) to [out=up, in=down] (1,1);}
\end{tz}
=
\lambda\,\begin{tz}
\path [use as bounding box] (0,0.5) rectangle +(1,-1.9);
\draw [surface, edge] (0,0) to [out=down, in=down, looseness=3] (1,0);
\end{tz}
&&
\textit{(f) }
\def\maskscl{1.6}%
\begin{tz}[scale=0.35, xscale=1.5, yscale=-1]
\path [use as bounding box] (0,0) rectangle (3,6);
\path [surface, even odd rule] (4,0) to (4,6) to (3,6) to [out=down, in=up] (2,4) to [out=down, in=up] (1,2) to (1,0) to (2,0) to [out=up, in=down] (3,2) to (3,4) to [out=up, in=down] (2,6) to (1,6) to (1,4) to [out=down, in=up] (2,2) to [out=down, in=up] (3,0)
 (0,0) to (0,6) to (4,6) to (4,0);
\draw [edge] (3,0) to [out=up, in=down] node [mask point,scale=\maskscl] (1) {} (2,2) to [out=up, in=down] node [mask point, scale=\maskscl] (2) {} (1,4) to (1,6);
\draw [edge] (3,4) to [out=up, in=down] node [mask point, scale=\maskscl] (3) {} (2,6);
\cliparoundone{1}{\draw [edge] (2,0) to [out=up, in=down] (3,2) to (3,4);}
\cliparoundtwo{2}{3}{\draw [edge] (1,0) to (1,2) to [out=up, in=down] (2,4) to [out=up, in=down] (3,6);}
%\tidythetop
\draw [white] (0,6) to (4,6) to (4,0) to (0,0);
\end{tz}
=\sqrt{|S|}\,
\begin{tz}[scale=0.35, xscale=1.5]
\path [use as bounding box] (0,0) rectangle (3,6);
\path [surface, even odd rule] (0,0) to (0,6) to (1,6) to [out=down, in=up] (2,4) to [out=down, in=up] (3,2) to (3,0) to (2,0) to [out=up, in=down] (1,2) to (1,4) to [out=up, in=down] (2,6) to (3,6) to (3,4) to [out=down, in=up] (2,2) to [out=down, in=up] (1,0);
\draw [edge] (1,0) to [out=up, in=down] node [mask point,scale=\maskscl] (1) {} (2,2) to [out=up, in=down] node [mask point,scale=\maskscl] (2) {} (3,4) to (3,6);
\draw [edge] (1,4) to [out=up, in=down] node [mask point,scale=\maskscl] (3) {} (2,6);
\cliparoundone{1}{\draw [edge] (2,0) to [out=up, in=down] (1,2) to (1,4);}
\cliparoundtwo{2}{3}{\draw [edge] (3,0) to (3,2) to [out=up, in=down] (2,4) to [out=up, in=down] (1,6);}
\end{tz}
\end{align}

\figurecaptionsuck
\caption{The shaded tangle calculus.\label{fig:reidemeister}}
\figurecaptionpostsuck
\end{figure}

%\end{document}